


\font\titlefont = cmr10 scaled\magstep 4
\font\sectionfont = cmr10
\font\littlefont = cmr5
\font\eightrm = cmr8

\def\ss{\scriptstyle}
\def\sss{\scriptscriptstyle}

\newcount\tcflag
\tcflag = 0  
\ifnum\tcflag = 0 \magnification = 1200 \fi  

\global\baselineskip = 1.2\baselineskip
\global\parskip = 4pt plus 0.3pt
\global\abovedisplayskip = 18pt plus3pt minus9pt
\global\belowdisplayskip = 18pt plus3pt minus9pt
\global\abovedisplayshortskip = 6pt plus3pt
\global\belowdisplayshortskip = 6pt plus3pt

\def\barsoff{\overfullrule=0pt}


\def\endignore{}
\def\ignore #1\endignore{}

\newcount\dflag
\dflag = 0


\def\monthname{\ifcase\month
\or January \or February \or March \or April \or May \or June%
\or July \or August \or September \or October \or November \or December
name \fi}

\newcount\dummy
\newcount\minute  
\newcount\hour
\newcount\localtime
\newcount\localday
\localtime = \time
\localday = \day

\def\advanceclock#1#2{ 
\dummy = #1
\multiply\dummy by 60
\advance\dummy by #2
\advance\localtime by \dummy
\ifnum\localtime > 1440 
\advance\localtime by -1440
\advance\localday by 1
\fi}

\def\settime{{\dummy = \localtime%
\divide\dummy by 60%
\hour = \dummy
\minute = \localtime%
\multiply\dummy by 60%
\advance\minute by -\dummy
\ifnum\minute < 10
\xdef\spacer{0} 
\else \xdef\spacer{}
\fi %
\ifnum\hour < 12
\xdef\ampm{a.m.} 
\else
\xdef\ampm{p.m.} 
\advance\hour by -12 %
\fi %
\ifnum\hour = 0 \hour = 12 \fi
\xdef\timestring{\number\hour : \spacer \number\minute%
\thinspace \ampm}}}



\def\endtitle{}
\def\title#1\endtitle{\vskip.5in\titlefont
\global\baselineskip = 2\baselineskip
#1\vskip.4in
\baselineskip = 0.5\baselineskip\rm}

\def\endauthors{}
\def\authors#1\endauthors{#1}

\def\endabstract{}
\def\abstract#1\endabstract{\vskip .3in%
\centerline{\sectionfont\bf Abstract}%
\vskip .1in
\noindent#1}

\newcount\nsection
\newcount\nsubsection

\def\section#1{\global\advance\nsection by 1
\nsubsection=0
\bigskip\noindent\centerline{\sectionfont \bf \number\nsection.\ #1}
\bigskip\rm\nobreak}

\def\subsection#1{\global\advance\nsubsection by 1
\bigskip\noindent\sectionfont \sl \number\nsection.\number\nsubsection)\
#1\bigskip\rm\nobreak}

\def\topic#1{{\medskip\noindent $\bullet$ \it #1:}}
\def\endtopic{\medskip}

\def\appendix#1#2{\bigskip\noindent%
\centerline{\sectionfont \bf Appendix #1.\ #2}
\bigskip\rm\nobreak}


\newcount\nref
\global\nref = 1

\def\ref#1#2{\xdef #1{[\number\nref]}
\ifnum\nref = 1\global\xdef\therefs{\noindent[\number\nref] #2\ }
\else
\global\xdef\oldrefs{\therefs}
\global\xdef\therefs{\oldrefs\vskip.1in\noindent[\number\nref] #2\ }%
\fi%
\global\advance\nref by 1
}

\def\listrefs{\vfill\eject\section{References}\therefs}


\newcount\nfoot
\global\nfoot = 1

\def\foot#1#2{\xdef #1{(\number\nfoot)}
\footnote{${}^{\number\nfoot}$}{\eightrm #2}
\global\advance\nfoot by 1
}


\newcount\nfig
\global\nfig = 1

\def\fig#1{\xdef #1{(\number\nfig)}
\global\advance\nfig by 1
}


\newcount\cflag
\newcount\nequation
\global\nequation = 1
\def\eqlabel{(1)}

\def\nexteqno{\ifnum\cflag = 0
\global\advance\nequation by 1
\fi
\global\cflag = 0
\xdef\eqlabel{(\number\nequation)}}

\def\lasteqno{\global\advance\nequation by -1
\xdef\eqlabel{(\number\nequation)}}

\def\label#1{\xdef #1{(\number\nequation)}
\ifnum\dflag = 1
{\escapechar = -1
\xdef\draftname{\littlefont\string#1}}
\fi}

\def\clabel#1#2{\xdef\eqlabel{(\number\nequation #2)}
\global\cflag = 1
\xdef #1{\eqlabel}
\ifnum\dflag = 1
{\escapechar = -1
\xdef\draftname{\string#1}}
\fi}

\def\cclabel#1#2{\xdef\eqlabel{#2)}
\global\cflag = 1
\xdef #1{\eqlabel}
\ifnum\dflag = 1
{\escapechar = -1
\xdef\draftname{\string#1}}
\fi}


\def\eeq{}

\def\eqnn #1\eeq{$$ #1 $$}

\def\eq #1\eeq{
\ifnum\dflag = 0
{\xdef\draftname{\ }}
\fi 
$$ #1
\eqno{\eqlabel \rlap{\ \draftname}} $$
\nexteqno}



\def\eol{& \eqlabel \rlap{\ \draftname} \crcr
\nexteqno
\xdef\draftname{\ }}

\def\eeol{& \eqlabel \rlap{\ \draftname}
\nexteqno
\xdef\draftname{\ }}

\def\eolnn{\cr
\global\cflag = 0
\xdef\draftname{\ }}

\def\eeolnn{\xdef\draftname{\ }}

\def\eqa #1\eeq{
\ifnum\dflag = 0
{\xdef\draftname{\ }}
\fi 
$$ \eqalignno{ #1 } $$
\global\cflag = 0}


\def\ie{{\it i.e.\/}}
\def\eg{{\it e.g.\/}}
\def\etc{{\it etc.\/}}
\def\etal{{\it et.al.\/}}

\def\cf{{\it c.f.\/}}


\def\anp#1#2#3{{\it Ann.\ Phys. (NY)} {\bf #1} (19#2) #3}

\def\jetpl#1#2#3#4#5#6{{\it Pis'ma Zh.\ Eksp.\ Teor.\ Fiz.} {\bf #1} (19#2) #3
[{\it JETP Lett.} {\bf #4} (19#5) #6]}

\def\npb#1#2#3{{\it Nucl.\ Phys.} {\bf B#1} (19#2) #3}
\def\plb#1#2#3{{\it Phys.\ Lett.} {\bf #1B} (19#2) #3}
\def\pla#1#2#3{{\it Phys.\ Lett.} {\bf #1A} (19#2) #3}

\def\prd#1#2#3{{\it Phys.\ Rev.} {\bf D#1} (19#2) #3}
\def\pr#1#2#3{{\it Phys.\ Rev.} {\bf #1} (19#2) #3}
\def\prep#1#2#3{{\it Phys.\ Rep.} {\bf C#1} (19#2) #3}
\def\prl#1#2#3{{\it Phys.\ Rev.\ Lett.} {\bf #1} (19#2) #3}

\def\zpc#1#2#3{{\it Zeit.\ Phys.} {\bf C#1} (19#2) #3}


\global\nulldelimiterspace = 0pt



\def\frac#1#2{{{#1} \over {#2}}\,}  
\def\hf{{1\over 2}}



\def\Dsl{\hbox{/\kern-.6700em\it D}} 
\def\dsl{\hbox{/\kern-.5300em$\partial$}}
\def\pxpsl{\hbox{/\kern-.5600em$p$}}
\def\ssl{\hbox{/\kern-.5300em$s$}}
\def\epssl{\hbox{/\kern-.5100em$\epsilon$}}
\def\delsl{\hbox{/\kern-.6300em$\nabla$}}
\def\lxpsl{\hbox{/\kern-.4300em$l$}}
\def\elxpsl{\hbox{/\kern-.4500em$\ell$}}
\def\kxpsl{\hbox{/\kern-.5100em$k$}}
\def\qxpsl{\hbox{/\kern-.5000em$q$}}
\def\sla#1{\raise.15ex\hbox{$/$}\kern-.57em #1}
\def\Pl{\gamma_{\sss L}}
\def\Pr{\gamma_{\sss R}}



\def\twi{\widetilde}

\def\roughly#1{\mathrel{\raise.3ex\hbox{$#1$\kern-.75em\lower1ex\hbox{$\sim$}}}}
\def\lsim{\roughly<}

\def\tw#1{\tilde{#1}}
\def\ol#1{\overline{#1}}





\def\Scl{{\cal L}}

\def\Sco{{\cal O}}


\def\ssa{{\sss A}}
\def\ssb{{\sss B}}

\def\ssl{{\sss L}}

\def\ssv{{\sss V}}
\def\ssw{{\sss W}}

\def\ssz{{\sss Z}}


\def\Re{{\rm Re\;}}
\def\Im{{\rm Im\;}}
\def\diag#1{{\rm diag}\left( #1 \right)}





\def\cc{{\rm c.c.}}


\def\ecm{{\it e}{\hbox{\rm -cm}}}

\barsoff


\def\ss{\scriptstyle}
\def\ssw{{\sss W}}
\def\ssz{{\sss Z}}
\def\ssv{{\sss V}}
\def\ssa{{\sss A}}
\def\SM{{\sss SM}}
\def\qw{Q_\ssw}
\def\tmw{\tw{m}_\ssw}
\def\tmz{\tw{m}_\ssz}

\def\em{{\rm em}}
\def\Mw{M_\ssw}
\def\Mz{M_{\sss Z}}
\def\mw{m_\ssw}
\def\mz{m_{\sss Z}}
\def\gf{G_{\sss F}}
\def\gsm{g^\SM}
\def\hsm{h^\SM}
\def\rht{{\sss R}}
\def\lft{{\sss L}}
\def\gwk{$SU_L(2) \times U_Y(1)$}
\def\lsm{\Scl_\SM}
\def\leff{\Scl_{\rm eff}}
\def\lnew{\Scl_{\rm new}}
\def\lhat{\hat{\Scl}}
\def\lnewht{\hat{\Scl}_{\rm new}}
\def\twe{\tw{e}}
\def\twg{\tw{g}}
\def\twh{\tw{h}}
\def\twV{\twi{V}}
\def\twm{\tw{m}}
\def\tws{\tw{s}_w}
\def\twc{\tw{c}_w}
\def\sw{s_w}
\def\cw{c_w}
\def\Fhat{\hat{F}}
\def\Ahat{\hat{A}}
\def\fhat{\hat{f}}
\def\What{\hat{W}}
\def\Zhat{\hat{Z}}

\def\clisq{\left(c_\lft^i\right)^2}
\def\srisq{\left(s_\rht^i\right)^2}

\def\hbar{\hbox{\raise4pt\hbox{-}\kern-.2700em h}}

\def\CP{$CP$}


\rightline{McGill-93/12, NEIPH-93-008, OCIP/C-93-6, UQAM-PHE-93/08,
UdeM-LPN-TH-93-155}
\rightline{November 1993.}
\vskip -.2in

\title
\centerline{MODEL-INDEPENDENT GLOBAL}
\centerline{CONSTRAINTS ON NEW PHYSICS}
\endtitle

\authors
\centerline{C.P. Burgess,${}^a$\footnote{*}{\eightrm Permanent Address:
Physics Department, McGill University, 3600 University St., Montr\'eal,
Qu\'ebec,  Canada, H3A 2T8.}
Stephen Godfrey,${}^b$ Heinz K\"onig,${}^c$ David London${}^d$ and Ivan
Maksymyk${}^d$}
\vskip .1in
\centerline{\it ${}^a$ Institut de Physique, l'Universit\'e de Neuch\^atel}
\centerline{\it 1 Rue A.L. Breguet, CH-2000 Neuch\^atel, Switzerland.}
\vskip .05in
\centerline{\it ${}^b$Ottawa-Carleton Institute for Physics, Physics
Department}
\centerline{\it Carleton University, Ottawa, Ontario, Canada, K1S 5B6.}
\vskip .05in
\centerline{\it ${}^c$ D\'epartement de  Physique, l'Universit\'e du
Qu\'ebec \`a Montr\'eal}
\centerline{\it C.P. 8888, Succ. A, Montr\'eal, Qu\'ebec, Canada, H3C 3P8.}
\vskip .05in
\centerline{\it ${}^d$ Laboratoire de Physique Nucl\'eaire, l'Universit\'e
de Montr\'eal}
\centerline{\it C.P. 6128, Montr\'eal, Qu\'ebec, Canada, H3C 3J7.}
\endauthors

\abstract
Using effective-lagrangian techniques we perform a systematic survey of
the lowest-dimension effective interactions through which heavy physics
might manifest itself in present experiments. We do not restrict ourselves
to special classes of effective interactions (such as `oblique'
corrections). We compute the effects of these operators on all currently
well-measured electroweak observables, both at low energies and at the $Z$
resonance, and perform a global fit to their coefficients. Despite the fact
that a great many operators arise in our survey, we find that most are
quite strongly bounded by the current data. We use our survey to
systematically identify those effective interactions which are {\it not}
well-bounded by the data -- these could very well include large new-physics
contributions. Our results may also be used to efficiently confront
specific models for new physics with the data, as we illustrate with an
example.
\endabstract


\vfill\eject
\section{Introduction}

Where is all the new physics? This, in a nutshell, has become the burning
question on most theorists' lips as experimental results from the 100 GeV
scale have poured in from LEP, SLC and the Tevatron. The higher the
precision of the experiments being performed, the better seems the
agreement with the standard electroweak model. And yet we know that
something new --- perhaps only the standard model Higgs --- must almost
certainly be found at or below several (tens of?) TeV, since at this scale
our description would otherwise fundamentally break down.

If, as now seems quite likely, any new particles are quite massive compared
to the electroweak gauge bosons, then their first observable effects can
still be sought through the virtual contributions they make to physics at
lower, but presently accessible, energies. While we wait for the
construction of accelerators powerful enough to directly produce these new
particles, theorists can usefully spend their time understanding where the
comparatively rare virtual contributions can be expected to take place. It
is particularly useful to be able to contrast the detailed predictions of
specific models for the physics at high energies with the more
model-independent predictions which can be obtained from an
effective-lagrangian viewpoint.

An effective lagrangian parametrizes in as model-independent a way as
possible the low-energy implications of new physics at a much higher scale,
$M$. This is done by constructing the most general set of effective
interactions that are consistent with the known low-energy particle content
and symmetries, and which can arise to a given order in $1/M$. The main
goal of an effective-lagrangian analysis is: ($i$) to determine how large
the effective couplings can be without contradicting existing experimental
information, and ($ii$) to find where to most fruitfully search for the
resulting interactions in future experiments.

This type of search for new physics using effective lagrangians has been
performed in the past, but has tended to be relatively limited in its
scope. Traditionally, either the implications of a single type of effective
interaction (such as an electric or chromoelectric dipole moment), or a
fairly small class of such operators (\eg\ anomalous gauge-boson
interactions), have been considered. The disadvantage of limiting the
investigation to a very few operators is that realistic models of the new
physics which underlies the effective lagrangian typically generate a host
of effective operators rather than just a few, and their effects for
well-measured observables can be correlated, or even cancel.
\ref\stu{M.E. Peskin and T. Takeuchi, \prl{65}{90}{964};
W.J. Marciano and J.L. Rosner, \prl{65}{90}{2963};
D.C. Kennedy and P. Langacker, \prl{65}{90}{2967}.}
\ref\peskin{M.E. Peskin and T. Takeuchi, \prd{46}{92}{381}.}
\ref\oblique{B. Lynn, M. Peskin, R. Stuart, in {\it Physics at LEP}, CERN
Report 86-02.}
Recent analyses \stu, \peskin\ of the implications of new physics for the
gauge-boson self-energies --- the so-called `oblique' corrections \oblique\
--- may also be viewed in this way since they can be described
\ref\holdom{M. Golden and L. Randall, \npb{361}{91}{3};
B. Holdom, \plb{259}{91}{329}.}
\holdom\ in an effective-lagrangian language in terms of a three-parameter
class of effective gauge-boson self-interactions. Although these latter
analyses have the virtue of considering the most general effective
interactions that might be generated by a given type of TeV-scale physics,
they are nevertheless limited in the scope of underlying models that they
can encompass by the very restriction to only oblique corrections.

In the present paper we wish to extend the confrontation of potential new
physics with the present electroweak data in a more comprehensive and more
systematic way, by analysing the data in terms of a much broader class of
effective interactions than has previously been considered. More
specifically, we consider all possible effective interactions which satisfy
the following three criteria:
\topic{1}
Since we wish to analyse the implications of the present data, we restrict
ourselves to effective interactions which involve only particles which have
already been observed. In particular, we do not assume the existence of a
light Higgs boson. For simplicity we do not consider operators involving
gluons, although their inclusion into our formalism is conceptually
straightforward.
\topic{2}
We work up to operator dimension five. That is to say, our effective
operators must have dimension (mass)$^d$, with $d\le 5$. We consider both
CP-preserving and CP-violating operators.
\topic{3}
We consider only effective interactions which contribute at tree level to
presently-measured observables.
\endtopic
In practice this means that we include all possible operators of dimension
$\le 5$ with the exception of anomalous three- and four-point electroweak
boson self-interactions, or interactions involving two fermions and two
electroweak bosons. Despite condition ({\it 3}) above, we do not ignore
loop-generated bounds completely, however. This is because we do consider
constraints on our list of operators which arise from their one-loop
contributions to particularly well-measured observables. (We give a more
precise justification of which observables are considered in the
appropriate sections.)

We present here explicit expressions for a wide class of observables in
terms of the couplings of these operators, and systematically constrain
their coefficients from the present data. Our results include as special
cases some previous analyses, and our formulae reduce to these in the
appropriate limits.

Although our results agree with previous workers in the cases of overlap,
we believe we have streamlined some of the technical details of the
calculations in comparison with the procedure of some other authors. Our
main improvement lies in our treatment of the new physics contributions to
measured quantities, particularly as regards how the standard model (SM)
predictions are altered due to the changes induced in the numerical values
that are inferred for the reference input parameters --- such as $\alpha$,
$\Mz$, or $\gf$. We perform this adjustment once and for all directly {\it
in the lagrangian}, thereby obviating the need to separately adjust each
observable as it is considered. In this way we dispose, at the outset, of
many terms which ultimately obscurely cancel in physical predictions in
many treatments.

We find that even with the above assumptions we must deal with a large
number of new-physics operators, of which many contribute to
flavour-changing neutral currents. Our formalism is sufficiently powerful
to deal with all of these. Surprisingly, however, we are still able to
meaningfully constrain the sizes of most of these operators by performing a
global fit to all charged- and neutral-current data. Our aim in doing such
an analysis is twofold. First, by considering all interactions, one may
discover that certain operators remain poorly constrained by current data.
Their effects might well be large, if only experiments would look for them.
We will, in fact, present several examples of such operators.

Our second purpose is to present a comprehensive set of constraints that
must be satisfied by all physics beyond the standard model. Any
model-builder has simply to compute the coefficients of these effective
operators in terms of the parameters of the model, and the bounds on these
coefficients can be obtained from our analysis. Of course, we have taken a
particularly conservative approach -- any reasonable model will have far
fewer parameters than we have operators, so the true constraints on that
model will in general be stronger than those presented here.

We illustrate the simplicity and power of our formalism by using it to
constrain a class of models which has been elsewhere directly fit to the
data. This example serves two purposes. Besides providing an illustration
of the comparative ease of performing the analysis with our general
formalism, we can also see how much weaker our bounds are than those that
are found with a direct fit to the parameters of the underlying model. We
find that although our approach leads to more conservative constraints on
these parameters, as it must, the limits we obtain are not much weaker than
those of the direct fit. Thus, for the models we consider, little
information is lost by the much simpler procedure of directly using the
analysis which we provide in this paper.

We organize our presentation in the following way. In the next section we
first illustrate our technique by reproducing the familiar oblique
correction analysis. We do so partly in order to demonstrate the simplicity
of our approach, but also as a vehicle for explaining the logic of our
analysis in this simplest possible case. These same techniques are then
applied to the general effective lagrangian in the following two sections.
In section (3) we describe the most general effective interactions which
satisfy our above criteria. We identify in this section how the powers of
$1/M$ which can be expected to premultiply each operator in our lagrangian
depends on the assumptions that are made concerning the nature of the
underlying  physics. This gives an indication of the circumstances under
which the interactions we have kept may be expected to dominate. The steps
required to make our lagrangian into an easily-used tool are then performed
in Section (4). Section (5) contains the main results of our analysis. Here
we perform a fit to all charged- and neutral-current experimental data to
constrain the new-physics parameters. We find limits on most such
parameters, although there are certain directions in parameter space which
remain unconstrained. Section (6) then applies these results to
illustrative example, namely the mixing of ordinary and exotic fermions.
Our conclusions are summarized in Section (7).

\section{`Oblique' Corrections Revisited}

In this section we work through the familiar case of oblique radiative
corrections \stu,\peskin. We do so in order to clearly demonstrate the
logic of our method in a simple context that is relatively unencumbered by
algebra. The reader interested in diving straight into the full calculation
can safely skip directly to Section (3).

\subsection{The Initial Lagrangian}

Following Refs.~\stu, \peskin, if we imagine that the hitherto undiscovered
new physics that lurks at the high scale, $M$, couples more significantly
to the electroweak gauge bosons than to the other known light particles.
The dominant effects of virtual loops of these heavy particles may
therefore be expected to arise among the self-couplings of these gauge
bosons. With the intuition --- justified, with some qualifications, in more
detail in later sections --- that the lowest-dimension interactions should
be least suppressed by inverse powers of the heavy mass, $1/M$, we imagine
supplementing the standard model by the following lowest-dimension
effective interactions:
\label\gaugekinetic
\eqa
 \leff & = \lsm(\twe_i) + \lnewht, \eolnn
\hbox{with}\qquad
    \lnewht &= - {A \over 4} \; \Fhat_{\mu\nu} \Fhat^{\mu\nu}
   - {B \over 2} \; \What_{\mu\nu}^\dagger \What^{\mu\nu} - {C \over 4} \;
   \Zhat_{\mu\nu} \Zhat^{\mu\nu} + {G \over 2} \; \Fhat_{\mu\nu}
   \Zhat^{\mu\nu} ,\eolnn
   & \qquad  - w \, \tmw^2 \; \What^\dagger_\mu \What^\mu - {z \over 2}
   \, \tmz^2 \; \Zhat_\mu \Zhat^\mu . \eeol \eeq
Here $\lsm$ represents the familiar SM lagrangian, after the top quark and
Higgs boson have been integrated out --- including loop effects to the
extent that experiments are sensitive enough to probe these.
$\Fhat_{\mu\nu}$ and $\Zhat_{\mu\nu}$ represent the usual abelian field
strengths, while the $\What_{\mu\nu}$ is required to be electromagnetically
gauge covariant: $\What_{\mu\nu} = D_\mu \What_\nu - D_\nu \What_\mu$ with
$D_\mu \What_\nu = \partial_\mu \What_\nu + ie \Ahat_\mu \What_\nu$.

The new-physics coefficients, $A$ through $z$, could be computed within any
given underlying theory and should be thought of as (presently unknown)
functions of the parameters of this underlying theory. The success of the
standard model is equivalent to the statement that all current experiments
are consistent with $A=B=C=G=w=z=0$.

Not all six of these parameters are physically significant however, since
only three independent combinations of them actually ever appear in
expressions for physical observables. Only three independent combinations
can have physical content because there is a three-parameter family of
changes to the original six parameters in $\lnewht$ that can be made by
redefining the fields, without altering the form of the SM lagrangian,
$\Scl_\SM$. The required redefinitions consist of rescalings of the SM
electroweak gauge potentials and Higgs doublet: $W^a_\mu$, $B_\mu$ and
$\phi$. A conventional parametrization of the three physical combinations
of the quantities $A$ through $z$ is given by Peskin and Takeuchi's
variables  $S$, $T$ and $U$. The connection is given explicitly by (we use
the notation $\sw =  \sin\theta_w$, $\cw = \cos\theta_w$ \etc):
\label\studefs
\eqa
\alpha S &= 4 \sw^2 \cw^2 \left( A - C - {\cw^2 - \sw^2 \over \cw
\sw} \; G \right) , \eolnn
   \alpha T &= w - z , \eol
   \alpha U &=  4 \sw^4 \left( A - {1\over \sw^2} B + {\cw^2\over \sw^2}
		\; C - 2 {\cw \over \sw} \; G \right) . \eeolnn
\eeq

There are two aspects of our notation that are particularly significant:
\topic{1} The carets that appear overtop of the initial lagrangian and
fields in eq.~\gaugekinetic\ refer to the fact that these fields are not
canonically normalized, since $\lnewht$ contains kinetic (and mixing) terms
for the gauge bosons, in addition to those that are already in $\lsm$.

\topic{2} The $\twe_i$ represent all of the parameters appearing in the SM
part of the total effective lagrangian, such as the Higgs Yukawa couplings
$\tw{y}_f$, the electromagnetic fine-structure constant $\tw{\alpha}$,
\etc. The tilde is meant to indicate that these parameters do not take
their ``standard" numerical values, such as $\alpha^{-1} = 137.035989$,
when they are inferred from experiment, since the expressions for
observables as a function of these parameters  are altered by the presence
of the new physics.
\endtopic

Our method now consists of diagonalizing and canonically normalizing the
gauge-boson kinetic terms, and then eliminating the parameters $\twe_i$ in
favor of  parameters, $e_i$, which take on the `standard' values. Once we
have done so, we have used up the freedom to redefine fields, and so we
find that the resulting couplings then depend only on the three physical
quantities  $S$, $T$ and $U$.  The resulting lagrangian, as we shall show,
can be readily used to calculate observables in terms of a SM result  plus
some linear combination of $S$, $T$ and $U$.

\subsection{Diagonalization and Canonical Normalization.}

It is a simple matter to canonically normalize and diagonalize the gauge
boson kinetic terms, the required field redefinitions being
\label\rescalingone
\eqa \Ahat_\mu &=  \left( 1 - {A\over 2} \right) \; A_\mu  +  G \; Z_\mu ,
\eol
\label\rescalingtwo
   \What_\mu &=  \left( 1 - {B\over 2} \right) \; W_\mu , \eol
\label\rescalingthree
   \Zhat_\mu &=  \left( 1 - {C\over 2} \right) \; Z_\mu . \eeol \eeq
Here and elsewhere we work only to linear order in the small coefficients
$A,B,\dots, z$. It is straightforward to keep higher-order terms, if desired.
After this transformation, the total kinetic and mass terms are of the
desired form:
\label\standardkin
\eq   - {1 \over 4} \; F_{\mu\nu} F^{\mu\nu} - {1 \over 2} \;
W_{\mu\nu}^\dagger W^{\mu\nu} - {1 \over 4} \;  Z_{\mu\nu} Z^{\mu\nu}
    - (1 + w - B) \, \tmw^2 \; W^\dagger_\mu W^\mu -
   \hf (1 + z - C)  \, \tmz^2 \; Z_\mu Z^\mu.  \eeq

These field transformations also alter the form of the SM electromagnetic,
charged-current and neutral-current couplings, which now become:
\label\zerothem
\eqa
\Scl_{\rm em} &=  - \twe \left( 1 - {A\over 2} \right) \sum_i  \ol{f_i}
   \gamma^\mu Q_i f_i  \; A_\mu , \eol \label\zerothcc
 \Scl_{\rm cc} &=  - \, {\twe  \over \tws \sqrt{2}} \left( 1 - {B\over 2}
\right) \sum_{ij} \twV_{ij} \;  \ol{f_i} \gamma^\mu \Pl f_j \;
W^\dagger_\mu + \cc, \eol
\label\zerothnc
    \Scl_{\rm nc} &= - \, {\twe \over \tws \twc} \left( 1 - {C\over 2}
\right) \sum_i \ol{f_i} \gamma^\mu \left[ T_{3i} \Pl - Q_i  \tws^2  + Q_i
\tws \twc \, G \right] f_i \; Z_\mu . \eeol
\eeq
In these expressions, $Q_i$ is the electric charge of fermion $f_i$,
normalized with $Q_e =-1$. $T_{3i}$ similarly represents the fermion's
third component of weak isospin. $\twV_{ij}$ represents the usual
Cabibbo-Kobayashi-Maskawa (CKM) matrix for quarks, and is the unit matrix,
$\delta_{ij}$, for leptons.

\subsection{Re-expressing the Lagrangian in Terms of `Standard' Parameters}

The lagrangian, as we have written it, depends on the three parameters
$\twe$, $\tmz$ and $\tws$ (as well as the fermion and Higgs masses $\twm_i$
and the CKM matrix elements $\twV_{ij}$). In SM electroweak physics, these
three parameters (plus the particle masses and CKM matrix elements) suffice
to describe all electroweak observables. We can eliminate $\twe$, $\tmz$
and $\tws$ in terms of three reference observables, and it is standard to
choose the best-measured observables for this purpose: the electromagnetic
fine-structure constant, $\alpha$, the physical $Z$ mass, $\Mz$, and the
Fermi constant, $\gf$, as measured in muon decay. Using the resulting
expressions in the formulae for any other observables then leads to
numerical predictions that can be made to any desired accuracy.

Once the standard model is supplemented by $\lnew$, however, the relation
between these three parameters and the reference observables changes. As a
result the value that is inferred from experiment for a parameter such as
$\twe$, will differ from what would be found for the corresponding
parameter --- call it simply $e$ --- purely within the standard model. Our
goal in this section is to compute this difference, for each of the basic
three electroweak parameters. It is sufficient for the present purposes to
do so at tree level in all interactions, since any loop effects are
negligible once multiplied by the already small new-physics parameters.

The program therefore consists of calculating the input observables,
$\alpha$, $\Mz$ and $\gf$, at tree-level in the new model as computed using
eqs.~\standardkin, \zerothem, \zerothcc\ and \zerothnc. These expressions
are then equated to the tree-level SM predictions for the same quantities.
The result is a system of three equations that can be inverted to obtain
$\twe_i$ in terms of their `standard' counterparts, $e_i$. These then may
be used for predicting any other observable.

Note that, for this choice of new physics (\ie\ oblique corrections only),
the relations $\twm_i=m_i$ and $\twV_{ij}=V_{ij}$ are unchanged. However
this is not true in the general case, as we shall see in subsequent
sections.

\topic{Electric Charge ($e$)}

\ref\nrqed{See \eg\ P. LePage, in {\it Quantum Electrodynamics}, edited by
T. Kinoshita (World Scientific, Singapore, 1990).}
The fine-structure constant as determined in electron--electron
scattering\foot\qheffect{ Actually, the fine-structure constant is
determined in nonrelativistic condensed-matter systems, such as in the
Quantum Hall Effect. However the quantity that is found in this way in the
very-low-energy, nonrelativistic effective theory, is ultimately matched
onto $\ss \alpha$ as is used at high energies by using electron--electron
scattering at energies near the electron mass \nrqed.} at very low energies
is given at tree level, using the interaction eq.~\zerothem, by
\label\chargenew
\eq 4\pi \alpha = \twe^2 \left( 1 - A \right). \eeq
On the other hand, the SM tree-level relation is simply
\label\chargesm
\eq 4\pi \alpha =  e^2.  \eeq
Comparing eq.~\chargenew\ and eq.~\chargesm\ gives the following relation:
\label\charge
\eq
\twe = e \left( 1 + {A\over 2} \right).
\eeq

\topic{$Z$ Mass ($\Mz$)}

At lowest order, the physical $Z$-boson mass, $\Mz$ is simply the square
root of the parameter, $\mz^2$, that appears as the coefficient of $\hf \,
Z_\mu Z^\mu$ in the SM lagrangian. At the same order, the $Z$ mass in the
new model is similarly given by
\eq \Mz^2 = \tmz^2  ( 1 + z - C). \eeq
Comparing these predictions we deduce
\label\zmass
\eq  \tmz^2 =  \mz^2  ( 1 - z + C).  \eeq

\topic{Fermi's Constant ($\gf$)}

Muon decay is mediated by the low-energy exchange of a $W$ boson. Thus, to
calculate the Fermi constant at tree-level in the new model, we use the
propagator suggested by eq.~\standardkin, and the charged-current
interaction expressed in eq.~\zerothcc. This results in
\label\gfnew
\eqa  {\gf \over \sqrt{2}} &=
{ \twe^2 (1 - B) \over 8 \tws^2 \tmw^2\; (1 + w - B)} \eolnn
   &= { \twe^2 \over 8 \tws^2 \twc^2 \tmz^2} \; (1 - w). \eeol
\eeq
Note that we are free to use SM relations, such as $\tmw = \tmz \twc$,
among the `twiddled', or standard-model, parameters. For comparison, the SM
tree-level prediction for $\gf$ is simply
\label\gfsm
\eq  {\gf \over \sqrt{2}}
   = { e^2 \over 8 \sw^2 \cw^2 \mz^2}.\eeq
We take this last expression as our {\it definition} of $\sw$.

Combining eqs.~\charge, \zmass, \gfnew\ and \gfsm, we obtain
\label\sintheta
\eq
\tws^2 = \sw^2 \left[ 1 + { \cw^2 \over \cw^2 - \sw^2} \;
( A - C - w + z) \right]
\eeq
as well as the following useful formulae, which we record in passing:
\label\useful
\eqa \twc^2 &= \cw^2 \left[ 1 - { \sw^2 \over \cw^2 - \sw^2} \;
(A - C - w + z) \right], \eol
 { \twe \over \twc \tws} &= {e \over \cw \sw}
\left[ 1 + { C + w - z \over 2} \right]. \eeolnn
\eeq
\endtopic

The above expressions achieve our goal of relating $\twe$, $\tws$ and
$\tmz$ to the standard parameters $e$, $s_w$ and $\mz$. The next step in
the process is to re-express the lagrangian itself in terms of these
standard parameters. To do so we simply substitute eqs.~\charge, \zmass,
\sintheta\ and \useful\ into the various lagrangian terms.

By construction the $Z$ mass term and electromagnetic interaction take
simple forms:
\eqa \Scl_{\sss Z} & = - \hf \, \mz^2 \; Z_\mu Z^\mu, \eolnn
\hbox{and}\qquad  \Scl_{\rm em} &= - e \sum_i \ol{f_i}\gamma^\mu Q_i f_i \;
A_\mu . \eeol
\eeq
By contrast, the $W$ mass term gives a more complicated expression
\label\wmass
\eqa &\mz^2 \cw^2 \left[1-B+C+w- z - {\sw^2\over \cw^2 - \sw^2}
   \; ( A - C - w + z) \right] \; W^\dagger_{\mu} W^{\mu}  \eolnn
   &= \mz^2 \cw^2 \left[ 1 - {\alpha S \over 2 ( \cw^2 - \sw^2)}  +
   {\cw^2 \; \alpha T \over  \cw^2 - \sw^2}+{\alpha U \over 4 \sw^2}\right]
 \; W^\dagger_{\mu} W^{\mu}, \eeol \eeq
and the charged-current interaction takes the form:
\label\ccint
\eqa
\Scl_{\rm cc} &= - \,  {e \over \sqrt{2} \sw} \left( 1 + \hf  \left[
   A - B - {\cw^2 \over \cw^2 - \sw^2} (A - C - w + z) \right] \right)
   \sum_{ij} V_{ij} \ol{f_i} \gamma^\mu \Pl f_j \; W^\dagger_\mu + \cc ,
\eolnn
&= - \, {e \over \sqrt{2} \sw} \left( 1 - { \alpha S \over 4 ( \cw^2 -
  \sw^2)} + { \cw^2\; \alpha T \over 2 (\cw^2 - \sw^2)}  +
{\alpha U \over 8 \sw^2} \right) \sum_{ij} V_{ij}\ol{f_i} \gamma^\mu \Pl
f_j \; W^\dagger_\mu + \cc. \eeol
\eeq
Note that here that all corrections due to $S,T,U$ are universal. The
strength of the charged-current interaction is therefore given by:
$h_{ij} = \hsm_{ij} + \delta h_{ij}$, with $\hsm_{ij} = V_{ij}$ and
\eq
\delta h_{ij} =  V_{ij} \left( - \; {\alpha S  \over 4
( \cw^2 - \sw^2)} + { \cw^2 \; \alpha T\over 2 (\cw^2 - \sw^2)} +
{\alpha U\over 8 \sw^2} \right).
\eeq
Finally, the neutral-current interaction becomes:
\label\ncint
\eqa
\Scl_{\rm nc} &= -{e \over \sw \cw} \left( 1 + {w - z\over 2} \right)
\sum_i \ol{f_i} \gamma^\mu \left[ T_{3i} \Pl \phantom{{\sw^2 \cw^2 \over
\cw^2 - \sw^2}} \right. \eolnn
  & \qquad \qquad \qquad \left. - Q_i \left( \sw^2 + {\sw^2 \cw^2 \over
\cw^2 - \sw^2} [A - C - w + z] - \sw\cw G \right) \right] f_j \; Z_\mu\eol
 &= -{e \over \sw \cw} \left( 1 + {\alpha T \over 2}\right) \sum_i
\ol{f_i} \gamma^\mu \left[ T_{3i} \Pl - Q_i  \left( \sw^2 + {\alpha S \over
   4 ( \cw^2 -  \sw^2)} - { \cw^2\sw^2 \; \alpha T \over \cw^2 - \sw^2}
    \right) \right]  f_i \; Z_\mu . \eeolnn \eeq
Here there are both universal and non-universal corrections due to $S,T,U$.
(We remark that, in the language of Ref.~\peskin, the factor multiplying
$Q_i$ in the weak couplings is simply $s_\ast^2$.) The neutral-current
couplings, $g_{i\lft}$ and $g_{i\rht}$, are therefore given by their SM
counterparts, $\gsm_{i\lft} = T_{3i} - Q_i \sw^2$ and $\gsm_{i\rht} = -Q_i
\sw^2$, plus the deviations:
\label\ncdeviation
\eq
\delta g_{i\lft(\rht)} = {\alpha T\over 2} \; \gsm_{i\lft(\rht)} -
 Q_i \left({ \alpha S \over 4 ( \cw^2 - \sw^2)} - { \cw^2\sw^2 \; \alpha T
 \over \cw^2 - \sw^2} \right).
\eeq

Eqs.~\wmass\ through \ncdeviation\ may now be used to predict the
implications for any desired observables.

\subsection{The Calculation of Observables}

The calculation of observables is now straightforward. As has been pointed
out before, since the constants which parametrize the new physics are
small, we may work to any desired loop order in the SM interactions, and to
tree level in the interactions which deviate from the standard model.

Consider, first, the mass of the $W$ boson. In the standard model this mass
may be predicted as a function of the three input parameters: $\Mw =
\Mw^\SM(\Mz,\alpha,\gf)$. With the new interactions this expression now
gets a new contribution which may be read from eq.~\wmass:
\label\wmassprediction
\eq
\left( \Mw^2 \right)_{\rm phys} = (\Mw^\SM )^2 \left[
1 - {\alpha S \over 2 ( \cw^2 - \sw^2)}  +  {\cw^2 \; \alpha T \over
\cw^2 - \sw^2}+{\alpha U \over 4 \sw^2}\right].
\eeq
Note that because we have eliminated $\twe$, $\tws$ and $\tmz$ in terms
of their untwiddled counterparts, the SM contribution in this formula takes
precisely its usual numerical value. The resulting expression is in
agreement with Ref.~\peskin.

The $\rho$-parameter, defined as the ratio of low-energy neutral- and
charged-current amplitudes, can be read off from from the universal
$S,T,U$-corrections to eqs.~\ccint\ and \ncint. Taking also into account
the corrections to the $W$-mass (eq.~\wmassprediction), one finds
\eq
\rho = 1 + \alpha T,
\eeq
as in Ref.~\peskin.

Finally, consider the LR asymmetry at the $Z$ pole. $A_{\lft\rht}$ is the
sum of the (radiatively-corrected) SM expression, plus the direct
tree-level contribution from the new interactions. This is
\label\lrasymmetry
\eq
A_{\lft\rht} = \left[
  { \left( g_{e\lft} \right)^2  - \left( g_{e\rht} \right)^2 \over
\left( g_{e\lft} \right)^2  + \left( g_{e\rht} \right)^2 } \right].
\eeq
Linearizing this expression about the SM value gives $\delta A_{\lft\rht}$.
Finally, adding this to the SM rate gives:
\eqa
A_{\lft\rht} & = A_{\lft\rht}^\SM +
{4 \, g_{e\lft}^\SM \, g_{e\rht}^\SM \over \left(
\left( g_{e\lft}^\SM \right)^2  +
\left( g_{e\rht}^\SM \right)^2 \right)^2 }
\left( g_{e\rht}^\SM \, \delta g_{e\lft} -
g_{e\lft}^\SM \, \delta g_{e\rht} \right) \eolnn
& = A_{\lft\rht}^\SM +
{4 \, g_{e\lft}^\SM \, g_{e\rht}^\SM \left(
g_{e\rht}^\SM - g_{e\lft}^\SM \right)
\over \left( \left( g_{e\lft}^\SM \right)^2  +
\left( g_{e\rht}^\SM \right)^2 \right)^2 }
\left({ \alpha S \over 4 ( \cw^2 - \sw^2)} - { \cw^2\sw^2 \; \alpha T
\over \cw^2 - \sw^2} \right), \eeol
\eeq
which again agrees with Ref.~\peskin.

Contrast the ease of application of our lagrangian with the procedure that
is often followed in much of the literature. There, authors instead
directly use the lagrangian expressed with the twiddled parameters,
$\twe_i$. The direct contribution, $\delta \Sco$, of  new physics to a
given observable, $\Sco$, is then added to the shift in the SM value for
that observable (due to the shift from $\delta e_i = \twe_i - e_i$) to get
the total new-physics effect:
\eq
\Sco = \Sco_\SM + \delta  \Sco + \sum_i \left( {\partial \Sco \over
\partial e_i} \right) \; \delta e_i ~ . \eeq
The savings in labour in our approach is more striking in the more general
lagrangian we consider in the remainder of the paper.

\section{The General Effective Lagrangian}

We now wish to repeat these steps without assuming the particular
lagrangian of eq.~\gaugekinetic. Since our conclusions can only be as
general as is the lagrangian with which we choose to work, the aim of the
present section is to justify our lagrangian's generality. We save its
re-expression in terms of the `standard' parameters, and its comparison
with observables for subsequent sections.

The only simplifying choice we make is to concentrate on the electroweak
sector only. The inclusion of nonstandard gluon couplings presents no
particular problems, and can be dealt with in our formalism in a
straightforward manner.

We start making real physical choices with our remaining two assumptions:
($i$) the particle content of our low-energy theory --- we take only
particles which have been detected to date --- and, ($ii$) the maximum
dimension of the effective interactions which we consider --- which we take
to be five. Although the first of these assumptions may not provoke much
argument, a justification of the second of these turns out to require some
thought. We therefore first present the terms that are permitted in our
effective lagrangian by the assumed low-energy particle content, before
returning to the question of the validity of the neglect of dimension-six
and higher terms in section (3.2).

\subsection{The Effective Interactions}

We wish to write down the most general effective interactions in
$\lnewht$ that are consistent with the particle and symmetry content
relevant to the energies to which the lagrangian is to be applied. Our
first task is then to decide on precisely what this low-energy particle
content is. Since our intended application here is to current experiments
whose accessible energy is of order 100 GeV or less, we take our particle
content to include only those which already have been detected, namely
most of the SM particles, including precisely three left-handed neutrinos.
We do {\it not} include the Higgs boson and the top quark, which we take to
have been integrated out, if they exist. Because of these missing
particles, the field content of our effective theory does not fill out
linear representations of the electroweak gauge group, and so this symmetry
{\it cannot} be linearly realized, and the resulting lagrangian must
eventually violate unitarity
\ref\cornwall{J.M. Cornwall, D.N. Levin and G. Tiktopoulos,
\prd{10}{74}{1145}.}
\cornwall\ at energies at most of order $4\pi v \sim$ a few TeV. In this
case it is simply a matter of convenience whether this gauge symmetry is
chosen to be present but nonlinearly realized, or simply ignored completely
\ref\equivalence{M.S. Chanowitz, M. Golden and H. Georgi,
\prd{36}{87}{1490}.}
\ref\cutoff{C.P.~Burgess and D.~London, \prl{69}{92}{3428};
\prd{48} {93}{4337}.}
\equivalence, \cutoff.\foot\old{This equivalence is an old --- in some
quarters recently forgotten --- result which dates right back to
Ref.~\cornwall\ and beyond.} Clarity of presentation leads us to choose the
second of these options here.

With these comments in mind, we may construct the most general lagrangian
that can arise to any order in $1/M$ for the known particles. Our starting
point is again the split:
\eqnn  \leff = \lsm(\tw{e}_i) + \lnewht. \eeq
with
\eq \lnewht = \sum_{d=2}^\infty \lhat_d. \eeq
\ref\cplist{The removal of such redundant interactions is discussed \eg\ in
C.P. Burgess and J.A. Robinson, in {\it BNL Summer Study on CP Violation}
S. Dawson and A. Soni editors, (World Scientific, Singapore, 1991). For a
more recent, detailed treatment see C. Arzt, preprint UM-TH-92-28
(unpublished).}
Here $\lhat_d$ contains all possible terms that have operator dimension
(mass)${}^d$. We wish to list explicitly all terms up to $\lhat_5$. We may
freely integrate by parts, and use the standard-model equations of motion
in order to simplify our lagrangian, since no operators that can be
eliminated in this way can have any physical effects \cplist.

A word should be said about new-physics operators involving neutrinos. Our
low-energy particle content does not include right-handed neutrinos. We can
nevertheless continue working with four-component spinors provided that we
take the neutrino spinors to be Majorana: $\nu =\nu^{\sss C}$, where
$\nu^{\sss C} = C{\overline\nu}^{\sss T}$, with $C$ the charge conjugation
matrix. This means that the parameters describing the interactions of
neutrinos are subject to more constraints than are those of the other,
electrically-charged, fermions. We identify these additional conditions,
case by case, for the various anomalous interaction terms presented below.

\topic{Dimension Two}
At dimension two we have the boson mass terms:
\eq  \Scl_2 =  - w \, \tw{m}_{\sss W}^2
   \; \What^\dagger_\mu \What^\mu - {z \over 2} \, \tw{m}_{\sss Z}^2 \;
	\Zhat_\mu \Zhat^\mu,  \eeq
a particular combination of which also appears in $\lsm$. Only the
combination $w-z$ of these two masses may therefore be detected through the
deviation it produces from the standard-model relation between gauge boson
masses (\cf\ eqs.~\wmass\ and \wmassprediction, for example).

\topic{Dimension Three}
The fermion mass terms arise at dimension three:
\eq \Scl_3 = -  \ol{\fhat} (\delta m_\lft \Pl  +  \delta m_\rht \Pr )
\fhat,
\eeq
where $f$ denotes a generic column vector in fermion generation space, and
$\delta m_\lft$ and $\delta m_\rht$ are matrices in this space.
Hermiticity of the action requires that $\delta m_\lft = \delta
m_\rht^\dagger$, and for neutrinos we have the additional condition $\delta
m_\rht^\nu = (\delta m_\lft^\nu)^*$. We choose our conventions so that
$\Scl_3$ is \CP-invariant if the matrices $\delta m_\lft$ and $\delta
m_\rht$ are real.

Apart from the neutrino masses, which are zero in the standard model, these
fermion mass terms are indistinguishable from the SM ones. They could
nevertheless become detectable in the event that a light Higgs particle
should be discovered. In this case such interactions could cause deviations
from SM relations, such as $y_f = m_f/v$, between the fermion-Higgs Yukawa
coupling and the fermion masses.

\topic{Dimension Four}
Dimension four contains two types of terms, ($i$) gauge-boson and fermion
kinetic terms, and ($ii$) gauge-boson -- fermion coupling terms.

We therefore have
\eqnn  \hat{\Scl}_4 = \hat{\Scl}_{\rm bkin} + \hat{\Scl}_{\rm fkin}
 + \hat{\Scl}_{\rm bff} + \hat{\Scl}_{\rm other}^{(4)} \eeq
with:
\eqa
\hat{\Scl}_{\rm bkin} &= - {A \over 4} \; \Fhat_{\mu\nu} \Fhat^{\mu\nu}
   - {B \over 2} \; \What_{\mu\nu}^\dagger \What^{\mu\nu} - {C \over 4} \;
   \Zhat_{\mu\nu} \Zhat^{\mu\nu} + {G \over 2} \; \Fhat_{\mu\nu}
   \Zhat^{\mu\nu} \eol
\hat{\Scl}_{\rm fkin} &= - \ol{\fhat} \gamma^\mu (I_\lft \Pl + I_\rht \Pr)
    D_\mu \fhat \eol
\label\bffterms
\hat{\Scl}_{\rm bff} &= - \, {\twe \over \tws \twc} \ol{\fhat}
   \gamma^\mu (\delta \hat{g}_\lft \Pl + \delta \hat{g}_\rht \Pr)
   \fhat \, \Zhat_\mu - {\twe \over \sqrt{2} \tws}
   \ol{\fhat} \gamma^\mu ( \delta \hat{h}_\lft \Pl + \delta \hat{h}_\rht
   \Pr) \tau_+ \fhat \, \What^\dagger_\mu + \cc \eeol
\eeq
We use a compact notation in these expressions, in which $I_\lft$,
$I_\rht$, \etc\ are matrices which act on the indices which label fermion
type (or flavour), and where $\tau_+$ is the $SU_\lft(2)$ raising operator.
The matrices $I_\lft$, $I_\rht$, $\delta\hat{g}_\lft$ and
$\delta\hat{g}_\rht$ must always be hermitian, with $I_\lft = I_\rht^*$ and
$\delta \hat{g}_{\lft} = - \delta \hat{g}_{\rht}^{*}$ holding in addition
for neutrinos. \CP-invariance follows if all of these couplings matrices
should be real. The derivative $D_\mu$ used in the fermion kinetic terms is
covariant with respect to the electromagnetic interactions. (If we also
considered nonstandard gluon couplings, we would demand covariance with
respect to the full unbroken gauge group, $SU_c(3) \times U_{\rm em}(1)$.)
Finally, as in Section (2), the ubiquitous carets indicate that the fields
are not yet canonically normalized.

\ref\bvrefs{For a review with references, see M. Mattis
\prep{214}{92}{159}.}
\ref\alitti{UA2 Collaboration, J. Alitti {\it et al.}, \plb{277}{92}{194}.}
$\lhat_{\rm other}^{(4)}$ contains all of the other dimension-four
operators which we do not consider here. There are two types of such terms,
although both involve only the electroweak gauge bosons. The first type
consists of a potential electroweak `$\Theta$-term' -- \ie\ a term
proportional to $\twi{W}_{\mu\nu} W^{\mu\nu}$. We ignore this here since it
produces completely negligible effects at zero
temperature.\foot\bviolation{See, however, the recent controversy
concerning the existence of potential weak-scale baryon-number violation in
TeV accelerators \bvrefs.} Also lumped into $\lhat_{\rm other}^{(4)}$ are
the dimension-four three- and four-point gauge-boson self-interactions. As
explained earlier, we have chosen not to include these here since they
cannot yet be well bounded at tree-level \alitti. This makes them
interesting in their own right, since it means that, so far as we know,
they could very well contain new physics.

\topic{Dimension Five}
At dimension five the following combinations can arise
\eqa \Scl_5 &= -\twe \ol{\fhat} \sigma^{\mu\nu}
   (\hat{d}_\lft \Pl + \hat{d}_\rht \Pr) \fhat \Fhat_{\mu\nu} -
{\twe \over \tws \twc} \ol{\fhat} \sigma^{\mu\nu} (\hat{n}_\lft \Pl
  + \hat{n}_\rht \Pr) \fhat \, \Zhat_{\mu\nu} \eol
& \qquad - {\twe \over \sqrt{2} \tws}
 \ol{\fhat} \sigma^{\mu\nu} (\hat{c}_\lft
   \Pl + \hat{c}_\rht \Pr) \tau_+ \fhat \, \What^\dagger_{\mu\nu} +
\lhat_{\rm other}^{(5)} + \cc  \eeolnn
\eeq
Again, all coefficients here --- $\hat{d}_{\sss L,R}$, $\hat{n}_{\sss
L,R}$, and $\hat{c}_{\sss L,R}$ --- are matrices in  flavour space, as is
the $SU_L(2)$ raising operator, $\tau_+$. It is required that
$\hat{d}_{\lft} = \hat{d}_{\rht}^{\dagger}$ and $\hat{n}_{\lft} =
\hat{n}_{\rht}^{\dagger}$ for hermiticity of the action, together with
restriction $\hat{d}_{\lft} = - \hat{d}_{\rht}^{*}$ and $\hat{n}_{\lft} = -
\hat{n}_{\rht}^{*}$ for neutrinos. \CP-conservation requires all of these
coupling matrices to be real. $\lhat_{\rm other}^{(5)}$ here includes all
four-point fermion-gauge-boson couplings, such as $ff W^\dagger W$, which
are also not yet probed in existing experiments.

\topic{Dimension Six}
Finally, there are a great many operators that can arise at dimension six
including a very long list of 4-fermion contact interactions. Their
inclusion would enormously complicate the present analysis, and so we
neglect them throughout what follows. We discuss in the next section the
circumstances under which the neglect of these dimension-six interactions
can be justified by their suppression by additional powers of $O(1/M^2)$.
\endtopic

\subsection{Power Counting}

What ultimately makes an effective-lagrangian analysis useful is the
property that only a limited number of effective interactions can arise to
any given order in the expansion in the inverse of the heavy mass, $M$, of
the new physics. Usually, powers of $1/M$ are simply counted by dimensional
analysis, with the coefficient, $c_n$, of an effective operator of
dimension (mass)${}^{d_n}$ being proportional to $M^{4 - d_n}$. (Some
\ref\effective{J. Polchinski, Lectures presented at TASI 92, Boulder, CO,
Jun 3-28, 1992, preprint NSF-ITP-92-132, hep-th/9210046}
exceptions to this common rule of thumb are discussed in Ref.~\effective.)
As has been stated earlier, we choose here to work only up to and
including effective interactions of dimension five.

If this were the whole story, then the neglect of dimension-six operators
could be simply justified as being due to their suppression by additional
powers of $1/M$ relative to those at dimension four and five. There are two
issues which complicate this simple picture, however.

First, it can happen that effective operators are more suppressed than
would be indicated by simple dimension counting. This can occur because of
the possibility of suppression by small dimensionless quantities, such as
small coupling constants in the underlying theory (like Yukawa couplings:
$y_f = m_f/v$), or by small mass ratios (like $v/M$, which is present if
$M$ is much larger than the electroweak-breaking scale,
$v$).\foot\nottoobig{It must be kept in mind here that since our low-energy
particle content does not fill out linear representations of $\ss SU_{\sss
L}(2) \times U_{\sss Y}(1)$, $\ss v/M$ cannot be smaller than roughly $\ss
1/4 \pi$.} If this type of additional suppression should arise for the
lower-dimension  terms which we keep, then their neglect relative to
unsuppressed dimension-six terms may no longer be justified.

Second, one might also worry that dimension-six operators may be suppressed
by fewer than two powers of $M$, such as if they were proportional to
$1/v^2$ or $1/vM$. As we shall see shortly, such coefficients are indeed
possible depending on the nature of the underlying physics that has been
integrated out. In such a case the neglect of dimension-six new-physics
operators in comparison to those of lower dimension need not be justified.

The bottom line is that the power of $v/M$ which appears in the coefficient
of a given effective interaction generically depends on the nature of
physics that is associated with the large scale, $M$. As a result, a
complete cataloguing of effective interactions according to their
suppression by $1/M$ cannot be made in an entirely model-independent way.
At some point this model-dependence may become a Good Thing: a comparison
of the sizes of various effective operators, should they ever be
discovered, may ultimately permit the diagnosis of the nature of the
underlying new  physics.\foot\usesft{For a related, and more detailed,
discussion of  heavy-mass dependence see Ref.~\cutoff.} We therefore
neglect dimension-six effective interactions, in the knowledge that an
element of model-dependence enters in this way into our conclusions. One
must simply check, when applying the bounds we obtain below to a particular
model, that this neglect is justified in the case of interest.

In order to more concretely illustrate what can be expected for the
strength of various effective interactions from differing types of
underlying physics at scale $M$, we next consider explicitly the
implications of two types of scenarios --- strongly and weakly coupled
electroweak symmetry-breaking physics. We do so partly to demonstrate
the existence of models for which four-fermi terms may be neglected,
and partly to contrast the sizes of the various terms in the effective
lagrangian for these two cases.

\topic{Strongly-Coupled New Physics}
It is possible that the symmetry-breaking sector of the electroweak theory
is strongly coupled, with only the three would-be Goldstone {\nobreak
bosons} (WBGB's) --- that is to say, the longitudinal $W$ and $Z$
polarizations --- appearing at experimentally accessible energies. In this
case the couplings of these WBGB's are completely dictated, at low
energies, to be those given by chiral perturbation theory
\ref\ccwz{S. Coleman, J. Wess and B. Zumino, \pr{177}{69}{2239};
E.C. Callan, S. Coleman, J. Wess and B. Zumino, \pr{177}{69}{2247};
J. Gasser and H. Leutwyler, \anp{158}{84}{142}.}
\ccwz. In the resulting effective lagrangian successive powers of the WBGB
fields are suppressed by inverse powers of the symmetry-breaking scale,
$v$. If the lagrangian were to be applied to energies $E \ll v$, then the
powers of $v$ that would be obtained in this way by dimensional analysis
would suffice for counting which interactions arise to a given order in
$E/v$.

In practice, however, applications are meant to be for higher energies, $E
\simeq v \ll M$. In this case a consistent expansion in powers of $E/M$ is
only possible if successive terms in the effective lagrangian are
suppressed by powers of $M$ rather than $v$. That is to say, an expansion
in powers of $1/M$ requires that some couplings in the effective theory
must be systematically suppressed by powers of $v/M$, compared to the
powers of $v$ that arise using straight dimensional analysis. This
suppression has been formulated in a precise way, based on experience with
chiral perturbation theory as applied to low-energy QCD, and is called
called ``Naive Dimensional Analysis'' (NDA)
\ref\nda{A. Manohar and H. Georgi, \npb{234}{84}{189}; H.Georgi and
L.Randall, \npb{276}{86}{241}; H. Georgi, \plb{298}{93}{187}.}
\nda. It states that a term having $b$ WBGB fields, $f$
(weakly-interacting) fermion fields, $d$ derivatives and $w$ gauge fields
has a coefficient whose size is:
\eq  c_n(M) \sim v^2 M^2 \; \left( { 1\over v} \right)^b \; \left(
{ 1\over M^{3/2}} \right)^f \; \left( { 1\over M} \right)^d \;
\left( { g\over M} \right)^w.
\eeq
In this expression the relation $M \lsim 4 \pi v$ must always be kept in
mind. (If the fermions were strongly interacting, as would be the case for
technifermions or for nucleons in QCD, then the factor is $1/v\sqrt{M}$ for
each fermion. This would lead to a coefficient of order $1/v^2$ for
dimension-six four-fermion interactions.) The implications of the above
estimate for the various effective interactions are listed in Column (2) of
Table (I).

\midinsert
$$\vbox{\tabskip=0pt \offinterlineskip
\halign to \hsize{\strut#& #\tabskip 1em plus 2em minus .5em&
\hfil#\hfil &#&\hfil$#$\hfil &#& \hfil$#$\hfil &#\tabskip=0pt\cr
\noalign{\hrule}\noalign{\smallskip}\noalign{\hrule}\noalign{\medskip}
&& Operator && \hbox{NDA} && \hbox{LRDA}&\cr
\noalign{\medskip}\noalign{\hrule}\noalign{\medskip}
&&  Gauge Boson Masses && g^2 v^2 && g^2 v^4/M^2 &\cr
&& Neutrino Masses && v^2/M &&  v^2/M  &\cr
&& Gauge Boson Kinetic Terms && g^2 v^2/M^2  &&  g^2 v^2/M^2  &\cr
&&  Dim 4 Gauge-Boson/Fermion Vertex && gv^2/M^2  &&  gv^2/M^2  &\cr
&&  Dim 5 Gauge-Boson/Fermion Vertex &&  gv^2/M^3  &&  g v/M^2  &\cr
&&  Dim 6 Four Fermion Terms &&  v^2/M^4  &&  1/M^2  &\cr
\noalign{\medskip}\noalign{\hrule}\noalign{\smallskip}\noalign{\hrule}
}}$$
\centerline{\bf Table (I)}
\medskip
\noindent {\eightrm We tabulate here the estimated sizes that would be
expected for the deviations from Standard Model among effective operators
of the lowest dimension, as is explained in the text. The two columns
contrast the implications of two types of assumptions concerning the nature
of the underlying physics, either Naive Dimensional Analysis (NDA), or
Linearly-Realized Dimensional Analysis (LRDA). We use the NDA rules for
weakly-coupled fermions in obtaining our estimate for four-fermion terms.}
\endinsert

\topic{Weakly-Coupled New Physics}
A completely opposite point of view is to suppose that the electroweak
symmetry-breaking physics is sufficiently weakly coupled to permit a
perturbative analysis. In this case one or more physical particles, besides
the WBGB's, would be expected to have masses of order $\lambda v$, where
$\lambda$ is a small dimensionless coupling. Being light, these particles
appear in the low-energy theory and, together with the WBGB's, fill out
linear realizations of the electroweak gauge group. The standard model
itself is an example along these lines, where the physical Higgs scalar
plays the role of this new light particle.

Besides the effects of their direct propagation, these new degrees of
freedom can appear  within the effective lagrangian through the powers of
$v/M$ that they contribute when their fields are replaced by their vacuum
expectation values ({\it v.e.v.}s). The precise power which appears in any
particular effective interaction therefore depends on the representations
in which the Higgs-like fields transform. The most plausible choice is one
or more doublets, with the standard hypercharge assignment, since this is
what is required to generate masses for the known fermions.

In this scenario the size of any non-Higgs interactions may be found by
finding the lowest-dimension interactions which contain the desired term,
replacing all Higgs fields by their {\it v.e.v.}s, and making up the rest
of the dimensions with powers of the heavy mass, $M$. We call the estimate
that is obtained in this way ``Linearly-Realized Dimensional Analysis''
(LRDA). This estimate is given for the effective operators of interest here
in Column (3) of Table (I).

\topic{A Comparison Between NDA and LRDA}
As is seen from Table (I), there are a number of differences between the
implications of NDA and LRDA for the lowest-dimension operators we are
considering.

Typically the linearly-realized gauge symmetry enforces relations amongst
the various coefficients of operators which involve a particular number of
fields or derivatives, depending on how these operators can be assembled
into linearly-realized multiplets. This is best illustrated with a few
examples.

Consider the contributions to the $W$- and $Z$-masses: $\Sco_\ssw =
W^\dagger_\mu W^\mu$ and $\Sco_{\sss Z} = \hf \, Z_\mu Z^\mu$. The
lowest-dimension operator which contains these terms is simply the
dimension-four Higgs kinetic term, $(D_\mu \phi)^\dagger (D^\mu \phi)$.
Here, because the gauge symmetry is linearly realized, the covariant
derivatives are \gwk-invariant. Thus, as in the standard model, replacing
$\phi$ by $v$ generates the particular combination $\cw^2 \,
\Sco_\ssw + \Sco_{\sss Z}$ with a coefficient that is of order $g^2 v^2$.
There are also dimension six contributions to the masses, such as
$(\phi^\dagger D_\mu \phi) \, (\phi^\dagger D^\mu \phi) / M^2$. This and
similar operators contribute to $\Delta\rho$ (that is, they spoil the mass
relation $\Mw = \Mz \cw$) by amounts that are of order $g^2 v^4/M^2$.
Therefore $\Delta \rho$ is automatically small in these theories provided
only that $v^2/M^2 \ll 1$. By contrast, if the symmetry-breaking  sector is
strongly interacting (NDA) generic contributions to both the $W$ and $Z$
boson masses are the same size, $O( gv)$, and so one requires an additional
custodial $SU(2)$ symmetry to explain the smallness of $\Delta\rho$.

For the other operators in Table (I), however, the NDA estimates are
typically {\it smaller} than or equal to those of LRDA. This need not
always be the case, as we have seen for the predictions for $\Delta \rho$,
above.
\endtopic

A glance at Table (I) also shows that, in LRDA, the effective operators we
are considering are all suppressed by at most two powers of $1/M$. It is
therefore consistent to neglect all operators which are suppressed by more
than $1/M^2$. While this rules out any operator of dimension seven or
higher, the necessity to include dimension six operators in general depends
on the nature of the underlying theory. As is witnessed by the power
counting of NDA, the suppression of four-fermion terms relative to those of
lower dimension is possible, even if these lower-dimension terms should be
$O(1/M^2)$.

\section{Transforming to Standard Form}

Having now determined which operators to keep at $O(1/M^2)$,  we must
recognize that not all of the parameters of our effective  lagrangian need
be physically significant.  As was the case for the oblique corrections in
the previous section, not all of the above interactions can represent a
physical deviation from the standard model, since some can be removed
without changing the form of $\lsm$ simply by rescaling and rotating the
fields. Only those that cannot be removed in this way without violating the
symmetries of the standard model can have physical consequences, since
these lead to deviations from the predictions that relate SM parameters,
such as appears in eq.~\wmassprediction\ in Section (2) above.

To determine the physical combinations we follow the logic set out in
Section (2): ($i$) first rescale all fields to put their kinetic and mass
terms into standard form, and ($ii$) eliminate the `tilde-ed' parameters in
the lagrangian in favour of the physical quantities that are extracted from
experiment. Only the algebra changes between this more general case, and
the simpler one studied in Section (2).

\subsection{Rescaling the Fields}

The diagonalization of the electroweak boson kinetic and mass terms is
identical to that found in eqs.~\rescalingone\ through \rescalingthree.
The fermion kinetic and mass terms are similarly diagonalized by the
transformation:
\eq  \fhat = \left[ \left(1 - {I_\lft \over 2} \right) \,
   U_\lft \, \Pl + \left(1 - {I_\rht \over 2} \right) \, U_\rht
   \, \Pr \right] \; f, \eeq
where the unitary matrices, $U_\lft$ and $U_\rht$, are chosen to ensure
that the mass matrix is diagonal with non-negative entries along the
diagonal:
\eqa
\diag{ \dots, m_i,\dots } &= U_\rht^\dagger \left[ \twm_\lft + \delta
   m_\lft - \hf \left( I_\rht^\dagger \twm_\lft + \twm_\lft I_\lft \right)
   \right] U_\lft \eolnn
 &= U_\lft^\dagger \left[ \twm_\rht + \delta m_\rht - \hf \left(
I_\lft^\dagger \twm_\rht + \twm_\rht I_\rht \right) \right] U_\rht. \eeol
\eeq
The matrices $\twm_{\sss L,R}$ which appear in these expressions denote the
left- and right-handed fermion mass matrices in the original fermion
basis.

After performing this redefinition, the standard-model and new-physics
contributions to the fermion electromagnetic coupling become
\label\newem
\eqa  \Scl_{\rm em} &= -\twe \left( 1 - {A \over 2} \right) \; \Bigl[
   \ol{f} \gamma^\mu Q \, f \; A_\mu + \ol{f} \sigma^{\mu\nu}
   ( d_\lft \Pl + d_\rht \Pr) \, f \; F_{\mu\nu} \Bigr], \eolnn
&\hbox{where:} \qquad d_\lft \equiv U_\rht^\dagger \hat{d}_\lft U_\lft,
\eol
&\phantom{\hbox{where:}} \qquad d_\rht \equiv U_\lft^\dagger \hat{d}_\rht
   U_\rht. \eeolnn \eeq
Note that for these interactions the unbroken gauge invariance only
permits the appearance of dipole-moment couplings, parametrized by the
matrices $d_{\sss L,R}$.

The neutral-current interactions similarly become:
\label\ncbeforedetwiddling
\eqa
\Scl_{\rm nc} &= - \, {\twe\over \tws \twc} \, \left( 1 - {C \over 2}
\right) \; \Bigl[ \ol{f} \gamma^\mu ( \twg_\lft \Pl + \twg_\rht \Pr) \, f
   \; Z_\mu + \ol{f} \sigma^{\mu\nu} ( n_\lft \Pl + n_\rht \Pr) \, f
   \; Z_{\mu\nu} \Bigr], \eol
 &\hbox{where:} \qquad \twg_\lft \equiv \gsm_\lft + \delta \twg_\lft =
   U_\lft^\dagger \left[ \gsm_\lft +
   \delta \hat{g}_\lft - \hf \left( I_\lft^\dagger \gsm_\lft + \gsm_\lft
I_\lft \right) \right] U_\lft + \tws \twc Q G, \eolnn
 &\phantom{\hbox{where:}}\qquad \twg_\rht \equiv \gsm_\rht + \delta
\twg_\rht = U_\rht^\dagger\left[ \gsm_\rht+
   \delta \hat{g}_\rht - \hf \left( I_\rht^\dagger \gsm_\rht + \gsm_\rht
I_\rht \right) \right] U_\rht + \tws \twc Q G, \eolnn
 &\hbox{and:} \qquad n_\lft \equiv U_\rht^\dagger \hat{n}_\lft U_\lft,
   \eolnn
  &\phantom{\hbox{and:}} \qquad  n_\rht \equiv U_\lft^\dagger
    \hat{n}_\rht U_\rht,  \eeolnn \eeq
which in general may involve flavour-changing neutral currents. As
discussed in Sec.~3.1, the left-handed and right-handed neutral-current
couplings of neutrinos are not independent, being related by $\twg_\rht^\nu
= -(\twg_\lft^\nu)^{\sss T}$.

Finally, the charged-current couplings become:
\label\newcc
\eqa
\Scl_{\rm cc} &= - \, {\twe\over \sqrt{2}\tws} \,
\left( 1 - {B\over 2}\right)
   \; \Bigl[ \ol{f} \gamma^\mu ( \twh_\lft \Pl + \twh_\rht \Pr) \,
   f' \; W^\dagger_\mu + \ol{f} \sigma^{\mu\nu} ( c_\lft \Pl + c_\rht \Pr)
\, f' \; W^\dagger_{\mu\nu} \Bigr] + \cc, \eolnn
 &\hbox{where:} \qquad \twh_\lft \equiv \hsm_\lft + \delta \twh_\lft =
   U_\lft^\dagger  \left[ \hsm_\lft + \delta \hat{h}_\lft - \hf \left(
   I_\lft^\dagger \hsm_\lft + \hsm_\lft I'_\lft \right) \right] U'_\lft,
\eol
 &\phantom{\hbox{where:}} \qquad \twh_\rht \equiv \hsm_\rht + \delta
\twh_\rht =
   U_\rht^\dagger \delta \hat{h}_\rht U'_\rht , \eolnn
 &\qquad \hbox{and:} \qquad c_\lft \equiv U_\rht^\dagger \hat{c}_\lft
U'_\lft, \eolnn
  &\qquad \phantom{\hbox{and:}} \qquad  c_\rht \equiv U_\lft^\dagger
   \hat{c}_\rht U'_\rht.  \eeolnn \eeq
In these expressions, $f$ represents a $u$-type quark or a neutrino, in
which case $f'$ is respectively either a $d$-type quark or a charged
lepton. Primes on the matrices $U'_{\sss L,R}$ and $I'_\lft$ are meant to
distinguish the matrices that are associated with $f'$ from those
associated with $f$. There are two qualitatively new features that arise
here: ($i$) the introduction of a right-handed current, and ($ii$)
modifications to the left-handed CKM matrix. We elaborate on these in more
detail in later sections.

The final remaining step is to determine the shift that is induced by the
new physics into the reference parameters in the lagrangian.

\subsection{Shifting to Physical Parameters}

Because of the present accuracy of the electroweak data, it suffices to
work only to linear order in the new-physics parameters of our effective
lagrangian. Keeping higher order terms is conceptually straightforward,
though algebraically more complicated.

None of the additional terms in this more general effective lagrangian
alter the connection between $\twe$ and $e$, or between $\tmz$ and $\mz$,
so these remain as given in eqs.~\charge\ and \zmass:
\eqa \twe &= e \left( 1 + {A\over 2} \right). \eolnn
\tmz^2 &=  \mz^2  ( 1 - z + C). \eeolnn
\eeq

The really new features arise for the definition of $\gf$ --- and so for
the expression for $\tws$ in terms of $\sw$ --- as well as for the
charged-current CKM matrices. This is because each of these quantities is
defined with reference to a charged-current fermion decay, and so their
determination is affected by the deviations of $h_{\sss L,R}$ from their SM
values. We consider these observables here in turn:
\topic{Fermi's Constant ($\gf$)}
We must compare the tree-level expression for muon decay as computed with
the new charged-current interactions, and read off the combination of
parameters in the decay rate that is to be identified as the Fermi
constant. The result is independent of the induced right-handed currents,
since these do not interfere with the left-handed currents to within the
accuracy we are interested. The same is true for the coefficients $c_{\sss
L,R}$. The quantity which does arise to linear order in the new physics is:
\label\newfermiconst
\eq  {\gf \over \sqrt{2}} = { \twe^2 \over 8 \tws^2 \twc^2 \tmz^2}
   \; (1 - w + \Delta_e + \Delta_\mu), \eeq
where:
\label\deltadef
\eq \eqalign{
 \Delta_f &\equiv \sqrt{\sum_i \left| 1 + \delta \twh_\lft^{\nu_i f}
\right|^2} \; -1 \cr
&=  \sum_{i} \Re(\delta \twh_\lft^{\nu_i f}) + \hbox{(higher order terms)}.
\cr}
\eeq
Note that only the real part of $\delta \twh_\lft^{\nu_i f}$ appears here.
This is because we are working to linear order only in the new-physics
parameters, and therefore the only operators which can enter into the above
expression are those which have SM counterparts with which they can
interfere. Since in our conventions the SM leptonic charged-current
couplings are purely real, $\Im(\delta \twh_\lft^{\nu_i f})$ can never
appear at linear order. Note also that, since we do not insist upon
lepton-number conservation, the sum is over all light neutrinos. In terms
of these variables, the analogue of eq.~\sintheta\ for $\tws^2$ is:
\eq \tws^2 = \sw^2 \left[ 1 + { \cw^2 \over \cw^2 - \sw^2} \;
	( A - C - w + z + \Delta_e + \Delta_\mu) \right], \eeq
where $\sw^2$ is defined as in Section (2): $\gf/\sqrt{2} \equiv e^2/(8
\sw^2\cw^2 \mz^2)$.
\topic{The CKM Matrix Elements ($V_{ij}$)}
As discussed above, the question of whether or not a new-physics operator
contributes to linear order in the expression for an observable depends
upon whether or not there is a corresponding SM operator with which it can
interfere. For CP-violating new-physics contributions to CKM matrix
elements, this appears to be problematic, since, according to this
argument, $\Im(\delta \twh_\lft^{ij})$ will appear only if the
corresponding SM CKM matrix element $V_{ij}$ is complex. However, the phase
of a single CKM matrix element is not physically meaningful -- any
particular matrix element can be made real by phase redefinitions of the
quark fields. It is only the phase of the product of four CKM matrix
elements $V_{ij} V_{ik}^* V_{lk} V_{lj}^*$ which has a physical meaning. In
other words, it is a phase-convention-dependent question whether
$\Re(\delta \twh_\lft^{ij})$ or $\Im(\delta \twh_\lft^{ij})$ (or both)
appears in the expression for a particular observable. It is possible to
express all observables in terms of the new-physics parameters in a
completely general way, with no assumptions as to the reality of the CKM
matrix elements, but this has the unfortunate effect of rendering the
formulae unduly cumbersome. It is therefore useful, for simplicity, to
choose a particular form for the SM CKM matrix. We use the approximate
\ref\wolf{L. Wolfenstein, \prl{51}{83}{1945}.}
parametrization \wolf
\label\wolfckm
\eq
V_{\sss CKM} \simeq \left(\matrix{
    1-{1\over 2}\lambda^2 & \lambda & A\rho\lambda^3 e^{-i\delta} \cr
  -\lambda(1+A^2\rho\lambda^4 e^{i\delta} )
& 1-{1\over 2}\lambda^2 - A^2 \rho\lambda^6 e^{i\delta} & A\lambda^2 \cr
   A\lambda^3 (1 - \rho e^{i\delta}) & -A\lambda^2 (1+\rho\lambda^2
e^{i\delta}) & 1 \cr}\right)~,
\eeq
in which $\lambda=0.22$ is the sine of the Cabibbo angle, the values of $A$
and $\rho$ are $\sim 1$, and $\delta$ is constrained to lie between 0 and
$\pi$ (due to the nonzero value of $\epsilon_{\sss K}$, $\delta$ very close
to 0 or $\pi$ is excluded). Note that, in this parametrization, all CKM
matrix elements save $V_{ub}$ and $V_{td}$ are essentially real. Therefore
we know in advance that the $\Im(\delta \twh_\lft^{ij})$, which can
contribute to CP-violating processes, will remain virtually unconstrained.

The relation between the $\twV_{ij}$ and the $V_{ij}$ depends crucially on
the manner in which the CKM matrix elements are measured experimentally.
For example, $V_{ud}$ is determined from the $\beta$-decay rate for
superallowed transitions in spinless nuclei. As such, these experiments
measure the nuclear matrix element of the vector part of the quark-level
transition $d \to u + e^- + \ol{\nu}$. If we read off the part of the
amplitude which appears in this matrix element we find:
\eq
{\gf \over \sqrt{2}} |V_{ud}| = {\twe^2 \over 8 \tws^2 \twc^2
   \tmz^2 } \; |(\twh_\lft^{ud} + \twh_\rht^{ud})| \;
(1 - w + \Delta_e).
\eeq
Using this result together with expression \newfermiconst\ for $\gf$ as
determined in muon decay gives:
\eq
|\twV_{ud}| \equiv |(\hsm_\lft )_{ud}| = |V_{ud}| \;
   \Bigl[ 1  + \Delta_\mu \Bigr] - \Re(\delta \twh_\lft^{ud} + \delta
\twh_\rht^{ud}).
\eeq
Analagous results hold for those elements of the CKM
matrix that are determined by measuring the hadronic matrix element of the
vector part of the quark-level transition $q_i \to q_j + e + \nu$. This is
true for $|V_{us}|$ as determined in $K_{e3}$ decays, or $|V_{cs}|$ as
measured in $D_{e3}$ decays. For these cases we have:
\label\vijpluselectron
\eq
| \twV_{ij} | \equiv \left| (\hsm_\lft )_{ij} \right| = | V_{ij} | \;
   \Bigl[ 1 + \Delta_\mu \Bigr] - \Re(\delta \twh_\lft^{ij} + \delta
   \twh_\rht^{ij}).
\eeq
On the other hand, the matrix element $V_{cd}$ is measured in the process
$\nu_\mu d \to c X$, which, to linear order in the new physics, is
sensitive only to the left-handed coupling. In this case,
\eq
| \twV_{cd} | = | V_{cd} | \; \Bigl[ 1  + \Delta_e \Bigr]
		- \Re( \delta \twh_\lft^{cd} ).
\eeq
In other words, there is no general expression for the relation between
$\twV_{ij}$ and $V_{ij}$ -- it must be calculated on a case-by-case
basis.
\endtopic

We may now use these parameters in the lagrangian. The terms of most
practical interest are the $W$ mass term, and the gauge-fermion couplings
of eqs.~\newem\ through \newcc. The coefficient of $W^\dagger_\mu W^\mu$
becomes:
\label\newwmass
\eq  \mw^2 = \mz^2 \cw^2 \left[ 1 - {\alpha S \over 2 ( \cw^2 - \sw^2)}  +
   {\cw^2 \; \alpha T \over  \cw^2 - \sw^2}  + { \alpha U \over 4 \sw^2}
   - {\sw^2 (\Delta_e + \Delta_\mu) \over \cw^2 - \sw^2} \right], \eeq
where $S$, $T$ and $U$ are still defined as in eq.~\studefs. The
electromagnetic interactions are straightforward to write down:
\label\newemcur
\eq  \Scl_{\rm em} = -e \Bigl[
   \ol{f} \gamma^\mu Q \, f \; A_\mu + \ol{f} \sigma^{\mu\nu}
   ( d_\lft \Pl + d_\rht \Pr) \, f \; F_{\mu\nu} \Bigr].
\eeq
The final form for the neutral-current interactions is:
\label\newnewnc
\eqa  \Scl_{\rm nc} &= -{e\over \sw \cw} \; \Bigl[ \ol{f} \gamma^\mu [(
   \gsm_\lft + \delta g_\lft) \Pl + ( \gsm_\rht + \delta g_\rht )\Pr] \, f
   \; Z_\mu + \ol{f} \sigma^{\mu\nu} ( n_\lft \Pl + n_\rht \Pr) \, f
   \; Z_{\mu\nu} \Bigr], \eolnn
 \delta g_{\sss L(R)}^{ij} &= \delta_{ij} \; {\gsm_{i,\lft(\rht)} \over
   2} \; \left( \alpha T - \Delta_e  - \Delta_\mu \right) \eol
   & \qquad - Q_i \, \delta_{ij}
   \; \left({ \alpha S \over 4 ( \cw^2 - \sw^2)} - { \cw^2\sw^2 \; \alpha T
    \over \cw^2 - \sw^2} + {\cw^2 \sw^2 (\Delta_e + \Delta_\mu) \over
   \cw^2 - \sw^2} \right) + \delta \twg_{\sss L(R)}^{ij}.\eeolnn \eeq
In the above expression for $\delta g_{\sss L(R)}$, the coefficient of
$\gsm_{\sss L(R)}$ represents a universal overall correction to the
strength of the interaction. The next term, proportional to the fermion
charge, $Q_i$, can be considered as a shift in the effective electroweak
mixing angle, $(\sw^2)_{\rm eff}$, as measured in neutral-current
experiments. The final term consists of any direct new contributions to the
current. Of these three types of contributions, this last term --- and only
this term --- can contain flavour-changing neutral currents (FCNC's).

Finally, the charged-current interaction becomes:
\label\newnewcc
\eqa
\Scl_{\rm cc} = & -{e\over \sqrt{2}\sw} \; \Bigl[ \ol{f} \gamma^\mu [(
 \hsm_\lft + \delta h_\lft ) \Pl + (\hsm_\rht + \delta h_\rht )\Pr] \, f'
 \; W^\dagger_\mu \eolnn
& \qquad\qquad\qquad \qquad\qquad\qquad
+ \ol{f} \sigma^{\mu\nu} ( c_\lft \Pl + c_\rht
\Pr) \, f' \; W^\dagger_{\mu\nu} \Bigr] + \cc, \eeol
\eeq
where, for leptons,
\label\leptonfac
\eqa
\delta h_\lft^{\nu_i \ell_j} &=  \delta_{ij} \left( - { \alpha S \over 4
( \cw^2 - \sw^2)} + { \cw^2\; \alpha T \over 2 (\cw^2 - \sw^2)}  + {\alpha
U \over 8 \sw^2} - { \cw^2\; (\Delta_e + \Delta_\mu) \over 2 (\cw^2 -
\sw^2)} \right) + \delta \twh_\lft^{\nu_i \ell_j}, \eolnn
\delta h_\rht^{\nu_i \ell_j} &= \delta \twh_\rht^{\nu_i \ell_j}~, \eeol
\eeq
while for quarks:
\label\quarkfac
\eqa
\delta h_\lft^{u_i d_j} &= \twV_{ij} \left( - { \alpha S \over 4
   ( \cw^2 - \sw^2)} + { \cw^2\; \alpha T \over 2 (\cw^2 - \sw^2)}  +
{\alpha U \over 8 \sw^2} - { \cw^2\; (\Delta_e + \Delta_\mu) \over 2 (\cw^2
- \sw^2)} \right) + \delta \twh_\lft^{u_i d_j}, \eolnn
   \delta h_\rht^{u_i d_j} &= \delta \twh_\rht^{u_i d_j}. \eeol
\eeq
As for the neutral currents, in the above equations the coefficients of the
$\delta_{ij}$ and $\twV_{ij}$ terms are universal corrections, while all
other corrections are non-universal. Also, as discussed previously, we have
not substituted for $\twV_{ij}$ in the above equations since there is no
general relation between $\twV_{ij}$ and $V_{ij}$ .

We may now apply these expressions to a number of relevant observables.

\section{Applications to Observables}

The ultimate goal of this analysis is to use current experimental data to
constrain the new-physics parameters. In this section we compute
expressions for a large number of observables in terms of our various
effective couplings. We also report on the results of detailed fits for
these couplings where this is appropriate.

Our starting point is the effective lagrangian we have constructed, which
consists only of the standard model supplemented by those effective
interactions which have the lowest few dimensions. It is important to keep
in mind the existence of a potentially infinite number of terms which we
have not written down, and which we imagine are suppressed compared to the
ones kept by additional powers of $1/M$. Because of the existence of these
other terms, when computing the implications for observables, it would be
{\it inconsistent} to work beyond linear order in our lowest-dimension
effective interactions, and to still neglect the higher-dimension operators
which we have not included. As a result we limit ourselves to working only
to linear order in the couplings of our effective lagrangian.

We consider only those observables to which our new-physics parameters
contribute at tree level for a slightly different reason. In this case any
contribution which is obtained by inserting an effective operator into a
loop can be cancelled by a small correction to the coefficient of the
operators which contribute to the same observable at tree level.
Alternatively, loop graphs tell us how the effective operators mix as they
are renormalized down from the high scale, $M$, where the new physics is
integrated out, to the lower scales where the observable in question is
measured.

Having said this, there is still one situation where working to higher
order in our effective couplings, or going beyond the tree-level
calculations, makes sense. This is in the case where measurements of an
observable are sufficiently precise to strongly exclude new-physics
contributions, even beyond linear order or tree-level. Although we cannot
ever rule out the possibility that a nonzero contribution from one of our
low-dimension operators at quadratic order (say) may cancel with a linear
contribution of an operator we have neglected, the likelihood of this
becomes more implausible the stronger the cancellation that is required. As
a result, we can use precision measurements to bound our interactions
beyond linear order in their coefficients, and beyond tree-level in their
contributions, provided that we are aware of this possibility of
cancellation.

In practice, sufficiently well-measured observables are usually associated
with processes that do not arise, or are highly suppressed, in the standard
model due to (approximate) conservation laws or selection rules. For the
present purposes we only work beyond linear order for observables which
involve flavour-changing neutral currents (FCNC's). These are highly
suppressed in the standard model, and so typically first arise to quadratic
order in our effective interactions. When computing these bounds we
therefore work to this order, but any limits we find that are not very
strong must be considered suspect, since they could easily be circumvented
through cancellations with higher-dimension operators. For all other
processes it suffices to work to linear order in the new-physics
parameters. At a practical level, this implies that most of the
coefficients of operators which are not of the SM form -- such as the
magnetic terms in eqs.~\newnewnc\ and \newnewcc, or of most CP-violating
interactions, will not be bounded in this analysis since they do not
interfere with the standard model.

Similarly, we only consider the loop-level contributions of our effective
operators to neutral-meson mixing, $\epsilon_{\sss K}$, anomalous magnetic
moments, and to particle electric dipole moments ($edm$'s), all of which
are measured (or bounded) with great precision. Again, weak limits should
not be taken too seriously, due to possible effects of cancellations
amongst the contributions of various operators.

For FCNC's, and well-measured quantities like $(g-2)_e$ and $(g-2)_\mu$, as
well as $edm$'s, only one (or, sometimes, two) observable is required to
bound each effective operator. In this case we simply quote the upper bound
that is required for the appropriate effective coupling. Most of the other
interactions can contribute to a great many quantities. In this instance we
perform a full fit to all of the observables using the entire effective
lagrangian. For comparison purposes, we report here on two types of fits.
In the first, called the `individual fit', only one parameter at a time is
allowed to be nonzero. This fit will obviously yield the most stringent
constraints on the parameter in question, since no possibility exists for
cancellations. The second procedure (the `simultaneous fit') allows all
parameters to vary simultaneously. Because of cancellations most
parameters are less constrained in this fit, and certain combinations
remain unconstrained entirely in this case. As we describe the various
observables, we also indicate which parameters are not bounded, and hence
can be excluded from the simultaneous fit.

\ref\exotic{P. Langacker and D. London, \prd{38}{88}{886}.}
\ref\nardi{E. Nardi, E. Roulet and D. Tommasini, \npb{386}{92}{239}.}
Much of the material in this section is adapted from Refs.~\exotic\ and
\nardi. Where numbers are given, we use $\alpha=1/128$ and $\sw^2=0.23$.

\subsection{Flavour-changing neutral currents}

As mentioned above, our only excursion past linear order in new physics
comes about in this section. In the standard model, there are no FCNC's at
tree level, and most loop-induced FCNC's are calculated to be extremely
small. Thus, FCNC's are a smoking gun for new physics, and it is useful to
investigate the prospects for their detection.

The terms in our effective lagrangian (see eqs.~\newemcur\ and \newnewnc)
which can lead to FCNC's are
\eqa
\label\fcnc
\Scl_{\rm fcnc} = & -{e\over \sw \cw} \; \Bigl[ \ol{f} \gamma^\mu
	(\delta g_\lft \Pl + \delta g_\rht \Pr) \, f \; Z_\mu
+ {1\over M} \; \ol{f} \sigma^{\mu\nu} ( n_\lft \Pl + n_\rht \Pr) \, f
   \; Z_{\mu\nu} \Bigr] \eolnn
& \qquad -{e\over M} \, \ol{f} \sigma^{\mu\nu} ( d_\lft \Pl + d_\rht
\Pr) \, f \; F_{\mu\nu} ~, \eeol
\eeq
where $\delta g^{ij} = \delta \twg^{ij}$ for $i \ne j$, and we introduce a
factor of $1/M$ in the `magnetic' terms for dimensional purposes. We discuss
the three types of FCNC terms ($\delta g_{\lft,\rht}$, $n_{\lft,\rht}$ and
$d_{\lft,\rht}$) in turn.

\topic{$\delta g_{\lft,\rht}$'s}
The strongest constraints on leptonic FCNC's come from the absence of the
decays $\mu\to 3e$ and $\tau\to 3\ell$. For this type of decay we find
\eq
\label\mueee
\Gamma(L \to 3\ell) = {\gf m_\lft^5\over 48 \pi^3} \left[ \left(
 \gsm_{\ell,\lft} \right)^2 + \left( \gsm_{\ell,\rht} \right)^2 \right]
\left[ \left\vert \delta \twg_\lft^{\ell L}\right\vert^2  + \left\vert
\delta \twg_\rht^{\ell L}\right\vert^2 \right], \eeq
\ref\pdb{Particle Data Group, Phys.~Rev.~{\bf D45} (1992) Vol.~45, part
II.}
where the masses of the final state particles have been ignored. Using the
experimental bounds on $\mu\to 3e$ and $\tau\to 3\ell$ \pdb, the limits
shown in Table (II) are obtained.

\ref\delphi{P. Abreu \etal, DELPHI Collaboration, \plb{298}{93}{247}.}
\ref\uaone{C. Albajar \etal, UA1 Collaboration, \plb{262}{91}{163}.}
\midinsert
$$\vbox{\tabskip=0pt \offinterlineskip
\halign to \hsize{\strut#& #\tabskip 1em plus 2em minus .5em&
\hfil#\hfil &#&\hfil#\hfil &#& \hfil#\hfil &#\tabskip=0pt\cr
\noalign{\hrule}\noalign{\smallskip}\noalign{\hrule}\noalign{\medskip}
&& Quantity && Upper Bound && Source &\cr
\noalign{\medskip}\noalign{\hrule}\noalign{\medskip}
&& $\vert\delta \twg_{\lft,\rht}^{e\mu}\vert$ && $2\times 10^{-6}$ &&
			$\mu\not\to 3e$ \pdb &\cr
&& \omit && $1\times 10^{-2}$ && $Z \not\to e\mu$ \delphi &\cr
\noalign{\smallskip}
&& $\vert\delta \twg_{\lft,\rht}^{e\tau}\vert$
 	&& $6\times 10^{-3}$ && $\tau \not\to 3\ell$ \pdb &\cr
&& \omit && $2 \times 10^{-2}$ && $Z \not\to e\tau$ \delphi &\cr
\noalign{\smallskip}
&& $\vert\delta \twg_{\lft,\rht}^{\mu\tau}\vert$
 	&& $6\times 10^{-3}$ && $\tau\not\to 3\ell$ \pdb &\cr
&& \omit && $2 \times 10^{-2}$ && $Z\not\to \mu\tau$ \delphi &\cr
\noalign{\smallskip}
&& $\vert\delta \twg_{\lft,\rht}^{ds}\vert$
 	&& $2\times 10^{-5}$ && $K_\lft\to\mu^+\mu^-$ \pdb &\cr
&& \omit && $3\times 10^{-4}$ && $\Delta m_{K_\lft K_{\sss S}}$ \pdb &\cr
\noalign{\smallskip}
&& $\left[{\rm Re} (\delta \twg_\lft^{ds} \delta \twg_\rht^{*ds} )
\right]^{1/2}$  && $8\times 10^{-5}$ && " &\cr
\noalign{\smallskip}
&& $\vert\delta \twg_{\lft,\rht}^{uc}\vert$
 	&& $4\times 10^{-4}$ && $D^0$-${\overline{D^0}}$ mixing \pdb &\cr
\noalign{\smallskip}
&& $\left[{\rm Re} (\delta \twg_\lft^{uc} \delta \twg_\rht^{*uc})
\right]^{1/2}$ && $1\times 10^{-4}$ && " &\cr
\noalign{\smallskip}
&& $\vert\delta \twg_{\lft,\rht}^{db}\vert$, $\vert\delta
\twg_{\lft,\rht}^{sb} \vert$ && $2\times 10^{-3}$ &&
$B\not\to\ell^+\ell^- X$ \uaone &\cr
\noalign{\medskip}\noalign{\hrule}\noalign{\smallskip}\noalign{\hrule}
}}$$
\centerline{\bf Table (II)}
\medskip
\noindent {\eightrm Constraints on the flavour changing neutral current
parameters $\ss \delta g_{\lft,\rht}^{ij} = \delta \twg_{\lft,\rht}^{ij}$,
for $\ss i \ne j$.}
\endinsert

There are also bounds from $Z \not\to \ell L$. The contribution to this
process is
\eq
\Gamma(Z\to \ell L) = {\alpha\Mz\over 6\sw^2\cw^2}
\left[ \left\vert \delta \twg_\lft^{\ell L} \right\vert^2
+ \left\vert \delta \twg_\rht^{\ell L} \right\vert^2 \right],
\eeq
which, when combined with the experimental limits in \delphi\ leads to the
constraints in Table (II).

For the $ds$ FCNC, the strongest constraint comes from the decay
$K_\lft\to \mu^+ \mu^-$.
\ref\nirsilver{Y. Nir and D. Silverman, \prd{42}{90}{1477}; D. Silverman,
\prd{45}{92}{1800}.}
Using the analogue of eq.~\mueee\ for the quarks in the Kaon system, and
following the analysis of Ref.~\nirsilver\ one finds
\eq
{BR(K_\lft \to \mu^+ \mu^-) \over BR(K^+ \to \mu^+\nu_\mu) } =
{\tau(K_\lft) \over \tau(K^+)} \;
{8 \left[ \left( \gsm_{\mu,\lft} \right)^2 + \left( \gsm_{\mu,\rht}
\right)^2 \right]  \left[ \left\vert \delta \twg_\lft^{ds}\right\vert^2
+ \left\vert \delta \twg_\rht^{ds}\right\vert^2 \right] \over
\vert V_{us}\vert^2}~,
\eeq
where $\tau(K)$ represents the corresponding $K$-meson lifetime.
\ref\geng{C.Q. Geng and J.N. Ng, \prd{44}{91}{1}.}
Ref.~\pdb\ gives $BR(K_\lft \to \mu^+ \mu^-) = (7.0\pm 0.82) \times
10^{-9}$, and the long-distance contribution from the $2\gamma$
intermediate state is found to be \geng\ $(6.83\pm 0.46)\times 10^{-9}$. In
light of this, we assume that the rate for $K_\lft \to \mu^+ \mu^-$ is
explained by the standard model, and require that the new physics
contribution be smaller than the experimentally measured value plus
$1.64\sigma$ (which corresponds to 90\% c.l.). This gives the bound in
Table (II).

\ref\lrmatrix{G. Beall, M. Bander and A. Soni, \prl{48}{82}{848}.}
There is also a constraint on the $ds$ FCNC from the $K_\lft$-$K_{\sss S}$
mass difference. We find
\label\klksfcnc
\eq
\Delta M_{\sss K} = {\gf\over\sqrt{2}} \left[
\left\vert \delta \twg_\lft^{ds}\right\vert^2
+ \left\vert \delta \twg_\rht^{ds}\right\vert^2
+ 2 (0.77) {\rm Re} (\delta \twg_\rht^{ds} \delta \twg_\lft^{*ds}) \right]
{4\over 3}f_{\sss K}^2 m_{\sss K} B_{\sss K}~,
\eeq
where we have used the results of \lrmatrix\ for the left-right matrix
element. We require that this contribution be less than the experimental
value ($+1.64\sigma$), leading to the constraints in Table (II). Note that
these limits are weaker than those from $K_\lft \to \mu^+ \mu^-$.

The constraints on $uc$ FCNC's are due to the absence of
$D^0$-${\overline{D^0}}$ mixing. Using eq.~\klksfcnc, adapted to the $D$
system, and taking $B_{\sss D}=1$, $f_{\sss D}=200$ MeV, we find the
constraints in Table (II).

Finally, the FCNC's involving the $b$-quark are constrained by using the
process $B\to\mu \mu X$. One has \nirsilver:
\eq
{BR(B \to \mu \mu X) \over BR(B \to \mu \nu_\mu X) } =
{ 4 \left[ \left( \gsm_{\mu,\lft} \right)^2 + \left( \gsm_{\mu,\rht}
\right)^2 \right] \left[ \left\vert \delta \twg_\lft^{db}\right\vert^2 +
\left\vert \delta \twg_\rht^{db} \right\vert^2 + \left\vert \delta
\twg_\lft^{sb}\right\vert^2 + \left\vert \delta \twg_\rht^{sb}
\right\vert^2 \right] \over \vert V_{ub}\vert^2 +
F_{ps} \vert V_{cb}\vert^2 }~,
\eeq
where $F_{ps}\simeq 0.5$ is a phase space factor. The constraints on the
FCNC parameters are given in Table (II), where we have used $BR(B \to \mu
\mu X) < 5 \times 10^{-5}$ \uaone.

\topic{$n_{\lft,\rht}$'s}
The $n_{\lft,\rht}$ FCNC's can be bounded in the same way as the $\delta
\twg_{\lft,\rht}$'s. For leptonic FCNC's we use the decays $\mu\to 3e$ and
$\tau\to 3\ell$. The contribution of the $n_{\lft,\rht}$'s to these decays
is found to be
\eq
\label\sigmamuee
\Gamma(L \to 3\ell) = {\gf m_\lft^5\over 30 \pi^3}
\left[ \left( \gsm_{\ell,\lft} \right)^2 + \left( \gsm_{\ell,\rht}
\right)^2 \right] {m_\lft^2\over M^2} \left[ \left\vert n_\lft^{\ell L}
\right\vert^2 + \left\vert n_\rht^{\ell L}\right\vert^2 \right].
\eeq
What is noteworthy here, and indeed in all of the following processes, is
the suppression factor $m^2/M^2$. From this we can deduce that low-energy
limits on FCNC's will not put terribly strong constraints on the
$n_{\lft,\rht}$'s. The bounds from $\mu\to 3e$ and $\tau\to 3\ell$ are
shown in Table (III). We take $M=1$ TeV.

\midinsert
$$\vbox{\tabskip=0pt \offinterlineskip
\halign to \hsize{\strut#& #\tabskip 1em plus 2em minus .5em&
\hfil#\hfil &#&\hfil#\hfil &#& \hfil#\hfil &#\tabskip=0pt\cr
\noalign{\hrule}\noalign{\smallskip}\noalign{\hrule}\noalign{\medskip}
&& Quantity && Upper Bound && Source &\cr
\noalign{\medskip}\noalign{\hrule}\noalign{\medskip}
&& $\vert n_{\lft,\rht}^{e\mu}\vert$ && $0.02$ &&
			$\mu\not\to 3e$ \pdb &\cr
&& \omit && $0.08$ && $Z\not\to e\mu$ \delphi &\cr
\noalign{\smallskip}
&& $\vert n_{\lft,\rht}^{e\tau}\vert$,
 	&& 3 && $\tau\not\to 3\ell$ \pdb &\cr
&& \omit && $0.09$ && $Z\not\to e\tau$ \delphi &\cr
\noalign{\smallskip}
&& $\vert n_{\lft,\rht}^{\mu\tau}\vert$,
 	&& 3 && $\tau\not\to 3\ell$ \pdb &\cr
&& \omit && 0.1 && $Z\not\to\mu\tau$ \delphi &\cr
\noalign{\smallskip}
&& $\vert n_{\lft,\rht}^{ds}\vert$ && 0.02 && $K_\lft\to\mu^+\mu^-$ \pdb
  &\cr
&& \omit && 0.3 && $\Delta m_{K_\lft K_{\sss S}}$ \pdb &\cr
\noalign{\smallskip}
&& $\vert n_{\lft,\rht}^{uc}\vert$ && 0.1
  	&& $D^0$-${\overline{D^0}}$ mixing \pdb &\cr
\noalign{\smallskip}
&& $\vert n_{\lft,\rht}^{db}\vert$, $\vert n_{\lft,\rht}^{sb}\vert$
  	&& 0.2 && $B\not\to\ell^+\ell^- X$ \uaone &\cr
\noalign{\medskip}\noalign{\hrule}\noalign{\smallskip}\noalign{\hrule}
}}$$
\centerline{\bf Table (III)}
\medskip
\noindent {\eightrm Constraints on the dimension-five coupling parameters
$\ss n_{\lft,\rht}^{ij}, i \ne j$ using a new-physics scale of $\ss M=1$
TeV.}
\endinsert

The contributions of the $n_{\lft,\rht}$ terms to $Z\to \ell L$ are
\eq
\Gamma(Z\to \ell L) = {\alpha\Mz\over 3\sw^2\cw^2} \; {\Mz^2\over M^2}
\left[ \left\vert n_\lft^{\ell L} \right\vert^2
+ \left\vert n_\rht^{\ell L} \right\vert^2 \right].
\eeq
{}From these one can deduce the limits shown in Table (III). Note that, for
FCNC's involving $\tau$'s, in contrast to the $\delta \twg_{\lft,\rht}$'s,
the constraints from the absence of leptonic FCNC in $Z$ decays are {\it
stronger} than those from low energy.

Turning to the $ds$ FCNC's, and adapting the results of Ref.~\nirsilver\ we
have
\eq
{BR(K_\lft \to \mu^+ \mu^-) \over BR(K^+ \to \mu^+\nu_\mu) } =
{192\over 15} \; { \tau(K_\lft) \over \tau(K^+) } \;
{ m_{\sss K}^2\over M^2} \; { \left[ \left( \gsm_{\mu,\lft} \right)^2 +
\left( \gsm_{\mu,\rht} \right)^2 \right] \left[ \left\vert n_\lft^{ds}
\right\vert^2 + \left\vert n_\rht^{ds}\right\vert^2 \right] \over
\vert V_{us}\vert^2}~,
\eeq
giving the bounds in Table (III). Extracting constraints from the
$K_\lft$-$K_{\sss S}$ mass difference is more problematic. The difficulty
is that, using the $n_{\lft,\rht}$ operators, new hadronic matrix elements
are obtained. Rather than trying to evaluate these, we will simply
estimate the contribution to $\Delta M_{\sss K}$ as
\eq
\Delta M_{\sss K} \sim {4\gf\over\sqrt{2}} {m_{\sss K}^2\over M^2}
\left[ \left\vert n_\lft^{ds}\right\vert^2 + \left\vert n_\rht^{ds}
\right\vert^2 \right] {4\over 3}f_{\sss K}^2 m_{\sss K} B_{\sss K}~,
\eeq
where we have taken the unknown matrix element to be of the order of the
left-left matrix element, and have ignored the left-right mixing term. With
this order-of-magnitude estimate, one obtains the limits shown in Table
(III). Note that these are much weaker than those due to
$K_\lft\to\mu^+\mu^-$.

The same difficulty is encountered in using $D^0$-${\overline{D^0}}$ mixing
to constrain the $uc$ FCNC's. Estimating the contribution to $\Delta
M_{\sss D}$ in the same way as was done for the Kaon system, we find the
constraints shown in Table (III).

For $B\to\mu \mu X$ we have
\eq
{BR(B \to \mu \mu X) \over BR(B \to \mu \nu_\mu X) } =
{192\over 30} \; {m_{\sss B}^2\over M^2} \;
{ \left[ \left( \gsm_{\mu,\lft} \right)^2 + \left( \gsm_{\mu,\rht}
\right)^2 \right] \left[ \left\vert n_\lft^{db}\right\vert^2 + \left\vert
n_\rht^{db}\right\vert^2 + \left\vert n_\lft^{sb}\right\vert^2 + \left\vert
n_\rht^{sb}\right\vert^2 \right] \over \vert V_{ub}\vert^2 + F_{ps} \vert
V_{cb}\vert^2 }~,
\eeq
which yields the constraints in Table (III).

It is noteworthy that the constraints on the couplings $n^{bs}_{\lft,\rht}$
from low-energy experiments are very weak. Contrary to the naive
expectation that the bounds on $B \to \mu \mu X$ would preclude any chance
of detecting $Z\to s{\bar b},{\bar s} b$ at LEP, we see here that the two
processes sample completely {\it different} operators. Should new physics
produce terms like $\ol{b} \sigma^{\mu\nu} ( n_\lft \Pl + n_\rht \Pr) \, s
\; Z_{\mu\nu}$, then such FCNC's could be seen at LEP without having been
ruled out in $B$ decays -- indeed, it would be very foolish to overlook
this possibility.

This example beautifully illustrates the power of the effective lagrangian
approach. By systematically listing all operators up to a given order in
$1/M$, one can discover terms which can give rise to physically observable
effects, which might not otherwise have been considered.

\topic{$d_{\lft,\rht}$'s}
The analysis leading to bounds on the $d_{\lft,\rht}$ FCNC's is similar to
that for the $n_{\lft,\rht}$'s, with the following important differences.
First, certain $d_{\lft,\rht}$'s can be bounded directly from the process
$f\to f' \gamma$. Second, due to the fact that the photon is massless,
decays such as $\mu\to 3e$ are {\it not} suppressed by powers of
$m_\mu/\mz$, as they are in the case of the $Z$-FCNC's. In fact, as we
shall see, there is a logarithmic enhancement of such decays. Finally,
there are no bounds on the $d_{\lft,\rht}$'s from FCNC's at the $Z$-peak,
since the contribution from photon exchange is very much suppressed in
these processes.

The strongest constraints on leptonic $d_{\lft,\rht}$ FCNC's come from the
experimental limits on the decays $\mu\to e\gamma$, $\tau\to e\gamma$ and
$\tau\to\mu\gamma$. The contribution of the $d_{\lft,\rht}$'s to these
decays is
\label\muegamma
\eq
\Gamma(L \to \ell\gamma) = \alpha \, {m_\lft^3\over M^2}
\left[ \left\vert d_\lft^{\ell L}\right\vert^2 +
\left\vert d_\rht^{\ell L}\right\vert^2 \right].
\eeq
Using the experimental limits from Ref.~\pdb\ gives the bounds listed in
Table (IV).

The leptonic $d_{\lft,\rht}$ FCNC's will also lead to the process $L\to
3\ell$. As noted earlier, since the photon is massless, there is no
suppression of this process relative to $L\to \ell\gamma$ due to the photon
propagator. The only suppression is due to an additional factor of
$\alpha$, as well as from the 3-body phase space, as compared to 2-body
phase space. And this is partially compensated for by a large logarithm,
due to the presence of an infrared mass singularity in the limit $m_\ell\to
0$. The contribution of the $d_{\lft,\rht}$'s to the process $\mu\to 3e$ is
found to be
\label\dmueee
\eq
\Gamma(\mu\to 3e) = {8\alpha^2\over\pi} \, {m_\mu^3\over M^2} \left[
{1\over 24} \log \left({m_\mu^2\over m_e^2}\right) - {1\over 18} \right]
\left[ \left\vert d_\lft^{e\mu}\right\vert^2 +
\left\vert d_\rht^{e\mu}\right\vert^2 \right].
\eeq
The contribution to the processes $\tau\to e\ell^+\ell^-$ and
$\tau\to\mu\ell^+\ell^-$ is obtained in the obvious way from the above
equation. This leads to the constraints shown in Table (IV). As mentioned
above, the constraints on the $d_{\lft,\rht}^{\ell L}$ from $L\not\to
3\ell$ are only slightly weaker than those arising from
$L\not\to\ell\gamma$.

\ref\bsg{E. Thorndike (CLEO Collaboration), talk given at the {\em 1993
Meeting of the American Physical Society}, Washington, D.C., April, 1993;
R. Ammar et.\ al.\ (CLEO Collaboration), \prl{71}{93}{674}.}
\ref\eamm{R.S. Van Dyck Jr., P.B. Schwinberg and H.G. Dehmelt,
\prl{59}{87}{26}.}
\ref\muamm{J. Bailey \etal, \plb{68}{77}{191}; \npb{150}{79}{1}.}
\ref\electronedm{K. Abdullah \etal, \prl{65}{90}{2347}.}
\ref\muonedm{J. Bailey \etal, \npb{150}{79}{1}.}
\ref\edmref{S.M. Barr and W.J. Marciano, in {\it CP Violation}, ed. C.
Jarlskog (World Scientific, Singapore, 1989), p.\ 455.}
\ref\neutronedm{K.F. Smith \etal, \plb{234}{90}{191};
I.S. Altarev \etal, \jetpl{44}{86}{360}{44}{86}{460}.}
\midinsert
$$\vbox{\tabskip=0pt \offinterlineskip
\halign to \hsize{\strut#& #\tabskip 1em plus 2em minus .5em&
\hfil#\hfil &#&\hfil#\hfil &#& \hfil#\hfil &#\tabskip=0pt\cr
\noalign{\hrule}\noalign{\smallskip}\noalign{\hrule}\noalign{\medskip}
&& Quantity && Upper Bound && Source &\cr
\noalign{\medskip}\noalign{\hrule}\noalign{\medskip}
&& $\vert d_{\lft,\rht}^{e\mu}\vert$ && $2\times 10^{-9}$ &&
			$\mu\not\to e\gamma$ \pdb &\cr
&& \omit && $3\times 10^{-9}$ && $\mu\not\to 3e$ \pdb &\cr
\noalign{\smallskip}
&& $\vert d_{\lft,\rht}^{e\tau}\vert$ && $5\times 10^{-5}$ &&
			$\tau\not\to e\gamma$ \pdb &\cr
&& \omit && $2\times 10^{-4}$ && $\tau\not\to 3\ell$ \pdb &\cr
\noalign{\smallskip}
&& $\vert d_{\lft,\rht}^{\mu\tau}\vert$ && $8\times 10^{-5}$ &&
			$\tau\not\to \mu\gamma$ \pdb &\cr
&& \omit && $3\times 10^{-4}$ && $\tau\not\to 3\ell$ \pdb &\cr
\noalign{\smallskip}
&& $\vert d_{\lft,\rht}^{ds}\vert$ && $2\times 10^{-7}$ &&
$K_\lft\to\mu^+\mu^-$ \pdb &\cr
&& \omit && $4\times 10^{-3}$ && $\Delta m_{K_\lft K_{\sss S}}$ \pdb &\cr
\noalign{\smallskip}
&& $\vert d_{\lft,\rht}^{uc}\vert$ && $6\times 10^{-3}$
  	&& $D^0$-${\overline{D^0}}$ mixing \pdb &\cr
\noalign{\smallskip}
&& $\vert d_{\lft,\rht}^{db}\vert$ && $3\times 10^{-5}$ &&
$B\not\to\ell^+\ell^- X$ \uaone &\cr
\noalign{\smallskip}
&& $\vert d_{\lft,\rht}^{sb}\vert$ && $2\times 10^{-4}$ && $b\to s\gamma$
\bsg &\cr
&& \omit && $4\times 10^{-5}$ && $B\not\to\ell^+\ell^- X$ \uaone &\cr
\noalign{\smallskip}
&& $|d^{ee}_\lft + d^{ee}_\rht|$ && $8 \times 10^{-6}$ && $a(e)$ \eamm
&\cr
\noalign{\smallskip}
&& $|d^{\mu\mu}_\lft + d^{\mu\mu}_\rht|$ && $1 \times 10^{-4}$ &&
$a(\mu)$ \muamm &\cr
\noalign{\smallskip}
&& $|d^{ee}_\lft - d^{ee}_\rht|$ && $8 \times 10^{-10}$ && Atomic $edm$'s
 \electronedm &\cr
\noalign{\smallskip}
&& $|d^{\mu\mu}_\lft - d^{\mu\mu}_\rht|$ && 0.05 && $(g_\mu -2)/2$ \muonedm
&\cr
\noalign{\smallskip}
&& $|d^{\tau\tau}_\lft - d^{\tau\tau}_\rht|$ && 5 &&
$e^+e^-\to\tau^+\tau^-$ \edmref &\cr
\noalign{\smallskip}
&& $|d^{{\nu_i}{\nu_i}}_\lft - d^{{\nu_i}{\nu_i}}_\rht|$ ($i=e,\mu$) &&
$5 \times 10^{-4}$ && $\nu e\to\nu e$ \edmref &\cr
\noalign{\smallskip}
&& $|d^{dd}_\lft - d^{dd}_\rht|,|d^{uu}_\lft - d^{uu}_\rht|$ &&
$6 \times 10^{-8}$ && neutron $edm$ \neutronedm &\cr
\noalign{\medskip}\noalign{\hrule}\noalign{\smallskip}\noalign{\hrule}
}}$$
\centerline{\bf Table (IV)}
\medskip
\noindent {\eightrm Constraints on the dimension-five coupling parameters
$\ss d_{\lft,\rht}^{ij}$, using a new-physics scale of $\ss M=1$ TeV.}
\endinsert

One constraint on the $ds$ FCNC's comes from the process $K_\lft\to
\mu^+\mu^-$. Adapting eq.~\dmueee\ to the process $s\to d\mu^+\mu^-$, and
using the results of Ref.~\nirsilver, we have
\eq
{BR(K_\lft \to \mu^+ \mu^-) \over BR(K^+ \to \mu^+\nu_\mu) } =
{3072 \, \alpha^2 \, \pi^2 \over m_{\sss K}^2 \, M^2 \, \gf^2} \;
{ \tau(K_\lft) \over \tau(K^+) } \; \left[ {1\over 24}
\log \left({m_s^2\over m_d^2}\right) - {1\over 18} \right]
{\left[ \left\vert d_\lft^{ds}
\right\vert^2 + \left\vert d_\rht^{ds}\right\vert^2 \right] \over
\vert V_{us}\vert^2}~.
\eeq
We take $m_s=150$ MeV and $m_d=5$ MeV, leading to the bounds in Table (IV).
There are also constraints from the $K_\lft$-$K_{\sss S}$ mass difference.
However, as was the case for the $n_{\lft,\rht}^{ds}$ FCNC's, we encounter
new hadronic matrix elements. Therefore, once again, we simply give a rough
estimate of the contribution to  $\Delta M_{\sss K}$:
\eq
\Delta M_{\sss K} \sim {e^2 \over 2M^2}
\left[ \left\vert d_\lft^{ds}\right\vert^2 + \left\vert d_\rht^{ds}
\right\vert^2 \right] {4\over 3}f_{\sss K}^2 m_{\sss K} B_{\sss K}~,
\eeq
This leads to the order-of-magnitude limits in Table (IV). As was the case
for the $n_{\lft,\rht}^{ds}$'s, these limits are much weaker than those due
to $K_\lft\to\mu^+\mu^-$.

The $uc$ FCNC's are constrained by $D^0$-${\overline{D^0}}$ mixing. Using
the same procedure as was done for the $K_\lft$-$K_{\sss S}$ mass
difference, we arrive at the bounds in Table (IV).

The process $B\to\mu \mu X$ constrains both the $db$ and $sb$ FCNC's:
\eq
{BR(B \to \mu \mu X) \over BR(B \to \mu \nu_\mu X) } =
{1536 \, \alpha^2 \, \pi^2 \over m_{\sss B}^2 \, M^2 \, \gf^2} \; \left[
{1\over 24} \log \left({m_b^2\over m_q^2}\right) - {1\over 18} \right]
{\left[ \left\vert d_\lft^{qb}
\right\vert^2 + \left\vert d_\rht^{qb}\right\vert^2 \right] \over
\vert V_{ub}\vert^2 + F_{ps} \vert V_{cb}\vert^2 }~,
\eeq
in which $q=d,s$. The bounds are shown in Table (IV). For the $sb$ FCNC,
there is also a limit due to the experimental measurement of $b\to
s\gamma$ \bsg. Using eq.~\muegamma\ we find the contribution of the
$d_{\lft,\rht}^{sb}$ to this process to be
\eq
BR(b\to s\gamma) = \tau_{\sss B} \,
\alpha \, {m_b^3\over M^2}
\left[ \left\vert d_\lft^{sb}\right\vert^2 +
\left\vert d_\rht^{sb}\right\vert^2 \right].
\eeq
Taking $\tau_{\sss B}=1.49~psec$ leads to the constraints in Table (IV).

\subsection{Anomalous Magnetic Moments}

Extremely precise measurements exist for the anomalous magnetic moments of
the electron and muon: $a_i = (g_i -2)/2 = (\mu_i/\mu_{\ssb_i}) - 1$, for
$i = e,\mu$ and $\mu_{\ssb_i} \equiv e_i/2 m_i$. The current best
experimental values for the electron and positron are \eamm:
\eq \eqalign{
a(e^-) &= 1~159~652~188.4~(4.3) \times 10^{-12} \cr
a(e^+) &= 1~159~652~187.9~(4.3) \times 10^{-12}, \cr} \eeq
while that for the muon is \muamm:
\eq \eqalign{
a(\mu^-) &= 1~165~937~(12) \times 10^{-9} \cr
a(\mu^+) &= 1~165~911~(11) \times 10^{-9}. \cr} \eeq
These are in good agreement with the corresponding SM (\ie\ QED)
predictions
\ref\ammtheory{T. Kinoshita, in {\it Quantum Electrodynamics}, edited by
T. Kinoshita (World Scientific, Singapore, 1990); T. Kinoshita and W.
Marciano, {\it ibid}.}
\ammtheory:
\eq \eqalign{
a^{\rm th}(e) &= 1~159~652~140~(5.3)~(4.1)~(27.1) \times 10^{-12} \cr
a^{\rm th}(\mu) &= 1~165~919~18~(191) \times 10^{-11} \cr} \eeq
The largest error in $a^{\rm th}(e)$ is due to the determination of
$\alpha$, a fact which presently limits using the comparison with $a^{\rm
th}(e)$ as a precision test of QED.

The quantities $d_{\lft,\rht}^{ee}$ and $d_{\lft,\rht}^{\mu\mu}$ contribute
directly to this observable, by an amount:
\eq
\delta a_i = {2 m_i \over M} \; (d_\lft^{ii} + d_\rht^{ii}) \eeq
where $i = e, \mu$. We obtain our bound by requiring that this contribution
be smaller than the corresponding 1.64$\sigma$ experimental error. Taking
$M= 1$ TeV, as before, produces the constraints shown in Table (IV).

\subsection{Electric Dipole Moments}

The difference between $d^{ff}_\lft$ and $d^{ff}_\rht$ contributes to the
corresponding particle's electric dipole moment, as defined as the
coefficient of the term in the particle's energy shift which is linear in
the applied field. For a fundamental fermion such a definition is
equivalent to defining $d_f$ as the coefficient of the following effective
electromagnetic interaction:
\eq
- \, {d_f  \over 2} \; \ol{f} i \sigma^{\mu\nu} \gamma_5 f \; F_{\mu\nu}~.
\eeq
In terms of the interactions in our effective lagrangian we therefore have
\eq \eqalign{
-i d_f &= {e\over M} \; (d_\lft^{ff} - d_\rht^{ff}) \cr
&= (2 \times 10^{-17} \, \ecm) \; (d_\lft^{ff} - d_\rht^{ff}). \cr}
\eeq
In this last line, as in Table (IV), we take the fiducial value  $M = 1$
TeV.

\ref\atomedm{S.A. Murthy \etal, \prl{63}{89}{965};
S.K. Lamoreaux \etal, {\it Phys.~Rev. Lett.} {\bf 59} (1987) 2275;
T.G. Vold \etal, \prl{52}{84}{2229}.}
Extremely good limits currently exist for the electric dipole moment of
various atoms \electronedm, \atomedm\ and for the neutron \neutronedm. The
atomic bounds permit the inference of a very strong bound on the $edm$ of
the electron \electronedm. Using these limits we arrive at the bounds given
in Table (IV). The constraints on the $edm$'s of light quarks are obtained
from the experimental limit on the neutron $edm$. For both the electron and
quark $edm$'s there is some uncertainty in extracting these bounds since
many operators in the  underlying theory can generate either atomic or
neutron $edm$'s. For electrons we quote here the bounds as given by the
experimental groups themselves. This is not done for the neutron, since
here there is the additional uncertainty associated with computing the
nucleon matrix element of the quark-level operator. To be conservative we
simply use the estimate
\eq
d_n \sim d_u \sim d_d,
\eeq
and quote a limit on $d_q$ which is ten times weaker than the measured
bound on $d_n$.

The $edm$'s of other particles may also be constrained. That for the muon
is directly limited by the experiment which measures $(g-2)_\mu$ \muonedm.
One may attempt to obtain a bound for the $\tau$, $\nu_e$ and $\nu_\mu$
electric moments from the observed absence of the effects that such moments
would produce in the reactions $e^+e^- \to \tau^+\tau^-$ or in $\nu e$
scattering \edmref. Since the $edm$ enters quadratically into these cross
sections, these bounds can only be inferred to the extent that
cancellations with other effective interactions can be ignored. As may be
seen from Table (IV), although this may be plausible for the neutrino
moments, it is not justified for the tau lepton.

More indirect limits on neutrino moments also exist in certain
circumstances \edmref. If neutrinos are Dirac (or pseudo-Dirac) particles
then right-handed sterile neutrinos likely exist and are light enough to be
produced from left-handed neutrinos, {\it via} the magnetic moment
interactions, in stars, supernovae and in the early universe. We do not
include these bounds here since we have excluded sterile right-handed
neutrinos from our low-energy particle content.

\vfill\eject
\subsection{Charged Currents}

We next turn to the bulk of the constraints on the effective lagrangian,
charged-current and neutral-current data. Since many of the effective
interactions can contribute to many observables, we evaluate the remaining
bounds by performing a global fit.

Some of the low-dimension effective interactions are not bounded to the
order we work. This is because many operators do not contribute at all to
linear order in their coefficients. This is true, in particular, for terms
which do not, on grounds of helicity conservation, interfere appreciably
with SM contributions. As a result we will not be bounding the magnetic
terms in eq.~\newnewcc. The same is true for the right-handed currents in
this equation, except insofar as they contribute to linear order to the CKM
matrix elements, and in $K\to 3\pi$ decays. We remind the reader that in
what follows we take $\alpha=1/128$ and $\sw^2=0.23$.

\topic{The $W$ Mass}
In the presence of new physics, the relationship between the $W$- and the
$Z$-mass is modified. Inspection of eq.~\newwmass\ gives the following
result:
\eqa
\Mw^2 =& (\Mw^2)_\SM \left[ 1 - {\alpha S \over 2 ( \cw^2 - \sw^2)} +
   {\cw^2 \; \alpha T \over  \cw^2 - \sw^2}  + { \alpha U \over 4 \sw^2}
   - {\sw^2 (\Delta_e + \Delta_\mu) \over \cw^2 - \sw^2} \right] \eolnn
=& (\Mw^2)_\SM [1 - 0.00723 \, S + 0.0111 \, T +0.00849 \, U
-0.426(\Delta_e+\Delta_\mu)] .\eeol
\eeq
Recall that the $\Delta_f$ are defined in eq.~\deltadef\ above, and since
we do not assume the conservation of lepton number, the sum in the
definition of $\Delta_f$ is over all light neutrinos.

\topic{CKM Unitarity}
The strongest experimental constraint on new couplings of the $W$ to quarks
comes from the unitarity of the CKM matrix. As discussed previously, the
relation between the parameters in the lagrangian, $\twV_{ij}$, and the
measured quantities, $V_{ij}$, is altered due to new physics. For $V_{ud}$
and $V_{us}$, the relation is as given in eq.~\vijpluselectron. This is not
the case for $V_{ub}$, which is measured using the endpoint spectrum of
semileptonic $B$ decays. However, in any event, because $V_{ub}$ is so
small, we drop terms of order $V_{ub}^2$. The three-generation relation,
$\sum_{i=1}^3 \vert \twV_{ui}\vert^2 = 1$, leads to
\label\firstrowunitarity
\eqa
\vert V_{ud}\vert^2 + \vert V_{us}\vert^2 + \vert V_{ub}\vert^2 &=
1 -  2 \Delta_\mu
+ 2 \vert V_{ud} \vert \Re( \delta \twh_\lft^{ud} + \delta \twh_\rht^{ud})
+ 2 \vert V_{us} \vert \Re( \delta \twh_\lft^{us} + \delta \twh_\rht^{us})
\eolnn
& ~~~~~~~~ + 2 \left[\Re(V_{ub}) \Re(\delta \twh_\lft^{ub})
+ \Im(V_{ub}) \Im(\delta \twh_\lft^{ub}) \right], \eeol
\eeq
where on the right-hand side we have replaced $\twV_{ij}$ by $V_{ij}$.
Note that the new-physics parameters $\Re( \delta \twh_\lft^{us})$,
$\Re(\delta \twh_\lft^{ub})$ and $\Im(\delta \twh_\lft^{ub})$ appear only
in the above expression; they contribute to no other charged-current
observables (at tree-level and to linear order). Therefore, in the
simultaneous fit, only the sum of terms $\vert V_{us} \vert \Re( \delta
\twh_\lft^{us}) + [\Re(V_{ub}) \Re(\delta \twh_\lft^{ub}) + \Im(V_{ub})
\Im(\delta \twh_\lft^{ub})]$ can ever be constrained, and we present the
bound on this combination only.

The second row of the CKM matrix is similar, except that $V_{cd}$ is
measured differently, as discussed in the section 4.2. We find
\label\secondrowunitarity
\eq
\vert V_{cd}\vert^2 + \vert V_{cs}\vert^2 + \vert V_{cb}\vert^2 =
1 + 2 \vert V_{cd} \vert \Re(\delta \twh_\lft^{cd})
+ 2 \vert V_{cs} \vert \Re(\delta \twh_\lft^{cs} + \delta \twh_\rht^{cs})
+ 2 |V_{cb}| \Re(\delta \twh_\lft^{cb}),
\eeq
where we have neglected all $\Delta_{e,\mu}$ terms, as they are much better
constrained in other processes. In the simultaneous fit of all parameters,
only the sum $\vert V_{cd} \vert \Re(\delta \twh_\lft^{cd}) + \vert V_{cs}
\vert \Re(\delta \twh_\lft^{cs} + \delta \twh_\rht^{cs}) + |V_{cb}|
\Re(\delta \twh_\lft^{cb})$ arises; the individual new-physics parameters
are unconstrained by our fit. As a consequence, as before, we present only
the bound on this sum when we perform the simultaneous fit.

\topic{Lepton Universality}
Lepton universality is tested in pion and tau decays. It is straightforward
to calculate
\eqa
R_\pi &\equiv {\Gamma(\pi\to e\nu) \over \Gamma(\pi\to\mu\nu) }
= R_\pi^\SM \left( 1 + 2 \Delta_e - 2 \Delta_\mu \right)~, \eolnn
R_{\tau} &\equiv {\Gamma(\tau\to e\nu\bar{\nu}) \over
\Gamma(\mu\to e\nu{\overline\nu}) } = R_\tau^\SM \left(1 + 2 \Delta_\tau -
2 \Delta_\mu \right)~, \eolnn
R_{\mu\tau} &\equiv {\Gamma(\tau\to\mu\nu{\overline\nu}) \over
\Gamma(\mu\to e\nu{\overline\nu}) } = R_{\mu\tau}^\SM \left(1 + 2
\Delta_\tau - 2 \Delta_e \right)~. \eeolnn
\eeq
Universality is also tested in leptonic Kaon decays, but the resulting
bounds are weaker than those given above.

\ref\mudecay{P. Langacker and D. London, \prd{38}{88}{907};
\prd{39}{89}{266}.}
\topic{Right-Handed Currents}
Right-handed leptonic charged currents can be constrained through the
Michel parameters in muon decay. However, it is necessary to go beyond
linear order in the new parameters, so we do not include these
measurements in our analysis. For a complete description of muon decay
including lepton-number-violating operators, see Refs.~\exotic, \mudecay.

\ref\donholstein{J.F. Donoghue and B.R. Holstein, \plb{113}{82}{382}.}
Hadronic right-handed currents can be constrained by considering PCAC
(partial conservation of axial-vector currents) predictions for $K_{\pi 3}$
decay relative to $K_{\pi 2}$ decay \donholstein. In terms of our
parameters, this gives
\eq
|\twh_\rht^{ud}| < {8\times 10^{-4}\over \vert V_{us} \vert}~,~~~~~
|\twh_\rht^{us}| < {8\times 10^{-4}\over \vert V_{ud} \vert}~.
\eeq
Following Ref.~\exotic, in our fit we consider these upper bounds as
$1\sigma$ errors.

\ref\cdhs{CDHS Collaboration, H. Abramowicz \etal, \zpc{12}{82}{225},
\zpc{15}{82}{19}.}
There are also constraints on right-handed currents in $d\leftrightarrow c$
and $s\leftrightarrow c$ transitions coming from the measurements of the
$y$ distributions in $\nu d,\nu s \to \mu^- c$ and ${\bar\nu}{\bar d},
{\bar\nu}{\bar s}\to \mu^+{\bar c}$ \cdhs. Here too, however, the
new-physics parameters appear first at quadratic order, so that the (rather
weak) bounds extracted in this way are somewhat unreliable, prone as they
are to cancellations from dimension-six operators. For this reason, we do
not include these constraints in our fits.

\subsection{Neutral Currents -- Low Energy}

As shown in section 5.1, flavour-changing neutral currents involving
charged particles are very well constrained, at least for the $\delta
\twg_{\lft,\rht}$ couplings. On the other hand, there are no bounds on
FCNC's in the neutrino sector, and we will therefore allow for this
possibility. In practice, however, since we are working to linear order in
the new physics, only the nonstandard flavour-conserving
$Z\nu{\overline\nu}$ vertex will be constrained -- the flavour-changing
couplings always appear quadratically in the expressions for the
observables.

\topic{The $\rho$-parameter}
As was discussed in section 2.4, the $\rho$-parameter, defined as the
relative strength of the low-energy neutral- and charged-current
interactions, can be read off from from the universal corrections to the
neutral-current and charged-current couplings ((eqs.~\newnewnc\ and
\newnewcc, respectively), taking also into account the corrections to the
$W$-mass (eq.~\newwmass). This gives
\eq
\rho = 1 + \alpha T,
\eeq
as before.

\topic{Deep-Inelastic $\nu$ Scattering}
$\nu q$ neutral current scattering is measured via the ratios
\eq
R_\nu = {\sigma(\nu N\to \nu X)\over \sigma(\nu N\to\mu^- X)}~,\qquad
R_{\overline\nu} = {\sigma({\overline\nu} N\to{\overline\nu} X)\over
\sigma({\overline\nu} N\to\mu^+ X)}~.
\eeq
The presence of new physics affects not only the neutral-current process in
the numerator, but also the reference charged-current process in the
denominator. In principle one must also worry about subsidiary quantities
such as the quark distribution functions and the charm threshold. However,
it has been argued in Ref.~\exotic\ that these are rather insensitive to
new physics effects. The logic of our discussion here follows the lines
laid out in this reference.

We wish to compute $R/R^\SM$, which we write as follows
\eq  {R \over R^\SM} = { \sigma(\ol{\nu} N \to \nu X) /
\sigma^\SM(\ol{\nu} N \to \nu X) \over \sigma(\ol{\nu} N \to \mu X) /
\sigma^\SM(\ol{\nu} N \to \mu X) }. \eeq
We next calculate the numerator and denominator of this expression.

The charged-current process is dominated by $u\leftrightarrow d$
transitions. We therefore compute the corrections only to this process
using the effective lagrangian. A subtlety arises, however, in that our
new effective interactions also appear in the reference charged-current SM
cross section. This is because the SM result must be taken as a function of
$V_{ud}$, as it is measured in superallowed beta decays, which itself
receives corrections from $\Re(\twh_{\lft,\rht}^{ud})$ \etc\ \ Whereas one
might expect these corrections to cancel with the corresponding terms in
$\sigma(\ol{\nu} N \to \mu X)$, this is not the case since $V_{ud}$ from
beta decay is corrected by the right-handed term, $\Re(\twh^{ud}_\rht)$,
while $\sigma(\ol{\nu} N \to \mu X)$ is not. As a result we find:
\label\cccorrection
\eqa
{ \sigma(\nu N\to\mu^- X) \over \sigma^\SM(\nu N\to\mu^- X) } &=
  { \left(1-2\Delta_e-2\Delta_\mu\right)  \left( \twV_{ud} + \delta
  \twh_\lft^{ud} \right)^2 \sum_i \left(\delta_{i\mu} + \delta
  \twh_\lft^{\nu_i \mu} \right)^2 \over \vert V_{ud} \vert^2} \eolnn
&= 1 + 2\Delta_\mu - 2\Delta_e - 2\; {\Re(\delta \twh_\rht^{ud}) \over
  \vert V_{ud} \vert}~. \eeol
\eeq
Note that all of the dependence on the oblique corrections, $S$, $T$ and
$U$, cancels between the corrections to the charged current couplings, and
those to the mass, $\Mw$, of the virtual $W$.

We now turn to the neutral-current part of the ratio: $\sigma(\ol{\nu} N
\to \nu X) /  \sigma^\SM(\ol{\nu} N \to \nu X)$. The easiest way to make
contact with the measurements is through the effective parameters,
$\epsilon_{\lft,\rht}(a)$. These parameters provide the conventional
parametrization of the effective neutrino-quark interaction that is probed
in deep-inelastic scattering:
\eq
-{\cal L}_{\rm eff}^{\nu q} = {4 \gf \over \sqrt{2}} {\bar\nu}_\lft
\gamma^\mu \nu_\lft \sum_{a=u,d,...} \left[ \epsilon_\lft(a)
{\overline q}_\lft^a \gamma_\mu q_\lft^a + \epsilon_\rht(a)
{\overline q}_\rht^a \gamma_\mu q_\rht^a \right],
\eeq
This is to be compared with the quark-flavour-diagonal piece of the
low-energy limit of our general effective lagrangian,
\label\ourlowlagrangian
\eq
-{\cal L}^{\nu q} = {8 \gf \over \sqrt{2}} \sum_{ij} {\bar\nu}_{i}
\gamma^\mu (\gsm_\lft + \delta \twg_\lft)^{ij} \Pl  \nu_{j}
\sum_{a=u,d,...} \ol{q}_a \gamma_\mu [(\gsm_\lft + \delta g_\lft)^{aa} \Pl
+ ( \gsm_\rht + \delta g_\rht )^{aa}\Pr] \, q_a~.
\eeq
We do not included a right-handed neutrino current in the above equation
since this cannot interfere with the SM contribution, and so cannot
contribute to linear order. For the same reason, even though FCNC's are
allowed, only the flavour-conserving piece $\delta
\twg_\lft^{{\nu_\mu}{\nu_\mu}}$ contributes to linear order.

Comparing these lagrangians, and dividing out by the square-root of the
charged-current correction factor, $\sqrt{\hbox{C.C.}} = 1 + \Delta_\mu -
\Delta_e - \Re(\delta \twh_\rht^{ud})/|V_{ud}|$ of eq.~\cccorrection, then
gives
\eqa
\epsilon_{\lft(\rht)}(a) &= {2 g^{\nu_\mu\nu_\mu}_\lft \;
g^{aa}_{\lft(\rht)} \over \sqrt{\hbox{C.C.}}} \eolnn
&= \gsm_{a,\lft(\rht)} \left[ 1 + \alpha T + 2 \delta
\twg_\lft^{\nu_\mu\nu_\mu} - 2 \Delta_\mu + {\Re(\delta \twh_\rht^{ud})
\over | V_{ud} |} \right] \eol
&\qquad - Q_a \, \left({ \alpha S \over 4 ( \cw^2 - \sw^2)} - { \cw^2\sw^2
\; \alpha T \over \cw^2 - \sw^2} + {\cw^2 \sw^2 (\Delta_e + \Delta_\mu)
\over \cw^2 - \sw^2} \right) + \delta \twg_{\lft (\rht)}^{aa} ~. \eeolnn
\eeq

The cross-section ratios, $R_\nu$ and $R_{\ol{\nu}}$, are finally given by
the following expressions \exotic: $R_\nu=g_\lft^2 + r g_\rht^2$ and
$R_{\overline\nu}= g_\lft^2 + g_\rht^2/\ol{r}$. Here $r=0.383$,
$\ol{r}=0.371$ are numbers, and the parameters $g_i^2$ (not to be confused
with the effective neutral-current couplings $g_{\lft,\rht}^{ab}$!) are
related to the $\epsilon_i(a)$ by $g_i^2 \equiv \epsilon_i(u)^2 +
\epsilon_i(d)^2$, with $i=L,R$. Combining these results gives the
quantities which we use in our fit:
\eqa
\left(g_\lft^2\right) &= \left( g^2_\lft \right)_\SM - 0.00269 \, S
 + 0.00663 \, T -1.452 \Delta_\mu  - 0.244 \Delta_e
\eolnn
&\qquad + 0.620 \, \Re(\delta \twh_\rht^{ud})
- 0.856 \, \delta \twg_\lft^{dd} + 0.689 \, \delta \twg_\lft^{uu}
+ 1.208 \, \delta \twg_\lft^{\nu_\mu\nu_\mu}~, \eolnn
\left(g_\rht^2\right) &= \left(g_\rht^2\right)_\SM
+ 0.000937 \, S - 0.000192 \, T + 0.085 \Delta_e - 0.0359 \Delta_\mu
\eolnn
&\qquad + 0.0620 \, \Re(\delta \twh_\rht^{ud})
+ 0.156 \, \delta \twg_\rht^{dd} - 0.311 \, \delta \twg_\rht^{uu}
+ 0.121 \, \delta \twg_\lft^{\nu_\mu\nu_\mu}~.  \eeolnn
\eeq

\topic{Neutrino-Electron Scattering}
Neutrino-electron scattering data is conventionally expressed in terms of
an effective vector- and axial-vector electron coupling, defined by the
following effective neutrino-electron interaction
\eq
-{\cal L}^{\nu_\mu e} = {2 G_F \over \sqrt{2}} {\bar\nu}_\lft \gamma^\mu
\nu_\lft \; {\bar e}\gamma_\mu \left( g_{e\ssv} - g_{e\ssa} \gamma_5
\right) e,
\eeq
which, when compared with our effective lagrangian (\cf\
eq.~\ourlowlagrangian\ above), gives
\eq
g_{e\ssv} = {2 g^{\nu_\mu\nu_\mu}_\lft \;\left( g^{ee}_\lft + g^{ee}_\rht
\right) \over \sqrt{\hbox{C.C.}}},\qquad g_{e\ssa} = {2
g^{\nu_\mu\nu_\mu}_\lft \;\left( g^{ee}_\lft - g^{ee}_\rht \right) \over
\sqrt{\hbox{C.C.}}}~.
\eeq

An additional complication arises here due to the fact that the
$\nu_\mu$-$e$ scattering cross sections are not all measured relative to
the same charged-current cross section \exotic. The high energy experiments
at CERN and Fermilab normalize to $\nu N \to \mu^- X$ as in deep-inelastic
scattering, so that the charged-current correction factor is
$1/\sqrt{\hbox{C.C.}}\Bigr|_{\sss HE} = 1 - \Delta_\mu + \Delta_e +
\Re(\delta\twh_\rht^{ud})/\vert V_{ud} \vert$, as before. The low-energy
experiments from BNL, on the other hand, normalize to the quasielastic
process $\nu_\mu n \to \mu^- p$, which gives a slightly different
correction factor: $1/\sqrt{\hbox{C.C.}}\Bigr|_{\sss LE} = 1 - \Delta_\mu +
\Delta_e$. Because the global averages of these measurements are dominated
by the high-energy experiments, we use $1/\sqrt{\hbox{C.C.}}\Bigr|_{\sss
HE}$ in our fits. We find
\eqa
g_{e\ssv} &= \left( g_{e\ssv}\right)_\SM + 0.00723 \, S  - 0.00541 \, T +
0.656 \Delta_e + 0.730 \Delta_\mu \eolnn
& \qquad + \delta \twg_\lft^{ee} + \delta \twg_\rht^{ee}  - 0.074 \, \delta
  \twg_\lft^{\nu_\mu\nu_\mu} - 0.037 \, \Re(\delta\twh_\rht^{ud})~, \eol
g_{e\ssa} &= \left( g_{e\ssa} \right)_\SM - 0.00395 \, T + 1.012 \Delta_\mu
  + \delta \twg_\lft^{ee} - \delta \twg_\rht^{ee} - 1.012 \, \delta
  \twg_\lft^{\nu_\mu\nu_\mu} - 0.0506 \, \Re(\delta\twh_\rht^{ud})~.\eeolnn
\eeq

\topic{Atomic Parity Violation/Weak-Electromagnetic Interference}
The low-energy lagrangian describing atomic parity violation is
conventionally parametrized as
\eq
-{\cal L}^{e q} = {\gf\over \sqrt{2}} \sum_i \left[ C_{1a} \,
{\bar e}\gamma_\mu \gamma_5 e \, {\overline q}_a \gamma^\mu q_a +
C_{2a} \, {\bar e}\gamma_\mu e \, {\overline q}_a \gamma^\mu \gamma_5 q_a
\right],
\eeq
in which
\eq
C_{1a} = 2 \left( g_\lft^{ee} - g_\rht^{ee} \right)
\left( g_\lft^{aa} + g_\rht^{aa} \right)~, ~~~~~
C_{2a}^\SM =  2 \left( g_\lft^{ee} + g_\rht^{ee} \right)
\left( g_\lft^{aa} - g_\rht^{aa} \right).
\eeq
Inserting our expressions for $\delta g_\lft$'s and $\delta g_\rht$ we find
\eqa
C_{1u} &= C_{1u}^\SM + 0.00482 \, S - 0.00493 \, T
+ 0.631 (\Delta_e + \Delta_\mu) \eolnn
&\qquad + 0.387 \,\delta \twg_\lft^{ee} - \delta \twg_\lft^{uu}
- 0.387 \,\delta \twg_\rht^{ee} - \delta \twg_\rht^{uu}~, \eolnn
C_{1d} &= C_{1d}^\SM - 0.00241 \, S +  0.00442 \, T
- 0.565 (\Delta_e + \Delta_\mu) \eolnn
&\qquad - 0.693 \,\delta \twg_\lft^{ee} - \delta \twg_\lft^{dd}
+ 0.693 \,\delta \twg_\rht^{ee} - \delta \twg_\rht^{dd}~, \eolnn
C_{2u} &= C_{2u}^\SM + 0.00723 \, S - 0.00544 \, T
+ 0.696 (\Delta_e + \Delta_\mu) \eolnn
&\qquad + \delta \twg_\lft^{ee} - 0.08 \,\delta \twg_\lft^{uu}
+ \delta \twg_\rht^{ee} + 0.08 \,\delta \twg_\rht^{uu}~, \eolnn
C_{2d} &= C_{2d}^\SM - 0.00723 \, S + 0.00544 \, T
- 0.696 (\Delta_e + \Delta_\mu) \eolnn
&\qquad - \delta \twg_\lft^{ee} - 0.08 \,\delta \twg_\lft^{dd}
- \delta \twg_\rht^{ee} + 0.08 \,\delta \twg_\rht^{dd}~. \eeol
\eeq
For heavy atoms, the matrix element of this effective interaction within
the atomic nucleus --- containing $N$ neutrons and $Z$ protons --- is
proportional to the `weak charge', $\qw$, defined by:
\eq
\qw(Z,N) = -2 \left[ (2Z+N) C_{1u} + (Z+2N) C_{1d} \right].
\eeq
For cesium, we find
\eqa
\qw({}^{133}_{55}Cs) &= \left[ \qw({}^{133}_{55}Cs) \right]_\SM
- 0.796 \, S -0.0113 \, T + 1.45 (\Delta_e + \Delta_\mu)
+ 147 \left( \delta \twg_\lft^{ee} - \delta \twg_\rht^{ee}\right) \eolnn
& \qquad + 422 \left( \delta \twg_\lft^{dd} + \delta \twg_\rht^{dd}\right)
+ 376 \left( \delta \twg_\lft^{uu} + \delta \twg_\rht^{uu}\right)~. \eeol
\eeq
Note that these expressions are automatically real, even in the presence of
CP violation, since the hermiticity of the lagrangian requires all of the
diagonal elements, $\delta\twg_{\lft,\rht}^{ii}$, to be real.

\subsection{Neutral Currents ($Z$ Peak)}

Our next class of observables concerns those that are measured in $e^+e^-$
collisions at the $Z^0$ resonance. Consider first the $Z$-boson partial
widths. Even in the presence of new physics, one has (neglecting fermion
masses)
\eq  \left[ \Gamma_{f_i} \right]_{\rm tree}  = {\alpha \Mz \over 6\sw^2
\cw^2} \left( |g_\lft^{ii}|^2  +  |g_\rht^{ii}|^2  \right). \eeq
The contributions from the nonstandard operators can be separated simply by
linearizing the above equation about the SM value. This gives
\label\partwidth
\eqa
\Gamma_{f} = & \Gamma^\SM_{f}
\left[ 1 + \alpha T - \Delta_e - \Delta_\mu + { 2 \gsm_{f,\lft} \, \delta
  \twg_\lft^{ff} + 2 \gsm_{f,\rht} \, \delta \twg_\rht^{ff} \over
  (\gsm_{f,\lft})^2 + (\gsm_{f,\rht})^2 } \right. \eolnn
& \qquad  \left. - \;{ 2 \gsm_{f,\lft} + 2 \gsm_{f,\rht} \over
  (\gsm_{f,\lft})^2 +
  (\gsm_{f,\rht})^2} \; Q_f \left( { \alpha S \over 4 ( \cw^2 - \sw^2)}  -
{ \cw^2\sw^2 \alpha T \over \cw^2 - \sw^2}
 + { \cw^2 \sw^2 (\Delta_e + \Delta_\mu) \over \cw^2 -\sw^2 } \right)
\right] .
\eeol
\eeq
Note that this expression holds for neutrinos as well as for charged
particles since the potentially-present neutrino FCNC's do not contribute
to linear order. Using eq.~\partwidth, we find the following partial widths
\eqa
\Gamma_{\ell^+\ell^-} &= \left( \Gamma_{\ell^+ \ell^-}\right)_\SM
\Bigl[ 1 - 0.00230 \, S + 0.00944 T - 1.209 (\Delta_e + \Delta_\mu)
\eolnn
& \qquad -4.29 \,\delta \twg_\lft^{\ell\ell} + 3.66 \,\delta
\twg_\rht^{\ell\ell} \Bigr] ~, \eolnn
\Gamma_{u{\bar u}} &= (\Gamma_{u{\bar u}})_\SM  \Bigl[ 1 - 0.00649 \, S +
0.0124 T - 1.59  (\Delta_e+\Delta_\mu) \eolnn  &\qquad + 4.82 \,\delta
\twg_\lft^{uu} - 2.13 \,\delta \twg_\rht^{uu} \Bigr]~, \eolnn
\Gamma_{d{\bar d}} &= (\Gamma_{d{\bar d}} )_\SM \Bigl[ 1 - 0.00452\, S +
  0.0110 T - 1.41 (\Delta_e + \Delta_\mu) \eolnn
&\qquad - 4.57 \,\delta \twg_\lft^{dd} + 0.828 \,\delta \twg_\rht^{dd}
\Bigr]~, \eolnn
\Gamma_{b{\bar b}} &= (\Gamma_{b{\bar b}})_\SM \Bigl[ 1 - 0.00452\, S +
  0.0110 T - 1.41 (\Delta_e + \Delta_\mu) \eolnn
&\qquad - 4.57 \,\delta \twg_\lft^{bb} + 0.828 \,\delta \twg_\rht^{bb}
\Bigr]~, \eolnn
\Gamma_{\rm had} &= (\Gamma_{\rm had})_\SM
\Bigl[ 1- 0.00518 \, S + 0.0114 T -
1.469 (\Delta_e + \Delta_\mu) \eolnn
& \qquad - 1.01 \Bigl( \delta \twg_\lft^{dd} + \delta \twg_\lft^{ss} +
\delta \twg_\lft^{bb} \Bigr) + 0.183 \Bigl( \delta \twg_\rht^{dd} + \delta
  \twg_\rht^{ss} + \delta \twg_\rht^{bb} \Bigr) ~,\eolnn
&\qquad + 0.822 \Bigl( \delta \twg_\lft^{uu} + \delta \twg_\lft^{cc} \Bigr)
- 0.363 \Bigl( \delta \twg_\rht^{uu} + \delta \twg_\rht^{cc} \Bigr)
\Bigr]\eolnn
\Gamma_{\nu_i{\overline\nu}_i}
&= (\Gamma_{\nu_i{\overline\nu}_i})_\SM \Bigl[ 1 +0.00781 T -
(\Delta_e+\Delta_\mu) + 4 \,\delta \twg_\lft^{\nu_i \nu_i} \Bigr] ~.\eeol
\eeq
The total width is then
\eqa
\Gamma_{\sss Z} &= (\Gamma_\ssz)_\SM \Bigl[ 1 - 0.00385 \, S +0.0105 \, T
  - 1.35 (\Delta_e+\Delta_\mu) + 0.574 \left( \delta \twg_\lft^{uu} +
  \delta \twg_\lft^{cc} \right) \eolnn
& \qquad - 0.254 \left( \delta \twg_\rht^{uu} + \delta \twg_\rht^{cc}
 \right)  + 0.268 \left( \delta \twg_\lft^{\nu_e\nu_e} + \delta
 \twg_\lft^{\nu_\mu\nu_\mu} + \delta \twg_\lft^{\nu_\tau\nu_\tau}
 \right) \eol
&\qquad  - 0.144 \left( \delta \twg_\lft^{ee} + \delta \twg_\lft^{\mu\mu} +
  \delta \twg_\lft^{\tau\tau} \right) + 0.123 \left( \delta \twg_\rht^{ee}
  + \delta \twg_\rht^{\mu\mu} + \delta \twg_\rht^{\tau\tau}
\right) \eolnn
& \qquad - 0.707 \left( \delta \twg_\lft^{dd} + \delta \twg_\lft^{ss} +
  \delta \twg_\lft^{bb} \right) + 0.128 \left( \delta \twg_\rht^{dd} +
  \delta \twg_\rht^{ss} + \delta \twg_\rht^{bb} \right) \Bigr] .
\eeolnn\eeq
Because the $\delta \twg_\lft^{\nu_e\nu_e}$ and $\delta \twg_\lft^{\nu_\tau
\nu_\tau}$ only contribute to our list of observables through the
$Z$-width, only their sum can be bounded in the simultaneous fit.

Various asymmetries are also measured at LEP. In terms of the new-physics
parameters, the expression for the left-right asymmetry, eq.~\lrasymmetry,
becomes,
\eqa
A_{\lft\rht} & = A_{\lft\rht}^\SM +
{4 \, \gsm_{e,\lft} \, \gsm_{e,\rht} \over
\left( (\gsm_{e,\lft})^2 + (\gsm_{e,\rht})^2 \right)^2 }
\left( \gsm_{e,\rht} \, \delta \twg_\lft^{ee} -
\gsm_{e,\lft} \, \delta \twg_\rht^{ee} \right) \eolnn
&= (A_{\sss LR})_\SM - 0.0284 \, S + 0.0201 \, T - 2.574 ( \Delta_e +
  \Delta_\mu) - 3.61 \, \delta \twg_\lft^{ee} - 4.238 \, \delta
\twg_\rht^{ee}~. \eeol
\eeq
Similarly, we obtain the following expressions for, $A_{\sss FB}(f)$, the
forward-backward asymmetries for $e^+e^- \to f\bar{f}$;
\eqa
A_{\sss FB}^{\ell^+\ell^-} &= 	{3\over 4} \; A_{\lft\rht}^{e^+e^-}
  A_{\lft\rht}^{\ell^+\ell^-} \eolnn
&= (A_{\sss FB})_\SM - 0.00677 \, S + 0.00480 \, T - 0.614  (\Delta_e
+ \Delta_\mu) \eolnn
& \qquad\qquad -0.430 \left( \delta \twg_\lft^{ee} + \delta
\twg_\lft^{\ell\ell} \right) -0.505 \left( \delta \twg_\rht^{ee} + \delta
\twg_\rht^{\ell\ell} \right), \eolnn
A_{\sss FB}(b{\bar b}) &= {3\over 4} \; \left( 1 - k_{\sss A} {\alpha_s
  \over\pi} \right) A_{\lft\rht}^{e^+e^-} A_{\sss LR}^{b\bar{b}} \eolnn
& = (A_{\sss FB}(b{\bar b}))_\SM -0.0188 \, S + 0.0133 \, T - 1.70 (\Delta_e
  +\Delta_\mu) \eolnn
& \qquad\qquad -2.36 \,\delta \twg_\lft^{ee} - 2.77 \,\delta \twg_\rht^{ee}
  -0.0322 \,\delta \twg_\lft^{bb} - 0.178 \,\delta \twg_\rht^{bb} \eolnn
A_{\sss FB}(c{\bar c}) &= {3\over 4} \; \left( 1 - k_{\sss A} {\alpha_s
  \over\pi} \right) \; A_{\lft\rht}^{e^+e^-}  A_{\sss LR}^{c\bar{c}} \eolnn
& = (A_{\sss FB}(c{\bar c}))_\SM - 0.0147 \, S + 0.0104 \, T - 1.333(
\Delta_e + \Delta_\mu) \eolnn
& \qquad\qquad - 1.69 \,\delta \twg_\lft^{ee} - 1.99 \,\delta \twg_\rht^{ee}
   + 0.175 \,\delta \twg_\lft^{cc}  + 0.396 \,\delta \twg_\rht^{cc} .
\eeol
\eeq
The factor $(1 - k_A \alpha_s/\pi)$ represents a QCD radiative correction,
as in Ref.~\peskin, for which we use the numerical value 0.93.

\ref\correlations{The OPAL Collaboration, CERN Report CERN-PPE/93-146
(1993) (unpublished).}
We can now determine the phenomenological constraints on the new-physics
parameters in our electroweak Lagrangian. The observables included in our
fit are listed in Table (V) along with their experimental value and the SM
predictions. The standard model values have been calculated with $m_t=150$
GeV and $M_H=300$ GeV. The LEP observables in Table (V) were chosen as they
are closest to what is actually measured and their uncertainties are
relatively weakly correlated. In our analysis we include the correlations
taken from OPAL results \correlations, but note that all LEP experiments
obtain similar results for the correlations.

\ref\lep{M. Swartz, Invited talk at the XVI International Symposium on
Lepton-Photon Interactions, Cornell University, Ithaca New York,
August 10-15, 1993.}
\ref\sld{K. Abe \etal, \prl{70}{93}{2515}.}
\ref\cdf{R. Abe {\sl et al.}, \prl{65}{90}{2243}.}
\ref\uatwo{J. Alitti {\sl et al.}, \plb{276}{92}{354}.}
\ref\paul{P. Langacker, to appear in the {\sl Proceedings of 30 Years of
Neutral Currents}, Santa Monica, February 1993.}
\ref\cesium{M.C. Noecker {\sl et al.},  \prl{61}{88}{310}.}
\ref\doug{D. A. Bryman, Comments Nucl. Part. Phys. {\bf 21}, 101 (1993)}
\ref\smpred{The standard model predictions come from P. Langacker,
Proceedings of the 1992 Theoretical Advanced Study Institute, Boulder CO,
June 1992 which includes references to the original  literature. We thank
P. Turcotte for supplying us with the standard model values for $g_\lft^2$
and $g_\rht^2$.}
\ref\atomictheory{V.A. Dzuba {\sl et al.}, \pla{141}{89}{147}; S.A.
Blundell, W.R. Johnson, and J. Sapirstein, \prl{65}{90}{1411}.}

\pageinsert
$$
\vbox{\offinterlineskip
\halign{&\vrule#&
   \strut\quad#\hfil\quad\cr
\noalign{\hrule}
height2pt&\omit&&\omit&&\omit&\cr
& Quantity && Value && Standard Model &\cr
height2pt&\omit&&\omit&&\omit&\cr
\noalign{\hrule}
height2pt&\omit&&\omit&&\omit&\cr
& $\Mz$ (GeV) && $91.187 \pm 0.007 $ \lep && input & \cr
& $\Gamma_Z$ (GeV) && $ 2.489 \pm 0.007 $ \lep && $2.490 [\pm 0.006]$ &\cr
& $R_e=\Gamma_{had}/\Gamma_{e{\bar e}}$ && $20.743 \pm 0.080$ \lep &&
$20.78 [\pm 0.07]$ & \cr
& $R_\mu=\Gamma_{had}/\Gamma_{\mu{\bar \mu}}$ && $20.764 \pm 0.069$ \lep &&
$20.78 [\pm 0.07]$ & \cr
& $R_\tau=\Gamma_{had}/\Gamma_{\tau{\bar \tau}}$ && $20.832 \pm 0.088$ \lep
&& $20.78 [\pm 0.07]$ & \cr
& $\sigma^h_p $ (nb) && $41.56 \pm 0.14$ \lep && $41.42[\pm 0.06]$ & \cr
& $R_b=\Gamma_{b\bar b}/\Gamma_{had}$ && $0.2200 \pm 0.0027$ \lep &&
	$ 0.2162 [\pm 0.0007]$ & \cr
& $A_{\sss FB}(e)$ && $0.0153 \pm 0.0038$ \lep && $0.0141$ & \cr
& $A_{\sss FB}(\mu)$ && $0.0132 \pm 0.0026$ \lep && $0.0141$ & \cr
& $A_{\sss FB}(\tau)$ && $0.0204 \pm 0.0032$ \lep && $0.0141$ & \cr
& $A_{pol}(\tau)$ && $0.142 \pm 0.017$ \lep && $ 0.137$ & \cr
& $A_e (P_\tau)$ && $ 0.130 \pm 0.025$ \lep && $ 0.137$ & \cr
& $A_{\sss FB}(b)$ && $0.098 \pm 0.006 $ \lep && $0.096$ & \cr
& $A_{\sss FB}(c)$ && $ 0.075 \pm 0.015$ \lep && $ 0.068$ &\cr
& $A_{\sss LR} $ && $ 0.100 \pm 0.044$ \sld && $ 0.137$ & \cr
height2pt&\omit&&\omit&&\omit&\cr
\noalign{\hrule}
height2pt&\omit&&\omit&&\omit&\cr
& $\Mw$ (GeV) && $79.91 \pm 0.39$ \cdf && $80.18 $ &\cr
& $\Mw/\Mz$ && $0.8798 \pm 0.0028 $ \uatwo && 0.8793 & \cr
& $g_\lft^2(\nu N\to \nu X) $ && $0.3003 \pm 0.0039$ \paul && $0.3021$ &
\cr
& $g_\rht^2 (\nu N\to \nu X) $ && $0.0323\pm 0.0033 $ \paul && $0.0302$ &
\cr
& $g_{e\ssa}(\nu e \to \nu e)$ && $ -0.508 \pm 0.015 $ \paul && $-0.506$
& \cr
& $g_{e\ssv} (\nu e \to \nu e)$ && $ -0.035\pm 0.017 $\paul && $-0.037$
& \cr
& $\qw(Cs)$ && $-71.04 \pm 1.58 \pm [0.88]$ \cesium && $-73.20$ & \cr
height2pt&\omit&&\omit&&\omit&\cr
\noalign{\hrule}
height2pt&\omit&&\omit&&\omit&\cr
& $ | V_{ud}|^2 + |V_{us}|^2 + |V_{ub}|^2 $ && $0.9992\pm 0.0014$ \paul &&
1 &\cr
& $ | V_{cd}|^2 + |V_{cs}|^2 + |V_{cb}|^2 $ && $1.043 \pm 0.40$ \pdb && 1
&\cr
& $R_\pi/R_\pi^\SM$
	&& $1.003 \pm 0.003$ \doug && 1 & \cr
& $R_\tau/R_\tau^\SM $  &&  $ 0.960\pm 0.024 $ \doug && 1 & \cr
& $R_{\mu\tau}/R_{\mu\tau}^\SM $  &&  $ 0.968\pm 0.024 $ \doug && 1
& \cr height2pt&\omit&&\omit&&\omit&\cr
\noalign{\hrule}}}
$$
\centerline{\bf Table (V)}
\medskip
\noindent {\eightrm Experimental values for the electroweak observables
included in the global fit. The $\ss Z^0$ measurements are the 1993 LEP
results taken from Ref.~\lep. The couplings extracted from neutrino
scattering data are the current world averages taken from Ref.~\paul. The
SM values are for $\ss m_t=150$ GeV and $\ss M_{\sss H}=300$ GeV \smpred.
We have not shown theoretical errors in the SM values due to uncertainties
in the radiative corrections, $\ss \Delta r$, and due to uncertainties in
$\ss \Mz$, as they are in general overwhelmed by the experimental errors.
The exception is the error due to uncertainty in $\ss \alpha_s$, shown in
square brackets. We include this error in quadrature in our fits. The error
in square brackets for $\ss \qw(Cs)$ reflects the theoretical uncertainty
in the atomic wavefunctions \atomictheory\ and is also included in
quadrature with the experimental error. All other quantities are as defined
in the text.}
\endinsert

The expressions for most of the observables in Table (V) have already been
discussed. Of the remaining observables $A_{pol}(\tau)$, or $P_\tau$, is
the polarization asymmetry defined by $A_{pol}(\tau) = (\sigma_\rht -
\sigma_\lft)/ (\sigma_\rht + \sigma_\lft)$, where $\sigma_{\sss L,R}$ is
the cross section for the reaction $e^+e^- \to \tau \ol{\tau}$ with  a
correspondingly polarized $\tau$ lepton; $A_e(P_\tau)$ is the joint
forward-backward/left-right asymmetry as normalized in Ref.~\paul. $A_{\sss
LR}$ is the polarization asymmetry which has been measured by the SLD
collaboration at SLC \sld. The expressions for $A_{pol}(\tau)$ and
$A_e(P_\tau)$ are the same as the expression we have already given for
$A_{\sss LR}$. The two remaining observables can be obtained using results
already given. In particular the parameter $R$ is defined as $R=
\Gamma_{had}/\Gamma_{l\bar{l}}$, and $\sigma^h_p =
12\pi\Gamma_{e\bar{e}}\Gamma_{had}/\Mz^2\Gamma_Z^2$ is the hadronic cross
section at the $Z$-pole.

We first consider the case in which only one of the parameters in our
Lagrangian is nonzero. The results of this fit are given in Column (2) of
Table (VI). In this case strong bounds on each of the parameters are
obtained since there is no possibility of cancellations. This procedure is
commonly used by most practictioners when bounding effective couplings.
Although the constraints obtained in this way are the tightest bounds
possible, they are clearly artificial in the sense that real underlying
physics would change more than one of the parameters. Ideally one could
calculate the effects of new physics on the parameters of the global
electroweak Lagrangian and then do a global fit on the specific parameters
of interest.

Conversely, a simultaneous fit to all of the effective parameters gives the
most conservative bounds, since cancellations can occur among different
parameters. We have performed such a fit. As mentioned in previous
sections, we have excluded some of the parameters in this simultaneous fit.
In particular there are a number of quantities that only appear in
particular linear combinations, and so only these combinations can be
bounded. Some examples are $ \vert V_{cd} \vert \Re(\delta \twh_\lft^{cd})
+ \vert V_{cs} \vert \Re(\delta \twh_\lft^{cs} + \delta \twh_\rht^{cs}) +
|V_{cb}| \Re(\delta \twh_\lft^{cb})$ in the unitarity of the CKM matrix and
$\delta \twg_\lft^{\nu_e\nu_e}+ \delta \twg_\lft^{\nu_\tau \nu_\tau}$ in
the $Z$ width. As more measurements become available the omitted parameters
will be able to be included in the simultaneous fit. The results of this
simultaneous fit are given in Table (VI).

There are a number of interesting features in Table (VI). What is perhaps
most surprising is that, despite the large number of parameters, most of
them are constrained, and the bounds are fairly tight. This reflects the
richness and complementarity of the experimental data. The most significant
result of our fit is that every single parameter is consistent with zero,
the standard model value. There is no evidence for physics beyond the
standard model.

One should be cautioned to not take the central values of this fit too
literally. With so many free parameters the central values obtained by the
fit are naturally not unique. We find that the errors seem to be stable so
that the best values lie within the error bounds irrespective of the search
strategy.

\pageinsert
$$
\vbox{\offinterlineskip
\halign{&\vrule#&
   \strut\quad#\hfil\quad\cr
\noalign{\hrule}
height2pt&\omit&&\omit&&\omit&\cr
& Parameter && Individual Fit && Global Fit &\cr
height2pt&\omit&&\omit&&\omit&\cr
\noalign{\hrule}
height2pt&\omit&&\omit&&\omit&\cr
& $S$ && $-0.10 \pm .16 $ && $-0.2\pm 1.0$ & \cr
& $T$ && $ +0.01 \pm .17 $ && $-0.02\pm 0.89$ & \cr
& $U$ && $-0.14 \pm 0.63$ && $+0.3\pm 1.2$ & \cr
& $\Delta_e$ && $-0.0008\pm .0010$  && $-0.0011 \pm .0041$ & \cr
& $\Delta_\mu$ && $+0.00047 \pm .00056$  && $+0.0005\pm .0039$ & \cr
& $\Delta_\tau$ && $-0.018 \pm 0.008$ && $-0.018\pm .009$ & \cr
& $\Re(\delta\twh^{ud}_\lft)$ && $-0.00041\pm .00072$ && $+0.0001\pm
.0060$& \cr
& $\Re(\delta\twh^{ud}_\rht)$ && $-0.00055\pm .00066$ && $+0.0003\pm
.0073$& \cr
& $\Im(\delta\twh^{ud}_\rht)$ && $0 \pm 0.0036$ && $-0.0036 \pm .0080$&
\cr
& $\Re(\delta\twh^{us}_\lft)$ && $-0.0018\pm .0032$ && --- & \cr
& $\Re(\delta\twh^{us}_\rht)$ && $-0.00088\pm .00079$ && $+0.0007\pm
.0016$& \cr
& $\Im(\delta\twh^{us}_\rht)$ && $0 \pm 0.0008$ && $-0.0004\pm .0016$& \cr
& $\Re(\delta\twh^{ub}_\lft)$,$\Im(\delta\twh^{ub}_\lft)$
&& $-0.09\pm .16$ && --- & \cr
& $\sum_1$ && --- &&$+0.005\pm .027$ & \cr
& $\Re(\delta\twh^{ub}_\rht)$ && --- && --- & \cr
& $\Re(\delta\twh^{cd}_\lft)$ && $+0.11 \pm .98$ && --- & \cr
& $\Re(\delta\twh^{cd}_\rht)$ && --- && --- & \cr
& $\Re(\delta\twh^{cs}_\lft)$ && $+0.022\pm .20$ && --- & \cr
& $\Re(\delta\twh^{cs}_\rht)$ && $+0.022\pm .20$ && --- & \cr
& $\Re(\delta\twh^{cb}_\lft)$ && $+0.5\pm 4.6$ && --- & \cr
& $\sum_2$ && --- && $+0.11\pm 0.98$ & \cr
& $\Re(\delta\twh^{cb}_\rht)$ && --- && ---  & \cr
& $\delta\twg^{dd}_\lft$ && $+0.0016\pm .0015$ && $+0.003\pm .012$ & \cr
& $\delta\twg^{dd}_\rht$ && $+0.0037\pm .0038$ && $+0.007\pm .015$ & \cr
& $\delta\twg^{uu}_\lft$ && $-0.0003\pm .0018$ && $-0.002\pm 0.014$ & \cr
& $\delta\twg^{uu}_\rht$ && $+0.0032\pm .0032$ && $-0.003 \pm .010$ & \cr
& $\delta\twg^{ss}_\lft$ && $-0.0009\pm .0017$ && $-0.003\pm .015$ & \cr
& $\delta\twg^{ss}_\rht$ && $-0.0052\pm .00095$ && $+0.002\pm .085$ & \cr
& $\delta\twg^{cc}_\lft$ && $-0.0011\pm .0021$ && $+0.001\pm .018$ & \cr
& $\delta\twg^{cc}_\rht$ && $+0.0028\pm .0047$ && $+0.009\pm .029$ & \cr
& $\delta\twg^{bb}_\lft$ && $-0.0005\pm .0016$ && $-0.0015\pm .0094$ & \cr
& $\delta\twg^{bb}_\rht$ && $+0.0019\pm .0083$ && $0.013\pm .054$ & \cr
& $\delta\twg^{\nu_e \nu_e}_\lft$ && $-0.0048\pm .0052$ && --- & \cr
& $\delta\twg^{\nu_\mu \nu_\mu}_\lft$ && $-0.0021\pm .0027$ &&
	$+0.0023 \pm .0097$ & \cr
& $\delta\twg^{\nu_\tau \nu_\tau}_\lft$ && $-0.0048\pm .0052$ &&--- & \cr
& $\delta\twg^{\nu_e \nu_e}_\lft + \delta\twg^{\nu_\tau \nu_\tau}_\lft$ &&
	--- && $-0.004 \pm .033$ & \cr
& $\delta\twg^{ee}_\lft$ && $-0.00029\pm .00043$ && $-0.0001\pm .0032$ & \cr
& $\delta\twg^{ee}_\rht$ && $-0.00014\pm .00050$ && $+0.0001\pm .0030$ & \cr
& $\delta\twg^{\mu\mu}_\lft$ && $+0.0040\pm .0051$ && $+0.005\pm .032$ & \cr
& $\delta\twg^{\mu\mu}_\rht$ && $-0.0003\pm .0047$ && $+0.001\pm .028$ & \cr
& $\delta\twg^{\tau\tau}_\lft$ && $-0.0021\pm .0032$ && $\; 0.000\pm .022$ &
\cr
& $\delta\twg^{\tau\tau}_\rht$ && $-0.0034\pm .0028$ &&
	$-0.0015\pm .019$ & \cr
height2pt&\omit&&\omit&&\omit&\cr \noalign{\hrule}}}
$$
\centerline{\bf Table (VI)}
\endinsert

\ref\rbdecays{M. Gronau and S. Wakaizumi, \prl{68}{92}{1814}; W.-S. Hou and
D. Wyler, \plb{292}{92}{364}.}
In the individual fit, three parameters remain unconstrained --
$\Re(\delta\twh^{cd}_\rht)$, $\Re(\delta\twh^{cb}_\rht)$ and
$\Re(\delta\twh^{ub}_\rht)$. (As explained in the text, there are in fact
(weak) constraints on $\Re(\delta\twh^{cd}_\rht)$, but they appear only at
quadratic order in this parameter, and so could be cancelled by
higher-dimension operators.) In addition, the constraints on
$\Re(\delta\twh^{cd}_\lft)$, $\Re(\delta\twh^{cb}_\lft)$, $\Re(\delta
\twh^{ub}_\lft)$ and $\Im(\delta \twh^{ub}_\lft)$ are quite weak. One
physical consequence of this observation is that the chirality of
the $b\to c$ and $b\to u$ transitions has really not been tested. In other
words, this is an ideal area to look for new physics. In fact, models have
recently been constructed \rbdecays\ in which $B$-decays are predominantly
right-handed.

In the simultaneous fit, three combinations of the nine parameters $\delta
\twg^{\nu_e\nu_e}_\lft$, $\delta\twg^{\nu_\tau\nu_\tau}_\lft$,
$\Re(\delta\twh^{us}_{\lft})$, $\Re(\delta\twh^{ub}_{\lft})$,
$\Im(\delta\twh^{ub}_{\lft})$, $\Re(\delta\twh^{cs}_{\lft,\rht})$,
$\Re(\delta\twh^{cd}_{\lft})$ and $\Re(\delta\twh^{cb}_{\lft})$ are also
unconstrained, since only three independent combinations enter into
well-measured observables. Apart from these exceptional cases, all the
other parameters are well bounded. In the individual fit, most of the
parameters are constrained at better than the 1\% level. In the
simultaneous fit the limits are only slightly weakened, to about 2-3\% for
most new-physics parameters. (Note that, although $S$, $T$ and $U$ appear
to be poorly constrained, their constraints in fact represent strong bounds
on new physics, since a factor of $\alpha$ has been divided out in their
definitions (see eq.~\studefs).)

The only case in which there is a discrepancy with the standard model is in
$\Delta_\tau$, which differs from zero by about 2$\sigma$. This is a
well-known problem, which is due to the apparent breaking of weak
universality in $\tau$ decays \doug. Many people remain skeptical that this
really is a sign of new physics, suggesting instead that the cause of the
problem is probably an incorrect measurement of the $\tau$ mass. However,
recent re-measurements
\ref\taumass{For a review, see H. Marsiske, SLAC-PUB-5977 (1992).}
of $m_\tau$ have not caused the effect to disappear \taumass.

\topinsert
\noindent {\eightrm Caption for Table (VI): Results of the fits of the
new-physics parameters to the data of Table (V). $\ss \sum_1$ and $\ss
\sum_2$ are defined as: $\ss \sum_1 \equiv \Re( \delta \twh_\lft^{us}) +
\left[\Re(V_{ub}) \Re(\delta \twh_\lft^{ub}) + \Im(V_{ub}) \Im(\delta
\twh_\lft^{ub}) \right]/ \vert V_{us}\vert $ and $\ss \sum_2 \equiv
\Re(\delta \twh_\lft^{cd}) + \vert V_{cs} \vert \Re(\delta \twh_\lft^{cs} +
\delta \twh_\rht^{cs})/\vert V_{cd} \vert + |V_{cb}| \Re(\delta
\twh_\lft^{cb})/\vert V_{cd} \vert $.}
\endinsert

Note also that, as expected, the $\Im(\delta \twh_\lft^{ij})$ remain
virtually unconstrained. Such operators can contribute to CP-violating
processes, and could very well be observed in studies of CP violation in
the $B$ system. This underlines the significance of CP-violating observables
as potent probes for new physics.

One of the interesting conclusions to be drawn from the results of the
simultaneous fit is that, although many of the hadronic charged-current
experiments are extremely precise, there is still a great deal of room for
new physics in this sector -- many of the $\delta\twh$'s are only weakly
constrained, if at all. This is due to the fact that, in the standard
model, the values of the CKM matrix elements are not predicted. Hence, the
only constraints we have are due to the unitarity of the CKM matrix. And,
since only the magnitudes of the CKM matrix elements involving the $u$- and
$c$-quarks have been measured, the only two constraints which can be used
are the normalization of the first two rows (eqs.~\firstrowunitarity\ and
\secondrowunitarity). This is not very restrictive. There are, however, a
number of ways to constrain new physics in the hadronic charged-current
sector more strongly. First, it would be useful to remeasure the known CKM
matrix elements, but using methods sensitive to different combinations of
the new-physics parameters. Second, measurements of CP violation in the $B$
system allow one to obtain the imaginary parts of the elements of the CKM
matrix. Using the unitarity of the CKM matrix, these can be used to extract
the magnitudes of the CKM matrix elements, which will help in
overconstraining the matrix and putting limits on new physics. Finally,
using the fact that the columns of the CKM matrix are orthonormal, accurate
measurements of the CKM matrix elements involving the top quark can be used
to constrain different combinations of the $\delta\twh$'s.

Since we have performed this analysis in a model-independent fashion, the
constraints presented here must hold for all physics beyond the standard
model, provided only that it agree on the low-energy particle content, and
that dimension six operators may be neglected. In any particular model of
new physics, one must simply compute the above new-physics parameters in
terms of the parameters of the model. The constraints can then be read off
from the Tables. As an example of how this works, we consider in the
following section the case of the mixing of ordinary and exotic fermions,
first studied in Refs.~\exotic\ and \nardi. Before doing so, however, we
briefly turn to possible constraints from loop-level processes.

\subsection{Loop Constraints}

In the previous subsections we found the constraints which current
tree-level experimental data put on our new-physics parameters. The bounds
on most of these parameters are quite stringent, though there are certain
new couplings which are constrained only weakly, if at all. In this
subsection we consider the limits which apply to the new-physics parameters
due to loop-level processes. For a given observable, we have already argued
that in general there can be cancellations between the loop-level
contributions of certain effective interactions, and the tree-level
contributions of other operators. The only possible case where a reasonably
reliable bound can be obtained is when the constraint on a new parameter
from such loop-induced processes is so strong that cancellations with the
higher-dimension operators would require significant fine-tuning. For this
reason we need only consider the loop-level contributions to observables
which are {\it extremely} well-measured.

Another reason to consider loop-level bounds is that, up to now, almost all
CP-violating operators have remained essentially unconstrained. (The only
exception are the constraints on the flavour-diagonal $d_{\lft,\rht}$'s
from $edm$'s.) Since the only observation of CP violation to date is the
parameter $\epsilon_{\sss K}$ in the Kaon system, which is a loop-level
process, it is interesting to investigate the implications this measurement
might have for CP-violating new-physics parameters.

We will therefore consider the contributions of the new-physics parameters
to four classes of loop-level observables: anomalous magnetic moments,
$edm$'s, neutral-meson mixing, and $\epsilon_{\sss K}$. It must be kept in
mind that the only reliable constraints from this analysis are those which
are extremely stringent -- weak bounds are suspect due to the possibility
of cancellation with effects from other operators. (This last point is
frequently glossed over when only one effective interaction is considered
at a time.) For the purposes of argument we will arbitrarily consider here
any bound which is greater than $10^{-3}$ to be too weak to preclude its
cancellation by other operators.

\topic{Anomalous Magnetic Moments}

\ref\einhorn{For bounds on effective operators due to their loop-level
contributions to anomalous magnetic moments, see also C. Arzt, M.B. Einhorn
and J. Wudka, preprint NSF-ITP-92-122 (unpublished).}
Although the measurements of $a_e$ and $a_\mu$ are extremely precise, they
turn out to be sensitive only to comparatively few of our effective
interactions \einhorn. The reason for this is fairly easy to see. Consider
first a dimension-four fermion-gauge boson interaction, such as $\delta
g_{\lft,\rht}^{ff}$ of eq.~\bffterms. These can contribute to a fermion
anomalous magnetic moment through Feynman graphs such as that of
\fig\magmomgraph
Fig.~\magmomgraph. An order-of-magnitude estimate for the contribution to
$a_i$, $i=e,\mu$ due to this graph is:
\eq
\delta a_i \sim \delta g^{ii}_{\lft,\rht} \; \left( {\alpha \over 4 \pi
\sw^2\cw^2} \right) \; \left( {m_i \over \mw} \right) \;
F\left( {m_i\over \mw} \right).
\eeq
Here the second term on the right-hand-side is the usual loop factor, and
the third term arises because $a_i$ is defined relative to the
corresponding Bohr magneton, $\mu_{\ssb_i} = e_i/2 m_i$. Largely due to the
suppression by the small electron or muon mass, the product of these two
terms is already very small: $\sim 2 \times 10^{-8}$ for the electron, and
$\sim 3 \times 10^{-6}$ for the muon. As a consequence, no useful bound on
the couplings $\delta g^{ee}_{\lft,\rht}$ or $\delta
g^{\mu\mu}_{\lft,\rht}$ is possible unless the remaining function,
$F(x_i)$, of the small mass ratio $x_i = m_i/\mw$ is not itself suppressed
by a power of $x_i$ for small $x_i$.

For the dimension-four interactions, helicity-conservation along the
fermion line shows that $F(x_i)$ is always suppressed by at least one power
of $x_i$, and so no useful bound for these operators is obtained in this
way. For the same reason current anomalous-magnetic-moment experiments are
not yet sensitive to ordinary SM weak-interaction effects.

The same need not be true for the dimension-five interactions. The only
effective couplings whose contribution to $a_e$ and $a_\mu$ is not further
suppressed by light fermion masses, together with the order-of-magnitude of
their corresponding bounds, are:
\eq \eqalign{
n^{ee}_{\lft,\rht}, c_\lft^{\nu_e e} &\lsim 5 \times 10^{-3} \cr
n^{\mu\mu}_{\lft,\rht}, c_\lft^{\nu_\mu \mu} &\lsim 0.08 .\cr}
\eeq

Given the ever-present possibility of cancellations that is inherent in
these loop-generated bounds, we do not consider these limits to be
particularly severe.

\topic{Electric Dipole Moments}

Some light-fermion $edm$'s are also extremely well bounded, so one might
expect these to also give significant bounds for operators which contribute
at the loop level. This turns out to be true, but only for those
comparatively few operators which can contribute to the electron or $u$-
and $d$-quark $edm$'s unsuppressed by small fermion
\ref\wbgthreeg{S. Weinberg, \prl{63}{89}{2333}.}
masses.\foot\threeg{Because of our neglect of gluon operators, we are
unable to consider some loop contributions to the neutron $\ss edm$, such
as those of Ref.~\wbgthreeg.} We consider here each type of effective
coupling separately.

The analysis for dimension-four interactions follows closely that for the
anomalous magnetic moments of the previous section. Helicity conservation
always implies a suppression by at least one factor of a light fermion
mass. The only bounds which we can infer in this way are:
\eq \eqalign{
\Im\left[ \delta g^{ee}_{\lft,\rht} \right], ~~\Im\left[ \delta
h_\lft^{\nu_e e} \right]
&\lsim 4 \times 10^{-3} \cr
\Im\left[ \delta g^{uu}_{\lft,\rht} \right], ~~\Im\left[ \delta
g^{dd}_{\lft,\rht} \right], ~~\Im\left[ \delta h_\lft^{ud} \right] &\lsim
0.08. \cr}
\eeq
The suppression of flavour changes in the SM by factors of $\lambda =
\sin\theta_c \simeq 0.2$ precludes obtaining significant bounds for other
quark operators -- \eg\ we find $\Im\left[ \delta h_\lft^{us} \right],
{}~~\Im\left[ \delta h_\lft^{cd} \right] \lsim 0.4$.

At dimension five there are three kinds of effective couplings:
$d_{\lft,\rht}^{ij}$, $c_{\lft,\rht}^{ij}$ and $n_{\lft,\rht}^{ij}$. It
turns out that no new bounds arise  for $d_{\lft,\rht}^{ij}$ from
loop-level $edm$'s, however. These operators might have  potentially
contributed through the Feynman diagram of
\fig\edmdfigs
Fig.~\edmdfigs, but the following argument shows that this graph leads to
no new limits. There are two cases to consider, depending on whether or not
the exchanged gauge boson is a photon, a $W$ or a $Z$. For the two neutral
bosons, the absence of SM flavour-changing vertices only permits
contributions from the same operators which are already directly bounded at
tree level, such as $d^{ee}_{\lft,\rht}$, and so no new bounds are
obtained. For the graph with a $W$ boson, the result must always be
suppressed by one factor of the mass of {\it both} the external and
internal fermions, and so gives too small a result to furnish a useful
bound.

It is the remaining couplings, $c_{\lft,\rht}^{ij}$ and
$n_{\lft,\rht}^{ij}$, that can receive nontrivial constraints from
loop-generated $edm$'s. We find that the only contributions which are
unsuppressed by too many powers of light masses and mixing angles are:
\eq
\label\edmbounds
\eqalign{
\Im\left[ c^{\nu_e e}_\lft \right], ~~\Im\left[ n^{e e}_{\lft,\rht} \right]
& \lsim 3 \times 10^{-7} \cr
\Im\left[ c^{ud}_{\lft,\rht} \right], ~~\Im\left[ n^{dd}_{\lft,\rht}
\right], ~~\Im\left[ n^{uu}_{\lft,\rht} \right] &\lsim 2 \times 10^{-5} \cr
\Im\left[ c^{us}_{\lft,\rht} \right], ~~\Im\left[ c^{cd}_{\lft,\rht}
\right] & \lsim 1 \times 10^{-4} \cr
\Im\left[ c^{ub}_{\lft,\rht} \right], ~~\Im\left[ c^{td}_{\lft,\rht}
\right] & \lsim 3 \times 10^{-3}. \cr} \eeq
Again, keeping in mind the potential for cancellations, we regard only the
first three of these as being of real significance.

\topic{Neutral Meson Mass Differences}

In the standard model, the short-distance contributions to neutral meson
($M^0$) mass differences ($\Delta M_{\sss M}$) are due to the box diagrams
which mix $M^0$ and ${\overline{M^0}}$. These SM box diagrams predict
values for the mass differences in the $K$-, $B$- and $D$-meson systems
which are in agreement with the experimental values, within significant
hadronic uncertainties. Because of these uncertainties, we can regard this
agreement as only to within an order of magnitude, and so in order to
obtain estimates of the loop-level bounds on the new physics parameters, we
therefore require that their contributions to $\Delta M_{\sss M}$ be less
than those of the SM.

\medskip
\noindent
$\Delta M_{\sss K}$: \hfil\break
\indent
The SM contribution to $\Delta M_{\sss K}$ is
\label\smdeltamk
\eq
\Delta M_{\sss K}^\SM = \left( {\gf \over \sqrt{2}}\;
{\alpha \over 6\pi\,\sw^2} \; f_{\sss K}^2 B_{\sss K} M_{\sss K} \right) \,
{m_c^2\over\mw^2} \, \Re(V_{cd}^* V_{cs})^2~.
\eeq

Consider now the case in which $\delta\twh^{ud}_\lft$ replaces one of the
SM $ud$ couplings. One finds a partial failure of the GIM mechanism in the
calculation of the box diagram, leading to the appearance of a logarithmic
enhancement:
\eq
\Delta M_{\sss K} \sim \left( {\gf \over \sqrt{2}}\;
{\alpha \over 6\pi\,\sw^2} \; f_{\sss K}^2 B_{\sss K} M_{\sss K} \right) \,
{m_c^2\over\mw^2} \log\left({\mw^2\over m_c^2}\right)
\, \Re(\delta\twh_\lft^{*ud} V_{cs} V_{cd}^* V_{cs})~.
\eeq
A comparison of this contribution with that of the SM leads to the bound
\eq
\vert \Re(\delta\twh_\lft^{*ud}) \vert \lsim \vert \Re(V_{ud}) \vert /
\log\left({\mw^2\over m_c^2}\right).
\eeq
Similar constraints exist for the new-physics parameters
$\delta\twh^{us}_\lft$, $\delta\twh^{cd}_\lft$, and $\delta\twh^{cs}_\lft$,
yielding
\label\lhbounds
\eqa
\vert \Re(\delta\twh_\lft^{ud}) \vert &\lsim 0.1, \eolnn
\vert \Re(\delta\twh_\lft^{us}) \vert &\lsim 0.03, \eolnn
\vert \Re(\delta\twh_\lft^{cd}) \vert &\lsim 0.03, \eolnn
\vert \Re(\delta\twh_\lft^{cs}) \vert &\lsim 0.1. \eeol
\eeq
As these constraints are quite weak, they cannot be considered at all
reliable, due to the possibility of cancellations with the contributions
from other operators.

Consider now the case of right-handed currents, in which
$\delta\twh^{ij}_\rht$ ($i=u,c$, $j=d,s$) is the new-physics parameter in
the box diagram. We find
\eq
\Delta M_{\sss K} \sim 7.7 \left( {\gf \over \sqrt{2}}\;
{\alpha \over 6\pi\,\sw^2} \; f_{\sss K}^2 B_{\sss K} M_{\sss K} \right) \,
{m_{int}^i \, m_{ext}\over\mw^2} \, \log\left({\mw^2\over m_c^2}\right)
\, \Re(\delta\twh_\rht^{*ij} V_{ij} V_{cd}^* V_{cs})~,
\eeq
where the factor 7.7 arises from the enhancement of the $LR$ matrix element
relative to the $LL$ matrix element \lrmatrix, and $m_{int}$ ($m_{ext}$) is
the mass of an internal (external) quark. Comparing this contribution with
that of the SM (eq.~\smdeltamk), we see that there are no significant
bounds on $\Re(\delta\twh^{ud}_\rht)$ and $\Re(\delta\twh^{us}_\rht)$, due
to the smallness of $m_u$. Taking $m_{ext}\sim M_{\sss K}/2$, the bounds on
$\Re(\delta\twh^{cd}_\rht)$ and $\Re(\delta\twh^{cs}_\rht)$ are of the same
order of magnitude as their left-handed counterparts (eq.~\lhbounds).

Finally, the contributions of the parameters $c_\lft$ and $c_\rht$ to
$\Delta M_{\sss K}$ should be of the same order as those of
$\delta\twh^{ij}_{\lft,\rht}$, with an additional suppression of a factor
of $m/M$, where $m$ is a light quark mass. Since the constraints on the
$\delta\twh^{ij}_{\lft,\rht}$ are relatively weak, there are thus no limits
on the $\Re(c_{\lft,\rht})$.

\medskip
\noindent
$\Delta M_{\sss B}$: \hfil\break
\indent
In the SM, $B^0$-${\overline{B^0}}$ mixing is dominated by the $t$-quark
contribution in the box diagram:
\label\smdeltamb
\eq
\Delta M_{\sss B}^\SM = \left( {\gf \over \sqrt{2}}\;
{\alpha \over 6\pi\,\sw^2} \; f_{\sss B}^2 B_{\sss B} M_{\sss B} \right) \,
x_t f(x_t) \, \Re(V_{td}^* V_{tb})^2~,
\eeq
in which $x_t\equiv m_t^2/\mw^2$ and $f(x_t)$ takes values $\sim 1$ for 100
GeV $< m_t < 200$ GeV.

We now estimate the new-physics contributions to $\Delta M_{\sss B}$. We
begin by considering the case in which one of the internal $t$-quark lines
is replaced by a $u$-quark, and $\delta\twh^{ud}_\lft$ replaces the SM $ud$
coupling. A calculation of the box diagram yields
\eq
\Delta M_{\sss B} = \left( {\gf \over \sqrt{2}}\;
{\alpha \over 6\pi\,\sw^2} \; f_{\sss B}^2 B_{\sss B} M_{\sss B} \right) \,
x_t f'(x_t) \, \Re(\delta\twh^{*ud}_\lft V_{ub} V_{td}^* V_{tb})~,
\eeq
in which $f'(x_t)$ is a different function from that in eq.~\smdeltamb. It
also takes values $\sim 1$ for 100 GeV $< m_t < 200$ GeV. A comparison of
this contribution with that of the SM produces the constraint
\eq
\vert \Re(\delta\twh^{*ud}_\lft V_{ub}) \vert
\lsim \vert \Re(V_{td}^* V_{tb}) \vert ,
\eeq
with similar expressions for $\delta\twh^{ub}_\lft$,
$\delta\twh^{cd}_\lft$ and $\delta\twh^{cb}_\lft$. Using the estimates of
the sizes of the CKM matrix elements given in eq.~\wolfckm, this gives
\label\lhbounds
\eqa
\vert \Re(\delta\twh_\lft^{ud}) \vert &\lsim 1, \eolnn
\vert \Re(\delta\twh_\lft^{ub}) \vert &\lsim O(\lambda^3) \sim 0.01, \eolnn
\vert \Re(\delta\twh_\lft^{cd}) \vert &\lsim O(\lambda) \sim 0.2, \eolnn
\vert \Re(\delta\twh_\lft^{cb}) \vert &\lsim O(\lambda^2) \sim 0.05. \eeol
\eeq
As was the case for $\Delta M_{\sss K}$, these constraints are weak and are
therefore not reliable.

For the $\delta\twh^{ij}_\rht$ ($i=u,c$, $j=d,b$), the contributions to
$\Delta M_{\sss B}$ are suppressed relative to those of the
$\delta\twh^{ij}_\lft$ by a factor $m_{int} M_{\sss B} / m_t^2$. This leads
to virtually no bounds on the $\delta\twh^{ij}_\rht$.

Finally, the contributions from the $c_{\lft,\rht}$ to $\Delta M_{\sss B}$
are suppressed, as in the Kaon system, by a factor of $m/M$ relative to
those of the $\delta\twh^{ij}_{\lft,\rht}$, leading to no constraints.

\medskip
\noindent
$\Delta M_{\sss D}$: \hfil\break
\indent
The analysis of $D^0$-${\overline{D^0}}$ mixing proceeds completely
analogously to that in the Kaon or $B$-system. As in these two systems, no
significant bounds are obtained on any of the new-physics parameters.

\topic{$\epsilon_{\sss K}$}
In the SM, $\epsilon_{\sss K}$ is calculated from the imaginary part of the
$K^0$-${\overline{K^0}}$ mixing box diagram. There are contributions from
diagrams with two internal $c$-quarks, and with one $c$- and one $t$-quark,
but the largest effect comes from the diagram with two internal $t$-quarks:
\eq
\epsilon_{\sss K} \simeq \left( {\gf \over \sqrt{2}}\;
{\alpha \over 6\pi\,\sw^2} \; f_{\sss K}^2 B_{\sss K} M_{\sss K} \right) \,
x_t f(x_t) \, \Im(V_{td}^* V_{ts})^2~.
\eeq
Note that, according to eq.~\wolfckm, $V_{td}$ has a large imaginary piece,
and $\Im(V_{td}^* V_{ts})^2 \sim \lambda^{10}$.

Consider now the diagram in which there is one internal $c$-quark and one
$t$-quark, and where the SM $cd$ coupling is replaced by
$\delta\twh_\lft^{cd}$. A calculation of the contribution of this diagram
to $\epsilon_{\sss K}$ yields
\eq
\sim \left( {\gf \over \sqrt{2}}\;
{\alpha \over 6\pi\,\sw^2} \; f_{\sss K}^2 B_{\sss K} M_{\sss K} \right) \,
x_t g(x_t) \, \Im(\delta\twh_\lft^{*cd} V_{cs} V_{td}^* V_{ts})~,
\eeq
where $g(x_t)$ is another function which takes values $\sim 1$ for the
allowed range of $m_t$. Comparing this contribution with that of the SM
yields $\Im(\delta\twh_\lft^{*cd} V_{cs} V_{td}^* V_{ts}) \lsim
\lambda^{10}$. There are similar expressions for the parameters
$\delta\twh_\lft^{ud}$, $\delta\twh_\lft^{us}$ and $\delta\twh_\lft^{cs}$.
These lead to the constraints
\label\epsbounds
\eqa
\vert \Re(\delta\twh_\lft^{ud}) \vert,
{}~~\vert \Im(\delta\twh_\lft^{ud}) \vert,
{}~~\vert \Re(\delta\twh_\lft^{cs}) \vert,
{}~~\vert \Im(\delta\twh_\lft^{cs}) \vert
&\lsim O(\lambda^4) \sim 2 \times 10^{-3}~,\eolnn
\vert \Re(\delta\twh_\lft^{us}) \vert,
{}~~\vert \Im(\delta\twh_\lft^{us}) \vert,
{}~~\vert \Re(\delta\twh_\lft^{cd}) \vert,
{}~~\vert \Im(\delta\twh_\lft^{cd}) \vert
&\lsim O(\lambda^5) \sim 5 \times 10^{-4}~. \eeol
\eeq
The second of these two constraints is perhaps sufficiently stringent to be
taken seriously. However, one must always be aware of the possibility of
evading such bounds via (fine-tuned) cancellations with the contributions
of other operators to $\epsilon_{\sss K}$.

On the other hand, the constraints on the $\delta\twh_\rht^{ij}$ ($i=u,c$,
$j=d,s$) are much weaker. As was the case in the calculation of $\Delta
M_{\sss K}$, there is a suppression of the contribution of the
$\delta\twh_\rht^{ij}$ to $\epsilon_{\sss K}$ by a factor $\sim m_{u,c}
M_{\sss K} / m_t^2$ relative to that of the corresponding
$\delta\twh_\lft^{ij}$. Even taking into account the enhancement of the
$LR$ matrix element \lrmatrix, $\Re(\delta\twh_\rht^{ij})$ and
$\Im(\delta\twh_\rht^{ij})$ are essentially unconstrained by
$\epsilon_{\sss K}$.

Similarly, there are no constraints on the $c_{\lft,\rht}$, whose
contributions to $\epsilon_{\sss K}$ are suppressed by a factor $m/M$.
\endtopic

To summarize, the only loop-level observables which yield significant
constraints on the new-physics parameters are the CP-violating electron and
neutron $edm$'s (eq.~\edmbounds). The bounds on these parameters are
$O(10^{-7}{\hbox{-}}10^{-5})$. There are also limits of
$O(10^{-4}{\hbox{-}}10^{-3})$ on other new-physics parameters from the
CP-violating quantity $\epsilon_{\sss K}$ (eq.~\epsbounds). However, one
cannot discount the possibility of evading these latter (weaker)
constraints through cancellations with contributions of other operators.

\section{Applications to Exotic-Fermion Mixing}

In this section we illustrate how the above constraints, which have been
obtained in a model-independent way, might be applied to a specific model
of new physics. The class of models we  consider here are those
containing exotic fermions. `Ordinary' fermions are defined as transforming
in the standard way under $SU_L(2)$ (left-handed ($LH$) doublets,
right-handed ($RH$) singlets). `Exotic' fermions have non-canonical
$SU_L(2)$ assignments. Here we restrict ourselves to $LH$ singlets and/or
$RH$ doublets. These exotic fermions can mix with the ordinary fermions
and, in so doing, change the couplings of the ordinary fermions to the
$W^\pm$ and $Z^0$. (In the effective-lagrangian language, these mixings
induce new operators.) The precision measurements described in the previous
sections have been used to put constraints on these mixings \exotic,
\nardi.

Our aim here is to simply show how the formalism introduced above could be
used to bound ordinary-exotic fermion mixing. We do not wish to perform a
complete update of the limits on such mixings. As a result, we keep the
description of the mixing formalism to a minimum. Those wishing more
details should refer to Ref.~\exotic. In addition, we do not present a
complete analysis of all the constraints, preferring instead to focus on a
few illustrative examples.

We begin by considering mixing between charged particles -- neutrinos are
be treated separately below. For each type of charged particle ($Q_{em} =
-1,-{1\over 3},{2\over 3}$), we put the $LH$ and $RH$ eigenstates of both
ordinary ($O$) and exotic ($E$) fermions into a single vector
\eq
\psi_{\lft(\rht)}^0=\pmatrix{\psi^0_{\sss O}\cr
                             \psi^0_{\sss E}\cr}_{\lft(\rht)},
\eeq
in which the superscript $0$ indicates the weak-interaction basis.
Similarly, the light ($l$) and heavy ($h$) mass eigenstates can be written
\eq
\psi_{\lft(\rht)}=\pmatrix{\psi_l\cr\psi_h\cr}_{\lft(\rht)}.
\eeq
The weak and mass eigenstates are related by a unitary transformation
\eq
\psi_a^0=U_a\psi_a,
\eeq
in which $a=L,R$. The matrix $U$ can be written in block form as
\eq
U_a=\pmatrix{A_a & E_a \cr F_a & G_a \cr}.
\eeq
Although $U_a$ is unitary, $A_a$ and $F_a$ are not by themselves unitary.
These matrices describe the overlap of the light eigenstates with the
ordinary and exotic fermions, respectively. We  henceforth restrict
ourselves to the light eigenstates only.

The effects of mixing on the couplings of the light fermions can now be
seen. In the weak basis, the charged-fermion neutral current can be written
\eq
{1\over 2}J_{\sss Z}^\mu =
{\overline\psi}_{l\lft} \gamma^\mu T_{3\lft} A_\lft^\dagger A_\lft
\psi_{l\lft} + {\overline\psi}_{l\rht} \gamma^\mu T_{3\lft} F_\rht^\dagger
F_\rht \psi_{l\rht}
- {\overline\psi}_{l} \gamma^\mu Q_{em} \sin^2\theta_{\sss W} \psi_{l}.
\eeq
The important implication of the above equation is that, since neither
$A_\lft$ nor $F_\rht$ is unitary, $A_\lft^\dagger A_\lft$ and
$F_\rht^\dagger F_\rht$ are not necessarily diagonal, and thus mixing
in general induces FCNCs among the light particles. In order to avoid these
problems, the assumption which is usually made is that each ordinary left-
and right-handed fermion mixes with its own exotic partner. In this case,
$A_\lft^\dagger A_\lft$ and $F_\rht^\dagger F_\rht$ {\it are} diagonal,
thus eliminating FCNC's.

With this assumption, one can write
\eq
\left(A_a^\dagger A_a\right)_{ij}=\left(c^i_a\right)^2 \delta_{ij}~,~~~~~
\left(F_a^\dagger F_a\right)_{ij}=\left(s^i_a\right)^2 \delta_{ij}~,~~~~~
{}~~~~~~a=L,R,
\eeq
in which $\left(s^i_a\right)^2 \equiv 1 - \left(c^i_a\right)^2 \equiv
\sin^2\theta_a^i$, where $\theta_{\lft(\rht)}^i$ is the mixing angle of the
$i^{th}$ $LH$ ($RH$) ordinary fermion and its exotic partner. Therefore, in
the presence of mixing the neutral current takes the following form:
\label\mixednc
\eqa
{1\over 2}J_{\sss Z}^\mu & = \sum_i\left[
{\overline\psi}_{i\lft} \gamma^\mu
\left(T_{3\lft}^i \clisq - Q^i_{em} \sin^2\theta_{\sss W} \right)
\psi_{i\lft} \right. \eolnn
& \left.~~~~~~~~~~~~+ {\overline\psi}_{iR} \gamma^\mu \left(T_{3\lft}^i
\srisq - Q^i_{em} \sin^2\theta_{\sss W} \right) \psi_{iR} \right], \eeol
\eeq
where the sum is over the known particles. Similarly, for quarks the
charged current is
\label\mixedcc
\eq
{1\over 2}J_{\sss W}^{\mu^\dagger} =
{\overline\psi}_{u\lft} \gamma^\mu V_\lft \psi_{d\lft}
+ {\overline\psi}_{u\rht} \gamma^\mu V_\rht \psi_{d\rht},
\eeq
in which $\psi_{u\lft}$ and $\psi_{d\lft}$ are column vectors of the light
$LH$ $u$-type and $d$-type quarks, respectively. The CKM matrix $V_\lft$ is
non-unitary in the presence of mixing. It can, however, be decomposed as
\label\mixedvckm
\eq
V_{\lft ij} = c_\lft^{u_i} c_\lft^{d_j} \twi{V}_{\lft ij},
\eeq
where, as before, $\twi{V}_\lft$ is the usual (unitary) CKM matrix. The
second term in eq.~\mixedcc\ is a $RH$ charged current. Like $V_\lft$, the
apparent $RH$ CKM matrix $V_\rht$ is non-unitary, but can be written
\eq
V_{\rht ij} = s_\rht^{u_i} s_\rht^{d_j} \twi{V}_{\rht ij},
\eeq
where $\twi{V}_\rht$ is unitary.

It is now straightforward to make contact with our general formalism. To do
so we first imagine integrating out all of the heavy particles in the model
which have not yet been discovered. This produces the low-energy effective
theory with which the earlier sections of this paper have been concerned.
At tree level the removal of heavy fermions is very easy: one simply
transforms to a basis of mass eigenstates, and sets all heavy fields equal
to zero. We are led in this way to interpret eqs.~\mixednc\ and \mixedcc\
as the resulting low-energy effective weak interactions. Other terms, such
as contributions to the oblique corrections, are generated once loop
effects are included. Although these contributions can be
phenomenologically interesting, for ease of presentation we do not pursue
them here. We focus instead on the tree-level case, and accordingly set
$A$=$B$=$C$=$G$=$w$=$z$=0, which leads to $S$=$T$=$U$=0.

The key observation to now make is that eqs.~\mixednc\ and \mixedcc\ are
the expressions for the effective charged and neutral currents after
diagonalization of the fermion fields, but {\it before} shifting to the
physical parameters. They should therefore be compared to
eqs.~\ncbeforedetwiddling\ and \newcc\ (remembering that $B$ and $C$ in
these equations are zero). This gives
\label\ncrel
\eq
\delta \twg_\lft^{ii} = - T_{3\lft}^i \left(s_\lft^i\right)^2 ~,~~~~~
\delta \twg_\rht^{ii} = + T_{3\lft}^i \left(s_\rht^i\right)^2 ~,
\eeq
and
\label\ccrel
\eq
\delta \twh_\lft^{u_i d_j} = -{1\over 2} \, V_{ij}
\left[ (s_\lft^{u_i})^2
+ (s_\lft^{d_j})^2 \right]~,~~~~~
\delta \twh_\rht^{u_i d_j} = s_\rht^{u_i} s_\rht^{d_j}
\twi{V}_{\rht ij}~.
\eeq

The formalism in the case of neutrinos is somewhat different. As before, we
denote all $LH$ neutrino states as $n_\lft$ and all RH states as
$n_\rht^c$. In the weak basis there are three types of LH neutrinos --
those with $T_{3\lft}=+1/2$ ($n^0_{\sss OL}$), those with $T_{3\lft}=-1/2$
($n^0_{\sss EL}$), and those which are $SU_L(2)$-singlets ($n^0_{\sss
SL}$). These can be put into a single vector,
\eq
n_\lft^0=\pmatrix{n_{\sss OL}^0 \cr n_{\sss EL}^0 \cr n_{\sss SL}^0}.
\eeq
The mass eigenstates can be classified according to whether the neutrinos
are `light' (\ie\ essentially massless) or `heavy':
\eq
n_\lft=\pmatrix{n_{l\lft}\cr n_{h\lft}\cr}.
\eeq
The unitary transformation which relates the weak and mass bases can be
written $n_\lft^0 = U_\lft n_\lft$, in which
\eq
U_\lft=\pmatrix{A & E \cr F & G \cr H & J \cr }_\lft~.
\eeq
This matrix, then, describes the mixing of ordinary and exotic neutrinos.
Note that we do not require that each ordinary neutrino mix with only one
exotic neutrino. This is because there is no experimental evidence against
FCNC's in the neutrino sector.

In the presence of fermion mixing, the leptonic charged current takes the
form
\label\mixedlepcc
\eq
{1\over 2}J_{\sss W}^{\mu^\dagger} =
{\overline n}_\lft \gamma^\mu A_\lft^{\nu^\dagger} c_\lft^e e_\lft
+ {\overline n}_\rht^c \gamma^\mu F_\rht^{\nu^\dagger} s_\rht^e e_\rht
\eeq
in which $e_{\lft(\rht)}$ represents a column vector of charged $LH$ ($RH$)
leptons. Following the previous analysis for the charged fermions, it is
straightforward to compare eqs.~\mixedlepcc\ and \newcc\ to obtain the
relations
\label\deltwhdef
\eq
\delta\twh_\lft^{\nu_i e_a} = (A_\lft^{\nu^\dagger})_{ia} \, c_\lft^{e_a} -
\delta_{ia} ~,~~~~~
\twh_\rht^{\nu_i e_a} = (F_\rht^{\nu^\dagger})_{ia} \, s_\rht^{e_a}
{}~.
\eeq
It is useful to write $A_\lft^\nu = 1 + \delta A_\lft^\nu$, where the
new-physics contribution $\delta A_\lft^\nu$ is assumed to be small. As a
result the quantity $\Delta_a = \Re \sum_i \delta\twh_\lft^{\nu_i e_a}$
which appears in all physical observables becomes:
\eq
\Delta_a = - \hf  \left(s_\lft^{e_a}\right)^2 + \Re \sum_i
\left( \delta A_\lft^{\nu^\dagger} \right)_{ia},
\eeq
to linear order in the new physics. These represent the correspondance
between our parameters and those of the mixing formalism before the shift
to the physical parameters.

Note, however, that it is conventional to parametrize the mixing in the
neutrino sector in terms of the mixing angles $(c_\lft^{\nu_a})^2 =
(A_\lft^\nu  A_\lft^{\nu^\dagger})_{aa}$, since these are the only
quantities which arise in the rates for realistic reactions in which the
final state neutrinos are unobserved. (There is also a piece coming from
the right-handed current in eq.~\mixedlepcc, but this is of higher order in
the mixing.) Recall that this is precisely the same reason that only the
combination $\Delta_a$ appears in our expressions in earlier sections.
Linearizing $(c_\lft^{\nu_a})^2$ in the new physics we have: $\Re \sum_i
(\delta A_\lft^{\nu^\dagger})_{ia} = - \hf \left(s_\lft^{\nu_{e_a}}
\right)^2 $, yielding the following correspondence
\label\deltarel
\eq
\Delta_a = - {1\over 2} \left[ \left(s_\lft^{e_a}\right)^2
+ \left(s_\lft^{\nu_{e_a}}\right)^2 \right]
\eeq
to leading order in the square of the mixing angles.

(In the original exotic-fermion mixing paper \exotic, mixing in the
neutrino sector is not assumed to be small. However, this does not
significantly change the above analysis. If the new-physics parameters
$(\delta A_\lft^{\nu^\dagger})_{ia}$ (and hence the $\delta\twh_\lft^{\nu_i
e_a}$) are allowed to be big, then one uses eq.~\deltwhdef\ and the exact
definition of $\Delta_a$ given in eq.~\deltadef\ to again arrive at
eq.~\deltarel.)

For the neutrino neutral current, the relations between the mixing angles
and our parameters are somewhat more complicated to derive, so for the sake
of brevity we do not include them here.

There is one other point we would like to re-emphasize. In Refs.~\exotic,
\nardi, the analysis of ordinary-exotic fermion mixing was done observable
by observable. This led to a certain amount of confusion since mixing
affects not only each observable, but also such parameters as $\gf$ and
$\sw^2$ which appear in the theoretical expressions for each process. While
it is true that the analyses in these papers ultimately dealt correctly
with these problems, our formalism avoids such headaches altogether by
incorporating all new-physics effects at the level of the lagrangian.

The translation from ordinary-exotic fermion mixing angles to our
parameters has been summarized in eqs.~\deltarel, \ccrel\ and \ncrel. It is
now a simple matter to bound the mixing angles using these relations and
the constraints in Table (VI). As mentioned already, the bounds obtained in
this way are in fact weaker than those which would be obtained in a direct
fit to the mixing angles themselves. This is simply because there are more
independent parameters in our fit. In this sense our results can be
considered the most conservative bounds possible. Nevertheless, the
constraints on the mixing angles are really quite restrictive.

One minor complication is that, while our parameters are allowed to be
apriori either positive or negative, the mixing angles
$\left(s_{\lft,\rht}^i\right)^2$ are necessarily $\ge 0$. This should be
taken into account in a proper fit (see Refs.~\exotic, \nardi). Ignoring
this detail, we find the following limits at 90\% c.l.~(defined as
1.64$\sigma$):
\eqa
\Delta_{e,\mu} : & ~~~
\left(s_\lft^e\right)^2,\left(s_\lft^{\nu_e}\right)^2 < 0.016 \eolnn
& ~~~ \left(s_\lft^\mu\right)^2,\left(s_\lft^{\nu_\mu}\right)^2 <
0.012 \eolnn
\delta \twh_\lft^{ud} : & ~~~
\left(s_\lft^u\right)^2,\left(s_\lft^d\right)^2 < 0.02 \eolnn
\delta g_{\lft,\rht}^{ii} : & ~~~ \left(s_\rht^e\right)^2 < 0.01 \eolnn
& ~~~ \left(s_\rht^\mu\right)^2 < 0.09 \eolnn
& ~~~ \left(s_\rht^u\right)^2 < 0.03 \eolnn
& ~~~ \left(s_\rht^d\right)^2 < 0.05 \eolnn
& ~~~ \left(s_\lft^s\right)^2 < 0.05 \eolnn
& ~~~ \left(s_\lft^b\right)^2 < 0.03 ~, \eeol
\eeq
where the numbers have been obtained using the constraints from the
simultaneous fit (Table (VI)), and we have indicated which of our
parameters has been used to obtain the limit on the mixing angle. We have
not presented all the limits since our purpose was simply to show how our
results could be used to bound a specific model of new physics. A
comparison of the above numbers with those found in eq.~\nardi\ reveals
that the bounds obtained in this way are very similar to those found in a
fit to the mixing angles themselves. Of course, our analysis applies to
{\it all} models of new physics, not just the particular case of the mixing
of ordinary and exotic fermions.

\section{Conclusions}

New physics can manifest itself in one of two ways -- either new particles
will be discovered, or their presence will be detected via the virtual
effects they induce in low-energy processes. Until the next generation of
accelerators comes on line, we will probably have to content ourselves with
the second possibility. Given this, it is fruitful to study, in as
model-independent a manner as possible, the various virtual effects which
might be detectable using today's colliders.

A useful framework in which to perform such an analysis is using an
effective lagrangian. It has the principal merit of being completely
systematic, so that one is sure that no potential low-energy effects of new
physics are accidentally missed. Here the new-physics operators can be
classified according to their dimension, \ie\ the number of powers of $1/M$
which are required by dimensional analysis. One subset of operators which
has already been studied consists of the new-physics contributions to
gauge-boson propagators -- the `oblique' corrections. In this paper we have
extended the analysis to include all operators of the same dimension,
including corrections to the $Z{\ol{f}}f$ and $W{\ol{f}}f$ vertices.

We have developed a formalism which can deal with all these new operators
in a relatively straightforward way. One of the main effects of new physics
is to shift the relationships between the input parameters to the standard
model---$\alpha$, $\gf$ and $\Mz$---and the measured values of these
quantities. We take these shifts into account in the lagrangian itself.
Having done this, it is no longer necessary to separately adjust each
observable as it is considered. This facilitates the calculation, and
removes a considerable amount of confusion from the analysis.

We find a great many operators which satisfy the following three
assumptions: (i) we concentrate on the electroweak sector alone; (ii) we
only keep interactions with dimension $\le 5$, both CP-preserving and
CP-violating; (iii) we consider only those operators which contribute at
tree level in well-measured processes. Despite the large number of
operators, most of these are well constrained by the current experimental
data. There are a few interesting exceptions:
\item{1.} Of the FCNC operators, dimension five terms of the form
${\ol{f}}\sigma^{\mu\nu} f' Z_{\mu\nu}$ are quite poorly bounded -- their
effects could easily be visible at LEP.
\item{2.} With a few exceptions (see eq.~\edmbounds), the constraints on
the other dimension-five operators --- the flavour-conserving neutral
current couplings, ${\ol{f}}\sigma^{\mu\nu} f Z_{\mu\nu}$, and the charged
current, ${\ol{f}}\sigma^{\mu\nu} f' W_{\mu\nu}$ --- are also quite weak.
\item{3.} There is still a great deal of room for new physics in the
hadronic charged-current sector. For example, the chirality of $b$ decays
has not yet been tested. There are a number of ways to constrain new
physics in this area -- remeasurements of the known CKM matrix elements
using different methods, CP violation in the $B$ system, and measurements
of the CKM matrix elements involving the $t$ quark are a few examples.
\item{4.} Universality violation in $\tau$ decays remains a puzzle.
\item{5.} Most CP-violating operators are virtually unconstrained. Their
effects might well be seen when CP violation in the $B$ system is studied.

All other operators are well constrained, particularly the neutral current
couplings, most to at least the 2-3\% level. The utility of such a global,
model-independent analysis is that it presents limits which must be
satisfied by {\it all} models of new physics. For any particular choice of
physics beyond the standard model, it is only necessary to compute the
coefficients of the above operators in terms of the parameters of that
particular model. The constraints presented in this paper then serve to
constrain that model. As an example of how this works, we considered mixing
of ordinary and exotic fermions. For this case we have shown that, indeed,
our constraints reproduce the results of previous analyses, but frequently
in a simpler way. It is our hope that this work will serve as a guide to
future model builders.

\bigskip
\centerline{\bf Acknowledgments}
\bigskip
S.G. and D.L. gratefully acknowledge helpful conversations and
communications with Paul Langacker, and thank Paul Turcotte for supplying
standard model values for $g^2_{\sss L}$ and $g^2_{\sss R}$. This research
was partially funded by funds from the N.S.E.R.C.\ of Canada and les Fonds
F.C.A.R.\ du Qu\'ebec.

\vfill\eject
\centerline{Figure Captions}
\bigskip

\topic{Figure (1)}
The Feynman diagram through which an anomalous fermion--$Z$-boson coupling
(blob) can contribute at one loop to the anomalous magnetic moment of the
electron or muon.

\topic{Figure (2)}
The Feynman diagram through which an effective fermion--photon coupling
(blob) can contribute at one loop to a light-quark or electron electric
dipole moment.
\endtopic

\listrefs

\bye